\documentclass[superscriptaddress,prx,notitlepage,longbibliography]{revtex4-1}
\pdfoutput=1
\usepackage{graphicx}
\usepackage{amsmath}
\usepackage{amssymb}
\usepackage{amsthm}
\usepackage{amsfonts}
\usepackage{comment}
\usepackage{placeins}
\usepackage[caption=false]{subfig}
\usepackage[colorlinks]{hyperref}
\usepackage[all]{hypcap}
\usepackage{tikz}
\usepackage{verbatim}
\usepackage{subfig}
\usetikzlibrary{arrows}
\usepackage{units}
\usepackage{soul}
\usepackage{algorithm}
\usepackage{algorithmic}
\usepackage{url}

\usetikzlibrary{external}
\newtheorem{theorem}{Theorem}
\newtheorem{lemma}[theorem]{Lemma}

\newcommand{\eq}[1]{Eq.~\hyperref[eq:#1]{(\ref*{eq:#1})}}
\renewcommand{\sec}[1]{\hyperref[sec:#1]{Section~\ref*{sec:#1}}}
\DeclareRobustCommand{\app}[1]{\hyperref[app:#1]{Appendix~\ref*{app:#1}}}
\newcommand{\tab}[1]{\hyperref[tab:#1]{Table~\ref*{tab:#1}}}
\newcommand{\fig}[1]{\hyperref[fig:#1]{Figure~\ref*{fig:#1}}}
\newcommand{\figa}[2]{\hyperref[fig:#1]{Figure~\ref*{fig:#1}#2}}
\newcommand{\figx}[2]{\hyperref[fig:#1]{Figure~\ref*{fig:#1}(#2)}}
\newcommand{\thm}[1]{\hyperref[thm:#1]{Theorem~\ref*{thm:#1}}}
\newcommand{\lem}[1]{\hyperref[lem:#1]{Lemma~\ref*{lem:#1}}}
\newcommand{\cor}[1]{\hyperref[cor:#1]{Corollary~\ref*{cor:#1}}}
\newcommand{\defn}[1]{\hyperref[def:#1]{Definition~\ref*{def:#1}}}
\newcommand{\alg}[1]{\hyperref[alg:#1]{Algorithm~\ref*{alg:#1}}}
\newcommand{\clm}[1]{\hyperref[claim:#1]{Claim~\ref*{claim:#1}}}

\newcommand{\sel}{\textsc{select} }
\newcommand{\prep}{\textsc{prepare} }

\def\avg#1{\mathinner{\langle{#1}\rangle}}
\def\bra#1{\mathinner{\langle{#1}|}}
\def\ket#1{\mathinner{|{#1}\rangle}}

\newcommand{\proj}[1]{\ket{#1}\!\!\bra{#1}}

\input{Qcircuit}

\begin{document}

\title{Encoding Electronic Spectra in Quantum Circuits with Linear T Complexity}

\date{\today}
\author{Ryan Babbush}
\email[Corresponding author: ]{babbush@google.com}
\affiliation{Google Inc., Venice, CA 90291, United States}
\author{Craig Gidney}
\affiliation{Google Inc., Santa Barbara, CA 93117, United States}
\author{Dominic W. Berry}
\affiliation{Department of Physics and Astronomy, Macquarie University, Sydney, NSW 2109, Australia}
\author{Nathan Wiebe}
\affiliation{Microsoft Research, Redmond, WA 98052, United States}
\author{Jarrod McClean}
\affiliation{Google Inc., Venice, CA 90291, United States}
\author{Alexandru Paler}
\affiliation{Institute for Integrated Circuits, Linz Institute of Technology, 4040 Linz, Austria}
\author{Austin Fowler}
\affiliation{Google Inc., Santa Barbara, CA 93117, United States}
\author{Hartmut Neven}
\affiliation{Google Inc., Venice, CA 90291, United States}

\begin{abstract}
We construct quantum circuits which exactly encode the spectra of correlated electron models up to errors from rotation synthesis. By invoking these circuits as oracles within the recently introduced ``qubitization'' framework, one can use quantum phase estimation to sample states in the Hamiltonian eigenbasis with optimal query complexity ${\cal O}(\lambda / \epsilon)$ where $\lambda$ is an absolute sum of Hamiltonian coefficients and $\epsilon$ is target precision. For both the Hubbard model and electronic structure Hamiltonian in a second quantized basis diagonalizing the Coulomb operator, our circuits have T gate complexity ${\cal O}({N + \log (1/\epsilon}))$ where $N$ is number of orbitals in the basis. This enables sampling in the eigenbasis of electronic structure Hamiltonians with T complexity ${\cal O}(N^3 /\epsilon + N^2 \log(1/\epsilon)/\epsilon)$. Compared to prior approaches, our algorithms are asymptotically more efficient in gate complexity and require fewer T gates near the classically intractable regime. Compiling to surface code fault-tolerant gates and assuming per gate error rates of one part in a thousand reveals that one can error correct phase estimation on interesting instances of these problems beyond the current capabilities of classical methods using only about a million superconducting qubits in a matter of hours.
\end{abstract}
\maketitle

\section{Introduction}

The ubiquitous problem of predicting material properties and chemical reactions from \emph{ab initio} quantum mechanics is among the most anticipated applications of quantum computing. The limitation of most classical algorithms for modeling the physics of superconductivity and molecular electronic structure arises from the seemingly exponential growth of entanglement required to accurately capture strong correlation in systems of interacting electrons. This apparent classical intractability was referenced by Feynman in his seminal work as a key motivation for why we need quantum computers \cite{Feynman1982,Feynman1986}. Fourteen years later, Lloyd formalized the concept of a universal quantum simulator \cite{Lloyd1996} and demonstrated an extension for treating systems of interacting electrons in second quantization \cite{Abrams1997}.

Since then, most work developing fermionic quantum simulation methods has focused on time evolution as a means of estimating Hamiltonian spectra and preparing eigenstates \cite{Abrams1999} via the quantum phase estimation algorithm \cite{Kitaev1995}. Beginning with the proposal of \cite{Aspuru-Guzik2005}, the idea that one should use phase estimation and adiabatic state preparation \cite{Farhi2001,Wu2002,BabbushAQChem} to extract quantum chemistry ground state energies became especially popular. More recently, experimental demonstrations \cite{Peruzzo2013,OMalley2016,Kandala2017,Siddiqi2017,Hempel2018} have focused research on the development of variational algorithms \cite{McClean2015,McClean2018} which are often \cite{Wecker2015a,Kivlichan2017}, but not always \cite{Romero2017,Dallaire-Demers2018}, inspired by time-evolution primitives.

Performing quantum phase estimation to sample Hamiltonian spectra requires a quantum circuit to implement an operation ${\cal W}(H)$ which has eigenvalues that are a known (and efficient-to-compute) function of the eigenvalues of $H$.
Most past work has analyzed phase estimation of circuits corresponding to dynamical Hamiltonian simulation, i.e.\ ${\cal W}(H) \approx e^{-i H \tau}$ for some duration $\tau$ \cite{Kitaev1995}. We denote by $f$ the cost of implementing a primitive circuit that is repeated to realize ${\cal W}(H)$; e.g., a Trotter step \cite{Suzuki1993} or Taylor series segment \cite{Berry2015}. We further define $g(\epsilon)$ as the number of times that one must repeat that primitive to ensure that error in the spectrum of $H$ encoded in the eigenphases of ${\cal W}(H)$ is at most ${\cal O}(\epsilon)$. Then, the cost of phase estimation is bounded by
\begin{equation}
\label{eq:pea_cost}
{\cal O}\left(\frac{f \cdot g\left(\epsilon\right)}{\epsilon} \left \| {\cal W}' (H)\right \|^{-1} \right)
\end{equation}
where $\| \cdot \|$ denotes the spectral norm and the operation ${\cal W}'(H)$ is that where we have taken the derivative of the function of the eigenvalues. That is, ${\cal W}'(H)$ has eigenvalues which are a function of the eigenvalues of $H$, and that function is the derivative of the function which gives the eigenvalues of ${\cal W}(H)$.
For the case of dynamical time-evolution
\begin{equation}
{\cal W}(H) \approx e^{-i H \tau}
\qquad \qquad
\left \|  {\cal W}'(H)\right \|^{-1} = \left \|-i \tau e^{-i H \tau} \right \|^{-1} = \frac{1}{\tau},
\end{equation}
implying that the cost of phase estimation is ${\cal O}(f \cdot g (\epsilon)/(\epsilon \tau))$ in this context.

Modern Hamiltonian simulation methods such as the signal processing algorithm \cite{Low2017} and qubitization \cite{Low2016} have achieved the provably optimal scaling that is possible for $g(\epsilon)$ within a query model that aims to synthesize $e^{-i H \tau}$:
\begin{equation}
\label{eq:lower_bound}
{\cal O}\left(\lambda \tau + \frac{\log\left(1/\epsilon\right)}{\log\log\left(1/\epsilon\right)}\right).
\end{equation}
The definition of $\lambda$ depends on the query model; e.g.\ in models for which the Hamiltonian is given as a weighted sum of unitaries, $\lambda$ is the sum of the absolute values of the weightings \cite{Low2016}. However, ${\cal W}(H) \approx e^{-i H \tau }$ is not the only encoding from which one may sample the spectrum of $H$ via phase estimation. Recent papers \cite{Berry2018,Poulin2017} have advocated performing phase estimation on a quantum walk operator corresponding to ${\cal W}(H) = e^{i \arccos(H / \lambda)}$ which can be realized exactly as a quantum circuit without approximations beyond those required for rotation synthesis \cite{Low2016}.
(This quantum walk operator also produces eigenvalues corresponding to $e^{-i \arccos(H / \lambda)}$, but we will ignore those for simplicity of the exposition here.)
Even within a black box query model one can achieve $g(\epsilon) = {\cal O}(1)$ if the goal is to implement $e^{i \arccos(H / \lambda)}$ rather than $e^{-i H / \lambda}$. Performing phase estimation on either circuit would provide the same information since the spectra of these operators are isomorphic.
In this case, the cost of phase estimation is ${\cal O}(f \cdot \lambda / \epsilon)$ which follows from \eq{pea_cost}, $g(\epsilon) = {\cal O}(1)$, and
\begin{equation}
{\cal W}(H) = e^{i \arccos(H / \lambda)}
\qquad \qquad
\left \|  {\cal W}'(H)\right\|^{-1}  = \left \| \frac{-i e^{i \arccos(H / \lambda)}}{\sqrt{\lambda^2 - H^2}} \right \|^{-1} \leq \lambda.
\end{equation}
This work develops methods with such scaling for modeling systems of correlated electrons so that $f = {\cal O}(N + \log(1/\epsilon))$.

Our focus on synthesizing unitaries for phase estimation, rather than time-evolution operators that could be used as variational primitives, will result in quantum circuits with millions of gates. Hence, we shall need quantum error correction. We focus on planar nearest-neighbor coupled arrays of qubits, which are being developed experimentally by multiple groups \cite{Duns17,Fu17,Bron17}. We shall use the surface code \cite{Brav98,Denn02,Raus07,Raus07d,Fowl12f}, as it has the highest gate threshold error rate for this geometry. Within this model of fault-tolerant quantum computation, the physical resources required for error correcting a quantum circuit are mostly determined by (i) the number of logical qubits and (ii) the number of T gates.

The focus on T gates arises because applying a single T gate consumes many logical qubits and takes significantly longer than applying any other operation \cite{Fowl12h}. Preparing a $\ket{\rm T}$ state to enable a T gate requires hundreds of thousands of physical qubits. If the goal is to minimize the number of qubits required to execute an algorithm, it makes sense to prepare $\ket{\rm T}$ states serially. Typically, it also takes over 100 rounds of error detection to prepare a $\ket{\rm T}$ state, leaving plenty of time to perform Clifford gates in parallel with this preparation, meaning the execution time of the complete algorithm can be approximated as the total number of T gates multiplied by the time to prepare each $\ket{\rm T}$ state. Thus, throughout this work we focus on T complexity as the primary cost model. We note, however, that for all algorithms presented or discussed in this work, the T complexity is within logarithmic factors of the gate complexity.

\begin{table*}[b]
\begin{tabular}{c|c|c|c|c|c||c}
Year
& Reference
& Basis
& Algorithm
& Oracle T Gates
& PEA Queries
& Total T Gates\\
\hline\hline
2005
& Aspuru-Guzik \textit{et al.}
\cite{Aspuru-Guzik2005}
& Gaussians
& Trotterization
& ${\cal O}(\textrm{poly}(N / \epsilon))$
& ${\cal O}(\textrm{poly}(N / \epsilon))$
& ${\cal O}(\textrm{poly}(N / \epsilon))$\\
2010
& Whitfield \textit{et al.}
\cite{Whitfield2010}
& Gaussians
& Trotterization
& ${\cal O}\left(N^4 \log \left(1/\epsilon\right)\right)$
& ${\cal O}\left(\textrm{poly}\left(N / \epsilon\right)\right)$
& ${\cal O}\left(\textrm{poly}\left(N / \epsilon\right)\right)$ \\
2013
& Wecker \textit{et al.}
\cite{Wecker2014}
& Gaussians
& Trotterization
& ${\cal O}\left(N^4\log \left(1/\epsilon\right)\right)$
& ${\cal O}\left(N^{6} / \epsilon^{3/2}\right)$
& ${\cal O}\left(\frac{N^{10} \log \left(1/\epsilon\right)}{ \epsilon^{3/2}}\right)$\\
2014
& McClean \textit{et al.} \cite{McClean2014}
& Gaussians
& Trotterization
& ${\cal O}\left(\sim N^2\log \left(1/\epsilon\right)\right)$
& ${\cal O}\left(N^6 / \epsilon^{3/2}\right)$
& ${\cal O}\left(\sim \frac{N^8 \log \left(1/\epsilon\right)}{ \epsilon^{3/2}}\right)$\\
2014
& Poulin \textit{et al.} \cite{Poulin2014}
& Gaussians
& Trotterization
& ${\cal O}\left(N^4 \log \left(1/\epsilon\right)\right)$
& ${\cal O}\left(\sim N^2 / \epsilon^{3/2}\right)$
& ${\cal O}\left(\sim \frac{N^6 \log \left(1/\epsilon\right)}{\epsilon^{3/2}}\right)$\\
2014
& Babbush \textit{et al.} \cite{BabbushTrotter}
& Gaussians
& Trotterization
& ${\cal O}\left(N^4 \log \left(1/\epsilon\right)\right)$
& ${\cal O}\left(\sim N / \epsilon^{3/2}\right)$
& ${\cal O}\left(\sim \frac{N^5 \log \left(1/\epsilon\right)}{\epsilon^{3/2}}\right)$ \\
2015
& Babbush \textit{et al.} \cite{BabbushSparse1}
& Gaussians
& Taylorization
& $\widetilde{\cal O}(N)$
& ${\cal O}\left(\frac{N^4\log(N/\epsilon)}{\epsilon \log \log (N/\epsilon)}\right)$
& $\widetilde{\cal O}(N^5 / \epsilon)$ \\
2016
& Low \textit{et al.} \cite{Low2016}
& Gaussians
& Qubitization
& $\widetilde{\cal O}\left(N\right)$
& {\scriptsize ${\cal O}\left(\frac{N^4}{\epsilon} + \frac{\log\left(N/\epsilon\right)}{\epsilon \log \log \left(N/\epsilon\right)}\right)$}
& $\widetilde{\cal O}\left(N^5 / \epsilon\right)$ \\
2017 &
Babbush \textit{et al.} \cite{BabbushLow}
& Plane Waves
& Taylorization
& $\widetilde{\cal O}\left(N\right)$
& ${\cal O}\left(\frac{N^{8/3}\log(N/\epsilon)}{\epsilon \log \log (N/\epsilon)}\right)$
& $\widetilde{\cal O}(N^{11/3}/\epsilon)$ \\
2017 &
Berry \textit{et al.} \cite{Berry2018}
& Plane Waves
& Qubitization
& $\widetilde{\cal O}\left(N\right)$
& ${\cal O}(N^{8/3}/\epsilon)$
& $\widetilde{\cal O}(N^{11/3}/\epsilon)$ \\

2018 &
Kivlichan \textit{et al.} \cite{Kivlichan2018}
& Plane Waves
& Trotterization
& {\scriptsize ${\cal O}\left(N^2 + N \log N\log \left(1/\epsilon\right)\right)$}
& ${\cal O}\left(\sim N^{3/2} / \epsilon^{3/2}\right)$
& ${\cal O}\left(\sim N^{7/2} / \epsilon^{3/2}\right)$ \\
2018 &
This paper
& Plane Waves
& Qubitization
& ${\cal O}\left(N + \log \left(1/\epsilon\right)\right)$
& ${\cal O}\left(N^{2}/\epsilon\right)$
& ${\cal O}\left(\frac{N^{3} + N^{2}\log \left(1/\epsilon\right)}{\epsilon}\right)$ \\
\hline
\end{tabular}
\caption{Progression of lowest T complexity algorithms for implementing a phase estimation unitary encoding eigenvalues of the electronic structure Hamiltonian in second quantization. $N$ is number of spin-orbitals and $\epsilon$ is target precision. Here and throughout the paper, ${\cal O}(\cdot)$ indicates an upper bound, $\widetilde{\cal O}(\cdot)$ indicates an upper bound ignoring polylogarithmic factors, and ${\cal O}(\sim \cdot)$ indicates empirical scaling extrapolated from numerics. ``Oracle T Gates'' refers to $f$ from \eq{pea_cost} and ``PEA Queries'' refers to the rest of the expression in \eq{pea_cost}. The scalings attributed to work on general methods of Hamiltonian simulation assume that one uses the best oracles for electronic structure available at that point in time; e.g., the scaling attributed to \cite{Low2016} assumes the use of oracles from \cite{BabbushSparse1} and the scaling attributed to \cite{Berry2018} assumes the use of oracles from \cite{BabbushLow}. While absent from this table since it did not asymptotically reduce T complexity, \cite{Reiher2017} was first to explicitly compile a quantum chemistry simulation to Clifford + T gates.}
\label{tab:chemistry}
\end{table*}

We focus on the two most-studied models of correlated electrons: the Fermi-Hubbard model and the molecular electronic structure Hamiltonian. The Hubbard Hamiltonian is an approximate model of electrons interacting on a planar lattice which some believe may qualitatively capture the behavior of high-temperature superconductivity in cuprates \cite{Hubbard1963}. The molecular electronic structure Hamiltonian is a realistic model of electrons interacting via the Coulomb potential with real kinetic energy in the presence of an external potential (which usually arises from atomic nuclei), in a finite-sized basis \cite{Helgaker2002}. We shall focus on simulating the electronic structure Hamiltonian in a basis diagonalizing the Coulomb potential \cite{BabbushLow,White2017,Kivlichan2018}. For both the Hubbard model and molecular electronic structure Hamiltonian we are able to provide circuits which simulate $e^{i \arccos(H / \lambda)}$ with T complexity ${\cal O}(N + \log (1/\epsilon))$ where $N$ is the number of orbitals in a second-quantized representation of the system. In \tab{chemistry} and \tab{hubbard} we compare the T complexity of past quantum simulation methods for these problems.

In \thm{chemistry} and \thm{hubbard} we concisely state the T complexity and ancilla requirements of our approach to phase estimation for both the electronic structure Hamiltonian and Hubbard model Hamiltonian, respectively. Both of these theorems are established throughout the paper, but especially in \eq{chem_cost}, \eq{chem_ancilla}, \eq{hubbard_cost}, and \eq{hubbard_ancilla}. In addition to bounding the T complexity of our algorithms we provide explicit circuits for their construction and compile all bottleneck primitives down to surface code fault-tolerant gates (topological braiding diagrams). Therefore, the fault-tolerant aspect of our analysis goes further than prior estimates in the simulation literature \cite{Jones2012}, the most rigorous of which stopped at estimates of T complexity for Trotter based electronic structure phase estimation \cite{Reiher2017} and for a variety of techniques used to effect time-evolution of the one-dimensional Heisenberg model \cite{Childs2017}. We show that one can perform fault-tolerant phase estimation on interesting instances of both Fermi-Hubbard and molecular electronic structure beyond the capabilities of known classical algorithms using roughly one million physical qubits in the surface code assuming an architecture with two-qubit error rates of about one part in a thousand.

\begin{table*}[t]
\begin{tabular}{c|c|c|c|c|c||c}
Year
& Reference
& Algorithm
& Ancillae
& Oracle T Gates
& PEA Queries
& Total T Gates\\
\hline\hline
1997
& Abrams \textit{et al.} \cite{Abrams1997}
& Trotterization
& ${\cal O}\left(1\right)$
& ${\cal O}\left(N \log \left(1/\epsilon\right)\right)$
& ${\cal O}\left(\textrm{poly}\left(N / \epsilon\right)\right)$
& ${\cal O}\left(\textrm{poly}\left(N / \epsilon\right)\right)$\\
2015
& Wecker \textit{et al.} \cite{Wecker2015b}
& Trotterization
& ${\cal O}\left(1\right)$
& ${\cal O}\left(N \log \left(1/\epsilon\right)\right)$
& ${\cal O}\left(N^2 / \epsilon^{3/2}\right)$
& ${\cal O}\left(\frac{N^3 \log \left(1/\epsilon\right) }{ \epsilon^{3/2}}\right)$\\
2015
& Babbush \textit{et al.} \cite{BabbushSparse1}
& Taylorization
& ${\cal O}\left(\frac{\log \left(N/\epsilon\right)}{\log \log \left(N/\epsilon\right)}\right)$
& ${\cal O}\left(N \log \left(1/\epsilon\right)\right)$
& ${\cal O}\left(\frac{N \log(N/\epsilon)}{\epsilon \log \log (N/\epsilon)}\right)$
& ${\cal O}\left(\frac{N^2 \log\left(N/\epsilon\right)\log \left(1/\epsilon\right)}{\epsilon \log \log \left(N/\epsilon\right)}\right)$ \\
2016
& Low \textit{et al.} \cite{Low2016}
& Qubitization
& ${\cal O}\left(\log N\right)$
& ${\cal O}\left(N \log\left(N / \epsilon\right)\right)$
& {\scriptsize ${\cal O}\left(\frac{N}{\epsilon} + \frac{\log(N/\epsilon)}{\epsilon \log \log (N/\epsilon)}\right)$}
& ${\cal O}\left(\frac{N^2 \log(N/\epsilon)}{\epsilon}\right)$ \\
2017 &
Berry \textit{et al.} \cite{Berry2018}
& Qubitization
& ${\cal O}\left(\log N\right)$
& ${\cal O}\left(N \log\left(N / \epsilon\right)\right)$
& ${\cal O}\left(N/\epsilon\right)$
& ${\cal O}\left(\frac{N^2  \log\left(N / \epsilon\right)}{\epsilon}\right)$ \\
2017 &
Poulin \textit{et al.} \cite{Poulin2017}
& Qubitization
& ${\cal O}\left(N\right)$
& ${\cal O}\left(N  + \log \left(1/\epsilon\right)\right)$
& ${\cal O}\left(N / \epsilon\right)$
& ${\cal O}\left(\frac{N^2 + N \log \left(1/\epsilon\right)}{\epsilon}\right)$ \\
2018
& Haah \textit{et al.} \cite{Haah2018}
& Qubitization
& ${\cal O}\left(\log N\right)$
& ${\cal O}\left(N \log\left(N / \epsilon\right)\right)$
& $\widetilde{\cal O}\left(1/\epsilon\right)$
& $\widetilde{\cal O}\left(N / \epsilon\right)$  \\
2018 &
Kivlichan \textit{et al.} \cite{Kivlichan2018}
& Trotterization
& ${\cal O}\left(\log N\right)$
& {\scriptsize ${\cal O}\left(N + \log N \log \left(1/\epsilon\right)\right)$}
& ${\cal O}\left(\sim 1 / \epsilon^{3/2}\right)$
& ${\cal O}\left(\sim N / \epsilon^{3/2}\right)$ \\
2018 &
This paper
& Qubitization
& ${\cal O}\left(\log N\right)$
& ${\cal O}\left(N + \log\left(1/\epsilon\right)\right)$
& $\widetilde{\cal O}\left(1 /\epsilon\right)$
& $\widetilde{\cal O}\left(N/\epsilon\right)$ \\
\hline
\end{tabular}
\caption{Progression of lowest T complexity algorithms for implementing a unitary encoding eigenvalues of the Hubbard model for phase estimation. $N$ is number of sites and $\epsilon$ is target precision. As before, ${\cal O}(\cdot)$ indicates an upper bound, $\widetilde{\cal O}(\cdot)$ indicates an upper bound ignoring polylogarithmic factors, and ${\cal O}(\sim \cdot)$ indicates empirical scaling extrapolated from numerics. ``Oracle T Gates'' refers to $f$ from \eq{pea_cost} and ``PEA Queries'' refers to the rest of the expression in \eq{pea_cost}. The scalings attributed to work on general methods of Hamiltonian simulation assume that one uses the best oracles for the Hubbard model available at that point in time. For instance, the scaling attributed to \cite{Low2016} and \cite{Berry2018} assume that one uses the \sel oracles from \cite{BabbushSparse1}, which also work for the Hubbard model. We assume that the work of \cite{Haah2018} would use oracles from \cite{Poulin2017}. The scalings attributed to this work assume our techniques are combined with those from \cite{Haah2018} even though we do not focus on that strategy in our fault-tolerant analysis since those methods seem less effective for finite problem sizes near the classically intractable regime. Nonetheless, we discuss how our methods can be combined to provide the stated complexity in \sec{lieb}.}
\label{tab:hubbard}
\end{table*}

\begin{theorem}\label{thm:chemistry}
Consider the electronic structure Hamiltonian in a basis of $N$ spin-orbitals which diagonalizes the Coulomb operator, $H = \sum_{p,q} T(p-q)\, a^\dagger_{p} a_{q}
+ \sum_{p} U(p)\, n_{p}
+ \sum_{p \neq q} V(p-q)\, n_{p} n_{q}$ where $\{a^\dagger_p, a_q\} = \delta_{pq}$ are fermionic raising and lowering operators. Furthermore, define $\lambda = \sum_{p q} \left | T(p-q) \right |
+ \sum_p \left | U(p) \right |
+ \sum_{p \neq q} \left | V(p-q) \right |$. Then, one can perform phase estimation to sample in the eigenbasis of $H$ with an additive error of at most $\epsilon$ in the eigenvalue using circuits with a number of T gates scaling as $24 \sqrt{2}\pi  N \lambda / \epsilon + {\cal O}((\lambda / \epsilon)\log(N / \epsilon) )$ and a number of ancilla qubits scaling as $ \log(\lambda^3 N^5/\epsilon^3)  + {\cal O}(1)$.
\end{theorem}

\begin{theorem}\label{thm:hubbard}
Consider the square planar Hubbard model with periodic boundary conditions in a basis of $N$ spin-orbitals, $H = - t \sum_{\avg{p,q}, \sigma} a^\dagger_{p,\sigma} a_{q,\sigma} + (u/2) \sum_{p, \alpha \neq \beta} n_{p,\alpha} n_{p,\beta}$ where $\{a^\dagger_{p,\alpha}, a_{q,\beta}\} = \delta_{pq} \delta_{\alpha\beta}$ are fermionic raising and lowering operators and the $\avg{p,q}$ notation implies a summation over nearest-neighboring orbitals on the periodic planar lattice. Furthermore, define $\lambda =  2 N t + N u / 2$.  Then, one can perform phase estimation to sample in the eigenbasis of $H$ with an additive error of at most $\epsilon$ in the eigenvalue using circuits with a number of T gates scaling as $10 \sqrt{2}\pi  N \lambda /\epsilon + {\cal O}(\lambda \log(N / \epsilon) / \epsilon)$ and a number of ancilla qubits scaling as $\log(\lambda N^3 / \epsilon) + {\cal O}(1)$.
\end{theorem}

In \sec{simulation} we overview the simulation strategy we will use to encode and sample eigenspectra via phase estimation. \sec{walk} discusses how one can synthesize $e^{i \arccos(H/\lambda)}$ within the linear combinations of unitaries query model requiring two oracle circuits: \sel and $\textsc{prepare}$. \sec{heisenberg} introduces a particularly precise variant of phase estimation which queries \sel and \prep oracles to estimate spectra with a precision exceeding the typical Holevo variance. \sec{error} analyzes the various sources of errors which we need to consider in this algorithm and then bounds the number of times we must query \sel and \prep in order to perform phase estimation.

\sec{primitives}, \sec{chemistry} and \sec{hubbard} focus on explicit constructions of \sel and $\textsc{prepare}$. \sec{primitives} introduces important primitives for both \sel and $\textsc{prepare}$. In \sec{streaming} we describe circuits applying controlled unitaries such as the mapping $\ket{\ell}\ket{\psi} \mapsto \ket{\ell} X_\ell \ket{\psi}$ with T complexity ${\cal O}(L)$ where $L$ is the number of possible values of $\ell$. In \sec{majorana} we show how to selectively apply a Majorana fermion operator, a primitive necessary for our implementation of \textsc{select} in later sections. In \sec{qrom} we use the result of \sec{streaming} to show a particularly efficient variety of Quantum Read-Only Memory (QROM) which we use for our \prep circuit. In \sec{subsample} we describe a general technique for implementing \textsc{prepare} in a fashion that keeps $\lambda$ as small as possible.

\sec{select_chem} and \sec{prepare_chem} discuss explicit constructions of \sel and \prep circuits for the electronic structure Hamiltonian. \sec{select_hub} and \sec{prepare_hub} discuss explicit constructions of \sel and \prep circuits for the Hubbard model Hamiltonian. \sec{chem_t_complexity} and \sec{hubbard_t_complexity} focus on quantifying the number of T gates and ancillae required by the algorithms described in \sec{chemistry} and \sec{hubbard}. These sections include investigations of the finite-size magnitude of the $\lambda$ and target precisions required to implement our algorithms for interesting problems.
In \sec{lieb} we discuss how our Hubbard model simulation techniques can be combined with recent results to achieve even lower scaling based on the locality of the Hubbard Hamiltonian.

Finally, \sec{resources} discusses compilation of these routines to surface code fault-tolerant gates and estimates the physical resources required for error-correcting these algorithms under realistic assumptions about hardware. We conclude in \sec{conclusion} with an outlook on future directions for quantum simulating correlated electron models.

\section{Phase estimating spectra of Hermitian linear combinations of unitaries}
\label{sec:simulation}

The primary contribution of this paper is to demonstrate a particularly efficient method of using quantum computation to sample the spectra of correlated electron Hamiltonians. Though details of our implementation are specialized to electronic systems, our high-level simulation strategy represents a general framework for spectral estimation. While aspects of this approach were introduced recently in \cite{Berry2018,Poulin2017}, the techniques involved emerged from a history of advances in Hamiltonian simulation prominently involving Szegedy quantum walks \cite{Szegedy2004}, the ``linear combination of unitaries'' (LCU) query model \cite{Childs2012}, and the method of Hamiltonian simulation known as ``qubitization'' \cite{Low2016}.

Oracular methods of Hamiltonian simulation assume that information about a Hamiltonian is provided by querying ``oracle'' circuits \cite{Aharonov2003}. These techniques aim to reduce the number of times one must query these oracles in order to effect the intended simulation to target accuracy. The techniques in this paper implement oracles from the LCU query model introduced in \cite{Childs2012}. As the name suggests, this approach begins from the observation that any Hamiltonian can be decomposed as a linear combination of unitaries,
\begin{equation}
\label{eq:lcu}
H = \sum_{\ell = 0}^{L - 1} w_\ell H_\ell
\quad \quad \quad
\textrm{s.t.} \quad \left(w_\ell \in \mathbb{R}\right) \land \left(w_\ell \geq 0\right) \quad \quad 
H_{\ell}^{2} = \openone
\end{equation}
where $w_\ell$ are scalars and $H_\ell$ are self-inverse operators which act on qubits; e.g., $H_\ell$ could be strings of Pauli operators. The convention in this paper is that the $w_\ell$ are real and non-negative, with any phases included in the $H_\ell$.

LCU simulation techniques are formulated in terms of queries to two oracle circuits.
The first oracle circuit, the ``preparation oracle'', acts on an empty ancilla register of ${\cal O}(\log L)$ qubits and prepares a particular superposition state related to the notation of \eq{lcu},
\begin{equation}
\label{eq:prepare}
\prep \equiv \sum_{\ell = 0}^{L - 1} \sqrt{\frac{w_\ell}{\lambda}} \ket{\ell}\!\!\bra{0}
\qquad \qquad
\prep \ket{0}^{\otimes \log L} \mapsto  \sum_{\ell = 0}^{L - 1} \sqrt{\frac{w_\ell}{\lambda}} \ket{\ell}
\equiv \ket{\cal L}
\qquad \qquad
\lambda \equiv \sum_{\ell = 0}^{L - 1} w_\ell \, .
\end{equation}
The quantity $\lambda$ is the same as that in \eq{lower_bound}, and turns out to have significant ramifications for the overall algorithm complexity. The second oracle circuit we require acts on the ancilla register $\ket{\ell}$ as well as the system register $\ket{\psi}$ and directly applies one of the $H_\ell$ to the system, controlled on the ancilla register. For this reason, we refer to the ancilla register $\ket{\ell}$ as the ``selection register'' and name the second oracle the ``Hamiltonian selection oracle'',
\begin{equation}
\label{eq:select}
\sel \equiv \sum_{\ell=0}^{L} \proj{\ell}\otimes H_\ell
\qquad \qquad
\sel \ket{\ell} \ket{\psi} \mapsto \ket{\ell} H_\ell \ket{\psi}.
\end{equation}
Note that the self-inverse nature of the $H_\ell$ operators implies that they are both Hermitian and unitary, which means they can be applied directly to a quantum state.

\subsection{Encoding Spectra in Szegedy Quantum Walks using Qubitization Oracles}
\label{sec:walk}

The essential simulation primitive deployed here (a quantum walk operator based on \sel and \textsc{prepare}) was first introduced as a subroutine to the qubitization approach for Hamiltonian time-evolution \cite{Low2016}. However, the direct use of this primitive for phase estimation was first suggested more recently in \cite{Berry2018}. In \sec{heisenberg} and \app{errAnal}, we go beyond existing work and prove that so long as the eigenphase of the walk operator is bounded away from zero (so that the Hamiltonian is not frustration-free) then this algorithm can in principle learn as quickly as traditional phase estimation using the Cram\'er-Rao bound.

\begin{figure}[h]
\centering
  \resizebox{\linewidth}{!}{\includegraphics{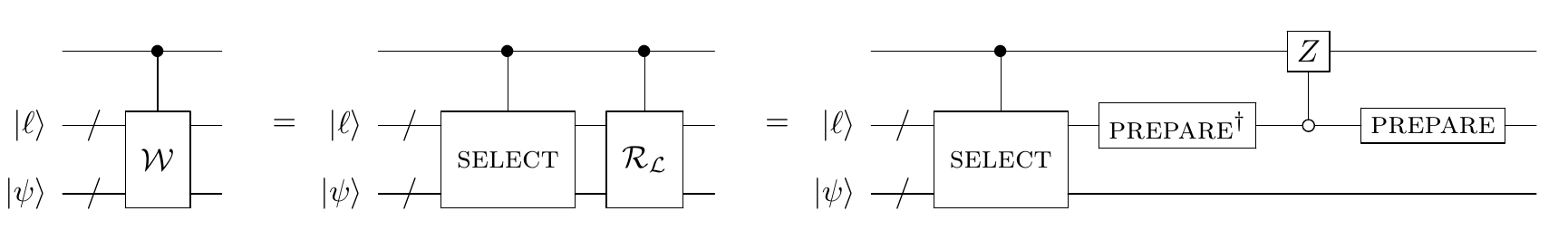}}
  \caption{\label{fig:w}
    A circuit realizing the Szegedy quantum walk operator ${\cal W}$ controlled on an ancilla qubit. The last three gates in the circuit on the right constitute the reflection ${\cal R_L}$ controlled on an ancilla. Note that the $Z$ gate with the 0-control is actually controlled on the zero state of the entire $\ket{\ell}$ register and not just a single qubit. Accordingly, implementation of that controlled-$Z$ has T complexity ${\cal O}(\log L)$ where $\log L$ is the size of the $\ket{\ell}$ register. However, that overhead is always negligible compared to the cost of the \prep and \sel operators in the constructions of this paper.}
\end{figure}

We begin our discussion with the observation that the state $\ket{\cal L}$ from \eq{prepare} encodes $H$ as a projection of $\sel$ onto $\ket{\cal L}$ 
\begin{equation}
\label{eq:diag1}
\left( \bra{\cal L}\otimes\openone \right) \sel \left(\ket{\cal L}\otimes\openone \right) = \frac{1}{\lambda} \sum_{\ell} w_\ell H_\ell  =\frac{ H}{\lambda}.
\end{equation}
This encoding is a general condition for qubitization \cite{Low2016}, but the LCU oracles \sel and \textsc{prepare}, as defined in \eq{select} and \eq{prepare}, are not necessarily the only constructions that meet this criterion; we refer to the broader family of circuits satisfying \eq{diag1} as ``qubitization oracles''. With this in mind, we discuss a walk operator ${\cal W}$ that encodes the spectrum of $H$ as a function of the eigenphases of ${\cal W}$, although the spectrum of ${\cal W}$ differs from that of the propagator $e^{- i H t}$. One such walk operator $\mathcal{W}$ may be constructed as
\begin{equation}
\label{eq:w}
\mathcal{W} \equiv {\cal R_L} \cdot \sel ,
\qquad \qquad
{\cal R_L} \equiv \left(2\proj{\cal L}\otimes \openone - \openone\right).
\end{equation}
This construction takes the form of a Szegedy walk \cite{Szegedy2004} since it is composed from a product of two reflection operations.
The operation ${\cal R_L}$ is manifestly a reflection operation, and it can be seen that $\sel$ is a reflection operation because
\begin{equation}
\label{eq:selectsquared}
\sel^2 = \left(\sum_{\ell} \proj{\ell} \otimes H_{\ell}\right)^2 = \sum_{\ell} \proj{\ell} \otimes H_{\ell}^2 = \openone .
\end{equation}
\fig{w} shows a circuit which implements ${\cal W}$ controlled on an ancilla.

The action of $\mathcal{W}$ partitions Hilbert space into a direct sum of two-dimensional irreducible vector spaces.  Through reasoning about these eigenspaces we can deduce the spectrum of $\mathcal{W}$ as well as the eigenvectors. In particular, we claim that the state $\ket{\cal L}\ket{k}$ and an orthogonal state $\ket{\phi_k}$ span the irreducible two-dimensional space that $\ket{\cal L}\ket{k}$ is in under the action of ${\cal W}$ for arbitrary eigenstate $\ket{k}$ of $H$ with eigenvalue $E_k$.  This state $\ket{\phi_k}$ is formally defined to be the component of $\mathcal{W} \ket{\cal L}\ket{k}$ that is orthogonal to $\ket{\cal L}\ket{k}$, which can be simplified using \eq{diag1} to
\begin{equation}
\label{eq:phi}
\ket{\phi_k} \equiv \frac{\left(\openone -\proj{\cal L}\otimes \proj{k}\right) \cdot \sel \ket{\cal L}\ket{k}}{\left\|\left(\openone -\proj{\cal L}\otimes \proj{k}\right) \cdot \sel \ket{\cal L}\ket{k}\right\|}= \frac{\left(\sel-
\frac{E_k}{\lambda}\openone\right)\ket{\cal L}\ket{k}}{\sqrt{1-\left(\frac{E_k}{\lambda} \right)^2}}.
\end{equation}
The matrix elements of $\mathcal{W}$ can be computed for this state. The upper diagonal matrix element follows from \eq{diag1},
\begin{equation}
\label{eq:diag}
\bra{k} \bra{\cal L}\mathcal{W} \ket{\cal L}\ket{k} = \bra{k} \bra{\cal L}\sel \ket{\cal L}\ket{k} = \frac{E_k}{\lambda}.
\end{equation}
The upper transition matrix element between $\bra{k}\bra{\cal L}$ and $\ket{\phi_k}$ is given from \eq{selectsquared} and \eq{phi} as
\begin{align}
\label{eq:offdiag}
\bra{k} \bra{\cal L}\mathcal{W} \ket{\phi_k} & = \bra{k}  \bra{\cal L} {\cal W} \frac{\left(\sel-
\frac{E_k}{\lambda}\openone\right)}{\sqrt{1-\left(\frac{E_k}{\lambda} \right)^2}}\ket{\cal L}\ket{k}
=  \frac{1-\left(\frac{E_k}{\lambda} \right)^2 }{\sqrt{1-\left(\frac{E_k}{\lambda} \right)^2}}=\sqrt{1-\left(\frac{E_k}{\lambda} \right)^2}.
\end{align}
Note that because phase estimation on ${\cal W}$ projects the system to an eigenstate of ${\cal W}$ and because ${\cal W}$ and $H$ share an eigenbasis, we are only concerned with the action of this operator for eigenstates.

\eq{diag} and \eq{offdiag} give the first row of the action of $\mathcal{W}$. The remaining entries can be calculated in a similar way, giving the action of ${\cal W}$ on this irreducible two-dimensional subspace as
\begin{equation}
\mathcal{W}\equiv \left(\begin{matrix} \frac{E_k}{\lambda}& \sqrt{1-\left(\frac{E_k}{\lambda} \right)^2} \\ -\sqrt{1-\left(\frac{E_k}{\lambda} \right)^2} & \frac{E_k}{\lambda} \end{matrix}\right)
=e^{i\arccos\left(E_k/\lambda\right)Y}
\end{equation}
where $Y$ is the Pauli-$Y$ operator constrained to this two dimensional space spanned by $\ket{\cal L}\ket{k}$ and $\ket{\phi_k}$. Finally, we can see that the phases of the eigenvalues of $\mathcal{W}$ in this subspace are $\pm \arccos(E_k/\lambda)$. Whereas the work of \cite{Low2016} focused on transforming the evolution under $\arccos(H)$ into evolution under $H$, the more recent work of \cite{Berry2018,Poulin2017} made the simple observation that by performing phase estimation directly on ${\cal W}$ one can obtain the spectrum of $H$ as
\begin{equation}
\textrm{spectrum}\left(H\right) = \lambda\cos\left(\arg \left(\textrm{spectrum}\left({\cal W}\right)\right)\right)\label{eq:spectrum}
\end{equation}
where $\arg$ is the argument function $\arg(e^{i \phi}) = \phi$.

\subsection{Heisenberg-Limited Phase Estimation of the Qubitized Quantum Walk}
\label{sec:heisenberg}

Since the original work of \cite{Kitaev1995}, many approaches have been proposed for estimating eigenphases of a unitary operator.  Whereas in the past iterative phase estimation approaches have been more popular in quantum simulation, here we propose using an entanglement based approach.  This has the virtue of requiring a number of applications of the unitary that saturates the Heisenberg limit. The ultimate precision that can be reached when one applies phase estimation by controlling a unitary when an ancilla is in $\ket{1}$ and applying the identity gate when the ancilla is in $\ket{0}$ is a Holevo variance of $\tan^2(\pi/(2^{m+1}+1))$ where the total number of applications of the unitary is $2^{m+1}-1$ and $m$ is the number of control qubits used. The Holevo variance is $\langle\cos(\hat\phi-\phi)\rangle^2-1$ where $\phi$ is the phase and $\hat\phi$ is the estimate of the phase given by the measurement. It is a convenient measure of variance for phase because it enables simple analytic results and is close to the mean-square error for narrowly peaked distributions.
The states for these optimal phase measurements were given in \cite{Luis1996}.
To apply them to phase estimation of a unitary, one can take the control qubits to be in this superposition state, rather than in a uniform superposition of computational basis states.

\begin{figure}[h]
\centering
  \resizebox{\linewidth}{!}{\includegraphics{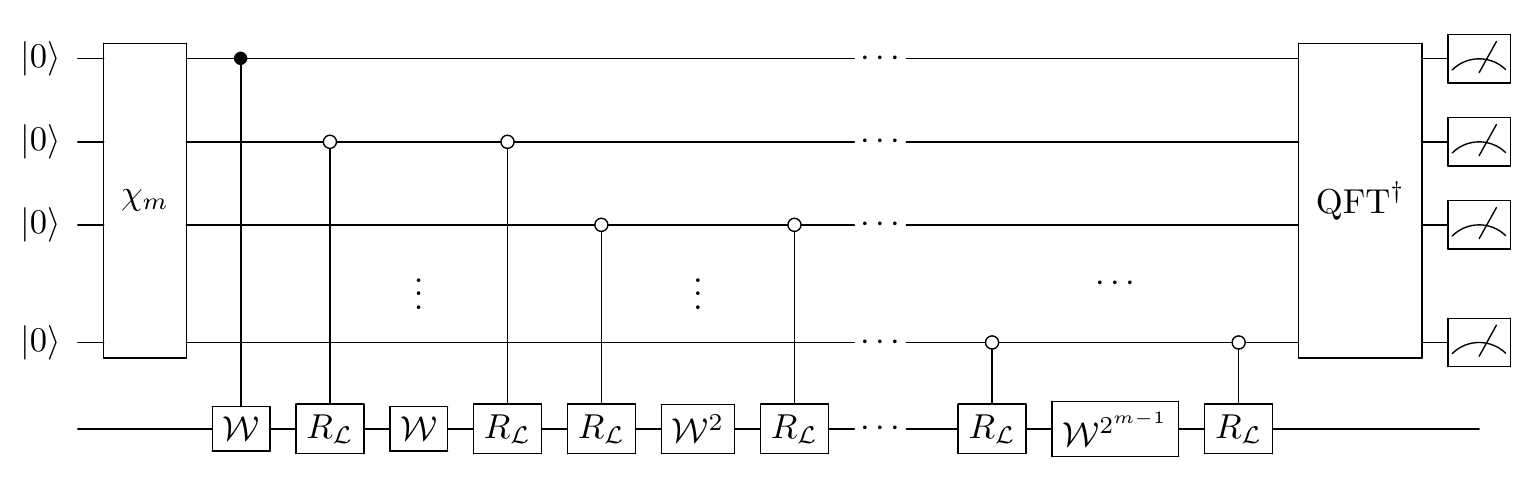}}
  \caption{
    \label{fig:quantum_walk}
   Heisenberg limited phase estimation circuit for learning the eigenphase of $\mathcal{W}$ with $m$ bits of accuracy with Holevo variance $\pi^2 / 2^{2(m+1)}$ where $R_{\mathcal{L}}$ is $(2\ket{\mathcal{L}}\!\bra{\mathcal{L}}\otimes \openone - \openone)$ and $\chi_m$ prepares the resource state from \eq{initial_pea_state}, which was shown to be optimal in \cite{Luis1996}. Both $\chi_m$ and the inverse quantum Fourier transform (QFT${}^\dagger$) have gate complexity $\widetilde{\cal O}(m)$ which is completely negligible compared to the overall gate complexity of phase estimation which scales as ${\cal O}(2^m)$. The controlled $R_{\cal L}$ and ${\cal W} = {\cal R_L} \cdot \sel$ gates are implemented as shown in \fig{w}. As a consequence of \eq{Winv} this circuit involves only $2^m - 1$ applications of ${\cal R_L}$ and as many applications of \textsc{select}. Note that the first unit of ${\cal R_L \cdot W \cdot R_L}$ is replaced by ${\cal W}$ controlled on the zero state of an ancilla in order to help disambiguate the outcomes of $\arccos(E_k / \lambda)$ and $\arccos(E_k / \lambda) + \pi$.}
\end{figure}

We will perform a slight optimization of that approach by applying the inverse unitary instead of the identity for the ancilla in the $\ket{0}$ state. Taking $\ket{\phi}$ to be an eigenstate of the unitary with eigenvalue $e^{i\phi}$, this means that
instead of applying $\ket{0}\ket{\phi}\rightarrow \ket{0}\ket{\phi}$ and $\ket{1}\ket{\phi}\rightarrow e^{i\phi}\ket{1}\ket{\phi}$ we apply $\ket{0}\ket{\phi}\rightarrow e^{-i\phi}\ket{0}\ket{\phi}$ and $\ket{1}\ket{\phi}\rightarrow e^{i\phi}\ket{1}\ket{\phi}$.
This doubles the effective phase difference and turns out to have the same complexity.
As shown in \fig{quantum_walk}, we accomplish the controlled inverse by removing controls from ${\cal W}^n$ and inserting controlled reflection operators ${\cal R_L}$ into the circuit which will cause us to apply either $({\cal W}^\dagger)^n$ or ${\cal W}^n$ depending on the state of the ancilla. We can see why this works by examining the relation
\begin{equation}
\label{eq:Winv}
{\cal R_L} \cdot {\cal W}^n \cdot {\cal R_L} = {\cal R_L}^2 \cdot \left(\sel \cdot {\cal R_L}\right)^n = \left(\sel \cdot {\cal R_L}\right)^n = \left({\cal W}^\dagger\right)^n
\end{equation}
which holds for any integer $n$ as a consequence of the self-inverse nature of ${\cal R_L}$ and ${\cal W}$.
Moreover, because ${\cal W}^n$ always ends with an ${\cal R_L}$ operation, each controlled ${\cal R_L}$ can be combined in the circuit with the ${\cal W}^n$, to yield a complexity no greater than the complexity of just performing the ${\cal W}^n$ operations.

This trick will result in measuring the phase modulo $\pi$. To eliminate the $\pi$ ambiguity, an additional controlled ${\cal W}$ can be performed without this trick.
That is shown as the first controlled operation in \fig{quantum_walk}.
For $m$ control qubits the Holevo variance is still $\tan^2(\pi/(2^{m+1}+1))$, but the complexity is reduced by approximately a half to $2^m$ applications of the unitary ${\cal W}$.

As seen in \fig{quantum_walk}, our modified phase estimation algorithm begins with a unitary ${\chi_m}$ which prepares the state
\begin{equation}
\label{eq:initial_pea_state}
\chi_m \ket{0}^{\otimes m} \mapsto \sqrt{\frac 2{2^m+1}}
\sum_{n=0}^{2^{m}-1}
\sin\left(\frac{\pi\left(n+1\right)}{2^{m}+1}\right)\ket{n}.
\end{equation}
To prepare this state with cost $\widetilde{\cal O}(m)$ we first perform Hadamards on $m+1$ qubits (initially in the $\ket{0}$ state) to give
\begin{equation}
\sqrt{\frac{1}{2^{(m+1)}}}\sum_{n=0}^{2^{m}-1}
\ket{n}\otimes (\ket{0}+\ket{1}).
\end{equation}
Next, we perform a series of $m$ controlled rotations with each of the first $m$ qubits as control and qubit $m+1$ as target. For control qubit $k$, the rotation on the target qubit $m+1$ is $e^{i\pi 2^k Z/(2^m+1)}$. Perform a further rotation of $e^{i\pi Z/(2^m+1)}$ on qubit $m+1$, and
the resulting state is
\begin{equation}
\sqrt{\frac{1}{2^{(m+1)}}}\sum_{n=0}^{2^{m}-1}\left(e^{i\pi (n+1)/(2^m+1)}
\ket{n}\otimes \ket{0}+e^{-i\pi (n+1)/(2^m+1)}
\ket{n}\otimes \ket{1}\right).
\end{equation}
Perform a Hadamard on qubit $m+1$ and measure in the computational basis.
Measuring $\ket{1}$ gives the state 
\begin{equation}
i\sqrt{\frac{1}{2^{m}}}\sum_{n=0}^{2^{m}-1}\sin\left(\frac{\pi\left(n+1\right)}{2^{m}+1}\right)\ket{n}.
\end{equation}
The probability of success is given by the normalization: $(1+2^{-m})/2$.
The scheme can be made unitary and deterministic via a single step of amplitude amplification.
Clearly, this preparation scheme  scaling as $\widetilde{\cal O}(m)$ will not dominate the cost of our overall phase estimation which scales as ${\cal O}(2^m)$, as we will discuss in the next section.

\subsection{Error Scaling and Query Complexity}
\label{sec:error}

Three sources of error will enter our simulation: error due to performing PEA to finite precision, $\epsilon_{\textsc{pea}}$, error due to approximate preparation of the Hamiltonian terms within the implementation of the $\prep$ oracle, $\epsilon_{\textsc{prep}}$, and the error in synthesizing the inverse QFT, $\epsilon_{\textsc{QFT}}$. We will choose to measure error through the root-mean-square error of the estimator used within phase estimation, i.e.
\begin{equation}
\Delta \phi \equiv \sqrt{\mathbb{E}\left[{\rm dist}(\phi_{\rm est},\phi_{\rm true})^2\right]},
\end{equation}
where the distance considered above is the angular distance between the estimated phase and the actual phase.

Provided phase estimation is performed on a unitary operation, the error in the estimate of the energy is at most the error in implementing the unitary \cite{Reiher2017}. We will break up the estimated phase as the sum of two contributions $\phi_{\rm est} = \phi + \epsilon_{\textsc{prep}}+\phi_{\rm true}$. Here $\phi$ is a random variable with zero mean $\mathbb{E}(\phi)$ and Holevo variance $\mathbb{V}_H(\phi)$ describing the output of phase estimation and $\epsilon_{\textsc{prep}}$ represents the systematic errors in the phase that arise because of gate synthesis.  In the limit of small variance, we can express this with high probability over the true phase as
\begin{equation}
\Delta \phi \approx \sqrt{\mathbb{E}[(\phi_{\rm est}-\phi_{\rm true})^2]}
\approx \sqrt{\mathbb{V}_H(\phi) +(\epsilon_{\textsc{prep}}+\pi\epsilon_{\textsc{QFT}})^2}
\approx \sqrt{\left(\frac{\pi}{2^{m+1}}\right)^2+(\epsilon_{\textsc{prep}}+\pi\epsilon_{\textsc{QFT}})^2},
\end{equation}
where $m$ ancillary qubits are used within the phase estimation algorithm. Note that such a division of the error is suboptimal since the cost involved in reducing the error for phase estimation is exponentially larger than that involved in increasing the accuracy of the circuit synthesis~\cite{Reiher2017}; however, we take the two errors to be equal for simplicity.

\eq{spectrum} implies that error in the energy is at most
\begin{equation}
\Delta E =  \lambda \Delta\cos(\phi)\le \lambda \Delta\phi\approx \lambda \sqrt{\left(\frac{\pi}{2^{m+1}}\right)^2+(\epsilon_{\textsc{prep}}+\pi\epsilon_{\textsc{QFT}})^2}.
\end{equation}
That suggests we can choose to estimate the phase to a number of bits given by
\begin{align}
\label{eq:pea_bits}
m = \left\lceil\log\left(\frac{\sqrt{2} \pi \lambda}{2\Delta E} \right)\right\rceil < \log\left(\frac{\sqrt{2} \pi \lambda}{\Delta E} \right)
\end{align}
and the target errors can be chosen as
\begin{equation}
\epsilon_{\textsc{prep}} \leq \frac{\sqrt{2}  \Delta E}{4 \lambda}
\qquad \qquad \epsilon_{\textsc{QFT}} \leq \frac{\sqrt{2}  \Delta E}{4 \pi \lambda}.
\label{eq:errdist}
\end{equation}
Thus, using the phase estimation procedure from \sec{heisenberg} we will need at most
\begin{equation}
\label{eq:sel_queries}
2^{m} < \frac{ \sqrt{2} \pi \, \lambda}{\Delta E}
\end{equation}
queries to the \sel oracle and at most twice as many queries to the \prep oracle in order to estimate spectra to within error $\Delta E$. Supposing that the circuit $\prep$ can be applied at gate complexity $P$ and the circuit $\sel$ can be applied at gate complexity $S$, the gate complexity of our simulation (ignoring for now the cost of $\chi_m$ and the cost of the QFT${}^\dagger$ since they scale as ${\cal O}(m)$) is then approximately bounded from above by
\begin{equation}
\label{eq:complexity}
\frac{\sqrt{2} \pi \lambda  \left(S + 2 P \right) }{\Delta E}.
\end{equation}
This paper will discuss implementations of \sel and \prep that minimize $S$ and $P$ without increasing $\lambda$.

To implement the inverse QFT which appears in \fig{quantum_walk} we will use the semiclassical algorithm described in \cite{Griffiths1996}. This version of the QFT requires just $m-1$ rotation gates and $m$ Hadamards when implemented on $m$ qubits. Thus, the error in each rotation must be at most $\epsilon_{\textsc{QFT}} / (\pi m )$, which implies that the inverse QFT will have T complexity scaling as ${\cal O}(m \log (m/\epsilon_{\textsc{QFT}}))$. As this is an additive cost to other parts of our phase estimation algorithm with T complexity scaling as ${\cal O}(2^m)$, the cost of performing the QFT within the required error budget can be safely neglected.

How errors in the coefficients of the implemented Hamiltonian propagate into $\epsilon_{\textsc{prep}}$ is slightly harder to bound owing to the fact that the error in the eigenphase is a nonlinear function of the error in the Hamiltonian implementation.  In particular the error can diverge for frustration-free Hamiltonians owing to the singularity of $\arccos$.  The main result, shown in~\app{errAnal}, is that \prep should be implemented so that if $\widetilde{w}_\ell$ is the effective coefficient of $H_\ell$ in the approximately implemented Hamiltonian then
\begin{equation}
\left | \widetilde{w}_\ell - w_\ell \right | \le \delta =\frac{\sqrt{2}\Delta E}{4 L\left( 1+\frac {\Delta E^2}{8\lambda^2}\right)}\left( 1-\frac{\left\|H\right\|^2}{\lambda^2}\right) .
\end{equation}

\section{Low T Complexity Primitives for LCU Oracles}
\label{sec:primitives}

In this section we will introduce three circuit primitives which are helpful for implementing \sel and \prep oracles with low T gate complexity. We use these primitives for electronic structure simulation, but expect them to be useful more generally. These primitives enable black box implementations of \sel and \prep for any problem with lower asymptotic complexity than prior constructions in the literature. They also have low T counts at finite size. We will use these constructions extensively in \sec{chemistry} and \sec{hubbard} of this paper.

In \sec{streaming}, we will introduce a technique for ``streaming'' bits of an iterator running over a unary register. One application is that this technique can be used to coherently apply operations controlled on a register with $\log L$ qubits in superposition (e.g.\ the selection register in \eq{prepare} and \eq{select}) using a number of T gates scaling as ${\cal O}(L)$, as opposed to ${\cal O}(L \log L)$ as one might normally expect. However, what is even more important is the versatile way that these constructions can be applied.

In \sec{majorana} we show how one can use the results of \sec{streaming} to implement a primitive corresponding to controlled application of a Majorana fermion operator. This primitive is used directly in our implementation of \sel in \sec{select_chem} and \sec{select_hub}.

In \sec{qrom}, we show a straightforward application of the techniques in \sec{streaming} which allow us to develop a particularly efficient quantum data lookup which we refer to as ``Quantum Read-Only Memory'' (QROM). In particular, for coherently querying a database with $L$ words our implementation of QROM has T complexity of $4L-4$ with no dependence on the word length, which is an asymptotic and constant-factor improvement over all prior literature. We have another paper in preparation which discusses QROM in more detail \cite{Gidney2018}.

In \sec{subsample}, we discuss a technique for initializing a state with $L$ unique coefficients (provided by a classical database) with a number of T gates scaling as $4 L + {\cal O}(\log (1/\epsilon))$ where $\epsilon$ is the largest absolute error that one can tolerate in the prepared amplitudes. This routine improves asymptotically over the gate complexity of prior constructions for a black box \textsc{prepare}. It also has the advantage of implementing \textsc{prepare} without increasing the value of $\lambda$ (from \eq{prepare}) which has been a frequent problem with other implementations of \prep \cite{BabbushSparse1,BabbushSparse2,BabbushLow} which attempted to obtain scaling sublinear in the number of terms in the linear combinations of unitaries decomposition.

\subsection{Unary Iteration and Indexed Operations}
\label{sec:streaming}

Many of the circuits in this paper rely heavily on a technique we refer to as ``unary iteration''.
The unary iteration process gradually produces, and then uncomputes, qubits indicating whether an index register is storing specific values (with respect to the computational basis). We call the process ``unary iteration'' because the indicator qubits are made available one by one (iteration), and they correspond to the one-hot (unary) encoding of the index register value. While these techniques were developed independently, we note that a scheme similar to unary iteration is also used for implementing \sel operations in \cite{Childs2017}. Compared to \cite{Childs2017} we lower the T-count from $6L - 4$ to $4L - 4$ and explain how to apply the scheme to sizes $L$ that are not powers of 2.

For an index register storing an index in the interval $[0, L)$, the space overhead of converting the index register into a unary register (as in \cite{Poulin2017}) would normally be $L$ qubits.
By comparison, our unary iteration technique is exponentially more efficient in space without any increased T complexity, requiring only $\log L$ ancillae.
Our unary iteration has a T-count of $4L - 4$ and can be parallelized if needed without increasing T-count. Despite its efficiency, unary iteration is still the dominant source of T complexity in our algorithms.
We use it for indexed operations, Majorana operators, reversible preparation of states, and database lookup; all of which have T-counts that scale like ${\cal O}(L)$.

To explain unary iteration, we will first focus on how it is used to implement a controlled indexed \textsc{not} operation:
\begin{equation}
\label{eq:selectx}
\ket{c} \ket{\ell}\ket{\psi} \mapsto \ket{c} \ket{\ell} (X_\ell)^c \ket{\psi}
\end{equation}
where $\ket{c}$ is the control register, $\ket{\ell}$ is the selection register, $\ket{\psi}$ is the system register, and the subscript $\ell$ on $X_\ell$ indicates that the \textsc{not} operation acts on qubit $\ell$ of the system register. From \eq{select} it should not be surprising that this primitive is helpful for our constructions of $\sel$.

\begin{figure}[h]
  \resizebox{.65\linewidth}{!}{
    \includegraphics{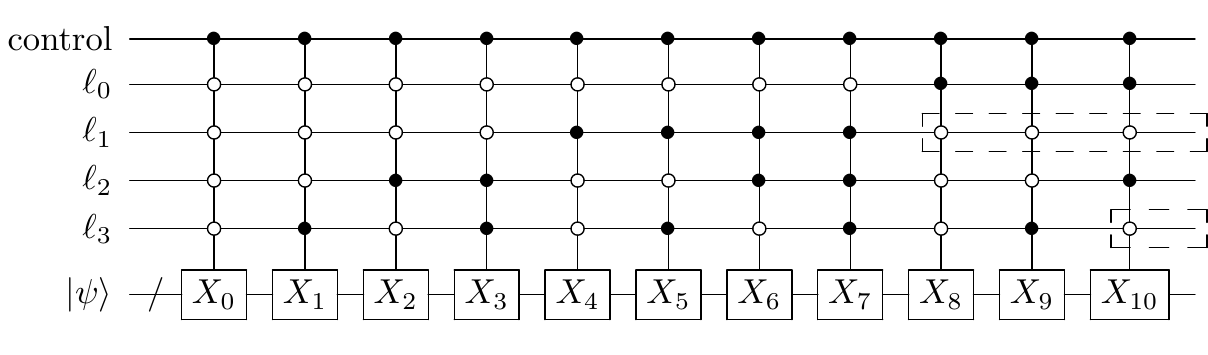}
  }
  \caption{\label{fig:total-control}
    Example total control circuit for performing a controlled indexed $X_\ell$ operation, with $0 \leq \ell < L = 11$.
    This is the (naive) starting point for producing a unary iteration circuit, before optimizations which asymptotically improve the T complexity.
    When indices outside the specified range do not occur, the highlighted runs of \textsc{off}-type controls reaching the right hand side of the circuit can be removed without affecting the circuit.
    (There are also other controls which could be removed, but for our purposes this would be counterproductive due to interfering with later optimizations.)}
\end{figure}

A simple (but suboptimal) way to implement \eq{selectx} would be to totally control the application of $X_\ell$ on all possible values that could occur in the register $\ket{\ell}$, as shown in \fig{total-control}. For instance, in order to apply $X_{158}$ when $|\ell\rangle = |158\rangle$, the total-control approach would place a $\textsc{not}$ gate targeting the qubit 158 in the system register $\ket{\psi}$, but with a control on each index bit. The control's type (\textsc{on} or \textsc{off}) would be determined by the binary representation of 158 ($158_{10} = 10011110_2$), so there would be a must-be-\textsc{off} control on the low bit of the index register (because the low bit of 158 in binary is 0), a must-be-\textsc{on} control on the next bit (because the next bit of 158 in binary is 1), and so forth. In order to cover every case, a separate $\textsc{not}$ gate with corresponding controls would be generated for every integer from 0 up to $L-1$. This would produce $L$ different \textsc{not} operations, each targeting a different qubit in the target register and each having a number of controls equal to the size of the index register (i.e.\ $\log L$). Thus, it takes ${\cal O}(L \log L)$ T gates to apply \eq{selectx} using this approach. Unary iteration will improve this T-count to $4L - 4$.

\begin{figure}[h]
\centering
  \resizebox{.6\linewidth}{!}{
      \includegraphics{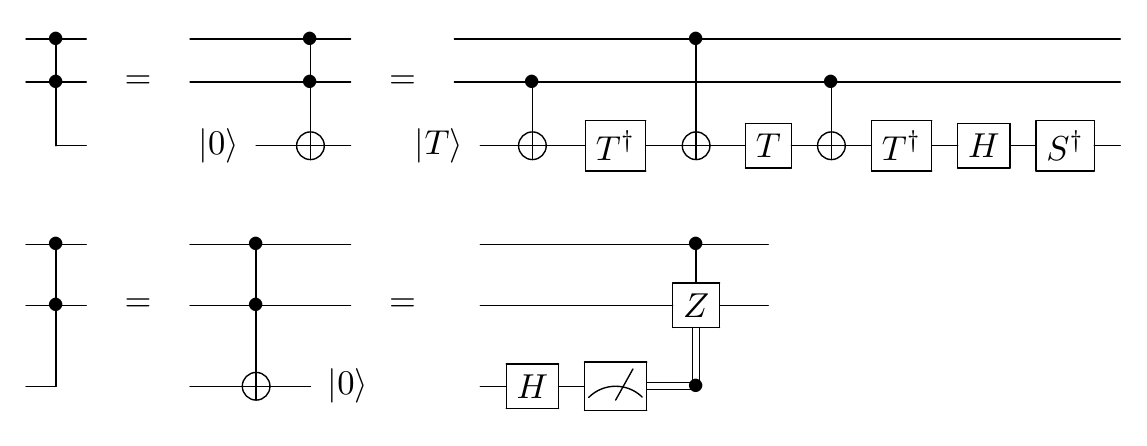}
  }
  \caption{\label{fig:temporary-and-notation}
    Computing and uncomputing \textsc{and} operations, defined in terms of Toffoli gates and in terms of Clifford+T gates \cite{GidneyAdder}.
    Computing an \textsc{and} consumes 4 $|\rm T\rangle$ states, and is equivalent to applying a Toffoli gate to a target qubit known to be $|0\rangle$.
    Uncomputing an \textsc{and} consumes no $|\rm T\rangle$ states, and is equivalent to applying a Toffoli gate to a target qubit guaranteed to end up in the $|0\rangle$ state.
    Drawing \textsc{and} operations as ``corners" instead of as $\oplus$ symbols is a visual cue that the target qubit will be (or was) \textsc{off} after (before) the operation.
    This is worth highlighting because it affects the T-count of synthesizing the operation and whether the target is available for reuse as an ancilla in later operations.
  }
\end{figure}

Consider that the controls for the operation targeting the qubit at offset $\ell=158$ are almost identical to the controls for the operation targeting the qubit at offset $\ell=159$. They differ only on the low bit of the index register, where 158 requires the bit to be \textsc{off} whereas 159 requires the bit to be \textsc{on}. If we combined the $\log L - 1$ other qubits of the index register into a single representative qubit that was set if and only if those controls were met, we could use this representative qubit once for the $\ell=158$ case and again for the $\ell=159$ case. Using the representative qubit twice, instead of computing it twice, decreases the total amount of work done. Unary iteration is the result of taking this kind of representative-reuse idea to its natural limit.

\begin{figure}[h]
  \resizebox{\linewidth}{!}{
    \includegraphics{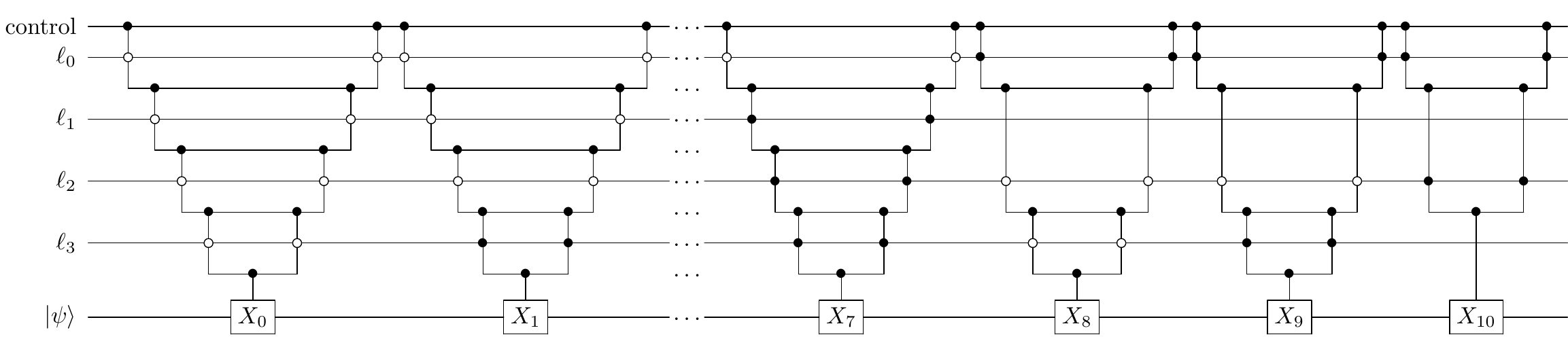}
  }
  \caption{\label{fig:expanded-controls}
    The ``sawtooth'' circuit resulting from removing unnecessary bits from \fig{total-control} and then adding \textsc{and} operations from \fig{temporary-and-notation} to combine the controls for performing a controlled indexed $X_\ell$ operation with $L=11$ possible targets.
  }
\end{figure}

\begin{figure}[tbh]
\centering
  \resizebox{.7\linewidth}{!}{
    \includegraphics{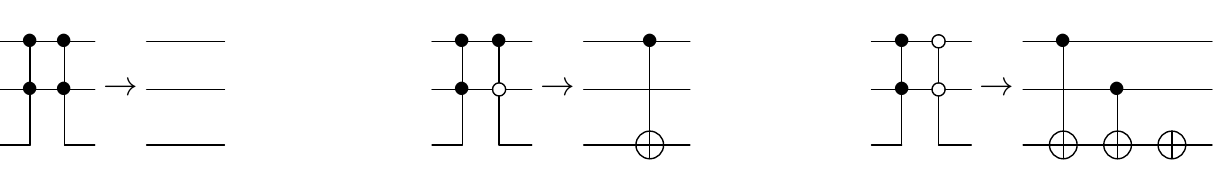}
  }
  \caption{\label{fig:combining-logical-ands}
    When two \textsc{and} operations are adjacent, the uncomputation-and-recomputation can be replaced by $\textsc{cnot}$ and $\textsc{not}$ operations.
    Each such merger saves 4 T gates.
  }
\end{figure}

\begin{figure}[tbh]
  \resizebox{\linewidth}{!}{
    \includegraphics{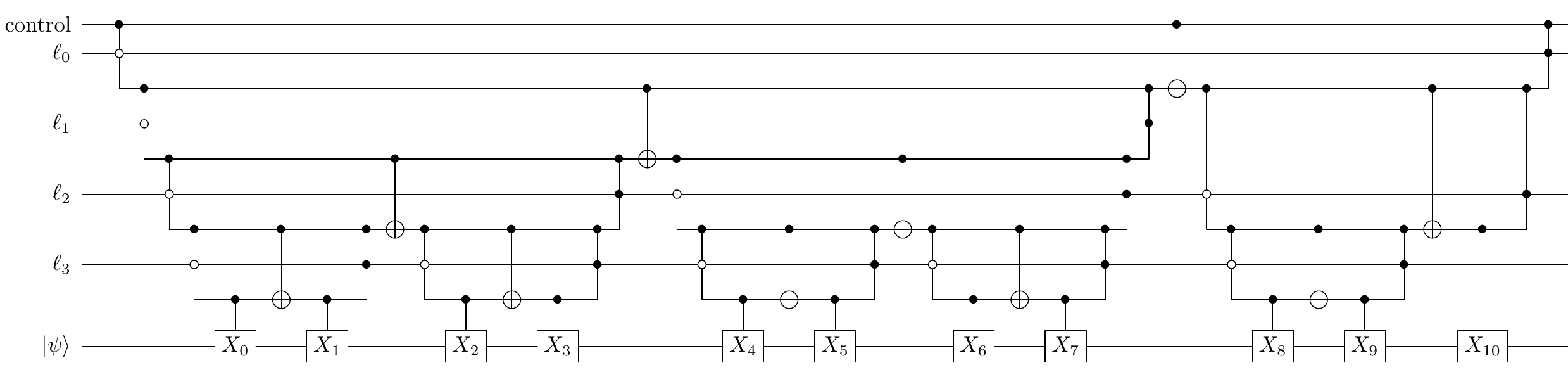}
  }
  \caption{\label{fig:unary-iteration-circuit}
    An $L=11$ unary iteration circuit which applies $X_\ell$ to the qubit $\ell$ in the system register $\ket{\psi}$, where $\ell$ is the value stored in the selection register. The circuit is obtained by merging \textsc{and} operations from \fig{expanded-controls} using the method shown in \fig{combining-logical-ands}. Computes 10 \textsc{and} operations, and so has a T-count of $10 \times 4 = 40 = 4L - 4$.
  }
\end{figure}

We define our unary iteration construction by starting with a total-control circuit, and then applying a fixed set of simple transformations and optimizations.
\fig{total-control} shows an example starting point, a total-control circuit for $L=11$. For unary iteration, we require that the index register never store an out-of-range value $\ell \geq L$.
For example, consider what occurs when the $X_{10}$ operation from \fig{total-control} is not conditioned on $\ell_0$ (the least significant bit of the index register).
This would cause an $X_{10}$ to be applied to the target when $\ell=11$, but this is fine since we know $\ell \neq 11$.
We use the $\ell < L$ condition to omit several controls from the circuit.
For each possible $\ell$, we will look at the $X_\ell$ operation and remove the control on the $b$'th index bit when the following two conditions are true: (i) the $b$'th bit of $L-1$ isn't set and (ii) setting the $b$'th bit of $\ell$ would change $\ell$ into a value larger than $L-1$.
Visually, this removes ``runs'' of must-be-\textsc{off} controls that manage to reach the right side of the circuit as highlighted in \fig{total-control}.

After removing the specified controls, we expand the remaining controls into nested \textsc{and} operations (the \textsc{and} operation is defined in \fig{temporary-and-notation}), always nesting so that lower controls are inside higher controls. For clarity, we consistently place the ancillae associated with an \textsc{and} operation just below its lowest input qubit. The result is the ``sawtooth" circuit shown in \fig{expanded-controls}. By iteratively optimizing adjacent \textsc{and} operations as shown in \fig{combining-logical-ands}, the sawtooth circuit from \fig{expanded-controls} is optimized into the circuit shown in \fig{unary-iteration-circuit}. This is our unary iteration circuit for $L=11$. The optimized circuit always ends up with $L-1$ \textsc{and} computations (even when $L$ is not a power of 2), each \textsc{and} takes 4 T gates to compute, and we have no other T-consuming operations in the circuit. Thus, the T-count of this construction is $4L-4$.

\FloatBarrier

\subsection{Selective Application of Majorana Fermion Operators}
\label{sec:majorana}

Now that we have described unary iteration, we can begin using it to construct primitives relevant for the \textsc{select} oracle. As discussed in detail in \sec{chemistry} and \sec{hubbard} below, our approach for implementing \sel will require that we have a circuit capable of selectively applying the Majorana fermion operator
\begin{equation}
\label{eq:majorana}
\ket{\ell} \ket{\psi} \mapsto \ket{\ell} \left(\frac{a^\dagger_\ell - a_\ell}{i}\right) \ket{\psi}
= \ket{\ell} Y_\ell \cdot Z_{\ell - 1} \cdots Z_{0} \ket{\psi}
\end{equation}
where the last equality holds under the Jordan-Wigner transformation \cite{Somma2002}. In this section we describe explicit circuits which accomplish the mapping of \eq{majorana}.

In \sec{streaming} we discussed selectively applying $X_\ell$ operations as a representative example of how one might use unary iteration. However, nothing intrinsic to the unary iteration construction requires that the indexed operation be so simple. For example, we could switch from applying $X_\ell$ to applying $Z_\ell$ halfway through the circuit. Or each $X_\ell$ could be replaced by multiple Pauli operations targeting multiple qubits. In general, each index could be associated with its own unique set of Pauli operators to be applied to various target qubits.

\begin{figure}[tbh]
\centering
\includegraphics[width=\linewidth]{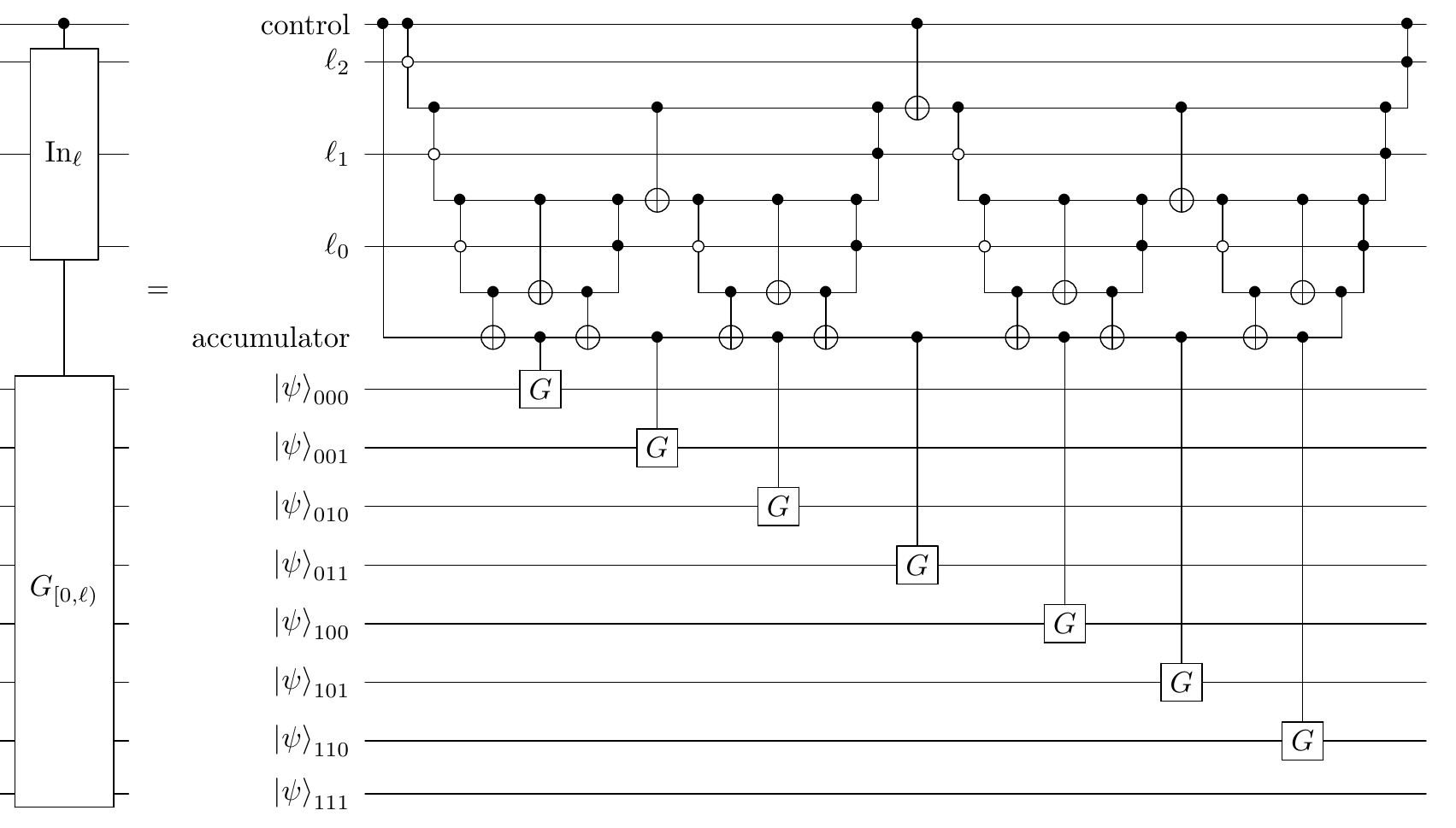}
\caption{\label{fig:ranged-operation}
  Ranged operation construction implementing $\ket{\ell} \ket{\psi} \rightarrow \ket{\ell} \prod_{k=0}^{\ell - 1} G_k \ket{\psi}$.
  Applies the $G$ operation to a range of values, instead of to a single value, by using an accumulator.
  The accumulator is guaranteed to be cleared after the final CNOT targeting it (drawn as a line merging into an ancilla qubit).
  This occurs because (unless \textsc{control} is not set and the accumulator simply stays unset) exactly one of the unary bits must have been set, and we targeted the accumulator with CNOTs controlled by each of those bits in turn. $G_{[p, q)}$ refers to $G$ being applied to every qubit index $k$ satisfying $p \leq k < q$.
}
\end{figure}

\begin{figure}[tbh]
\centering
\includegraphics[width=\linewidth]{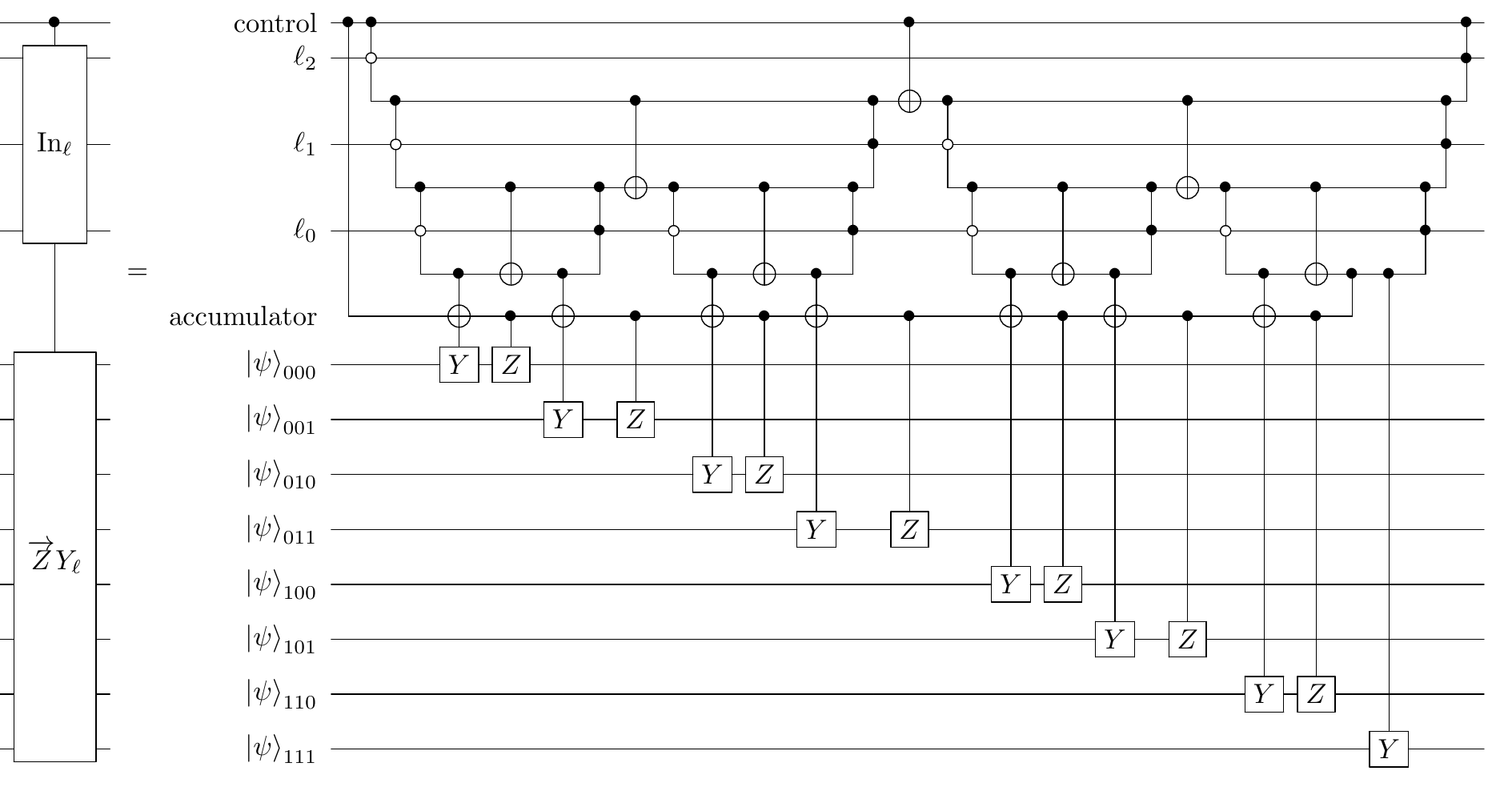}
\caption{
  \label{fig:majorana}
  Application of a selected Majorana fermion operator, $\ket{\ell} \ket{\psi} \mapsto \ket{\ell} Y_\ell \cdot Z_{\ell - 1} \cdots Z_{0} \ket{\psi}$ as described in \eq{majorana}. This is accomplished by performing a ranged operation (as shown in \fig{ranged-operation}) and an indexed operation (similar to what is shown in \fig{unary-iteration-circuit}) with a single pass through the selection register $\ket{\ell}$.
  It has a T-count of $4L - 4$, where $L$ is the number of integer values that can be held by the selection register $\ket{\ell}$.}
\end{figure}

We can also apply transformations to our quantum unary iterators (analogous to transformations of classical iterators). Iterators can be mapped, filtered, zipped, aggregated, batched, flattened, grouped, etc. For instance, given a classical stream of bits, one can aggregate over it with the $\oplus$ operation. This produces a new iterator, which iterates over bits equal to the parity of the values-so-far from the original iterator. It is possible to apply this xor-aggregation idea to the quantum unary iteration process.
Introduce an ``accumulator'' qubit and, as each iterated unary qubit is produced, \textsc{cnot} it into the accumulator. In effect, if the index register is storing $\ell$ then the accumulator will stay \textsc{off} until the $\ell$'th qubit toggles it \textsc{on}.
The accumulator will then stay \textsc{on} until the end of the iteration process, where it is uncomputed by a \textsc{cnot} from the control qubit. By conditioning $X_\ell$ on the accumulator qubit, instead of on the original unary qubits, efficient {\em ranged} operation such as $\ket{\ell} \ket{\psi} \rightarrow \ket{\ell} G_\ell \cdot G_{\ell+1} \cdots G_{L - 1} \ket{\psi}$ are produced. We show an example of an accumulator-based ranged operation in \fig{ranged-operation}.

By using both the accumulator qubit and the original unary qubits, we can apply a ranged indexed operation and an indexed operation in a single unary iteration which gradually sweeps over the possible target qubits.
The resulting combined operation, shown in \fig{majorana}, is a crucial part of our $\sel$ circuit, effecting the transformation of \eq{majorana}.
\FloatBarrier

\subsection{Quantum Read-Only Memory (QROM) for Low T Complexity Data Lookup}
\label{sec:qrom}

\begin{figure}[h]
\centering
  \resizebox{.8\linewidth}{!}{
    \includegraphics{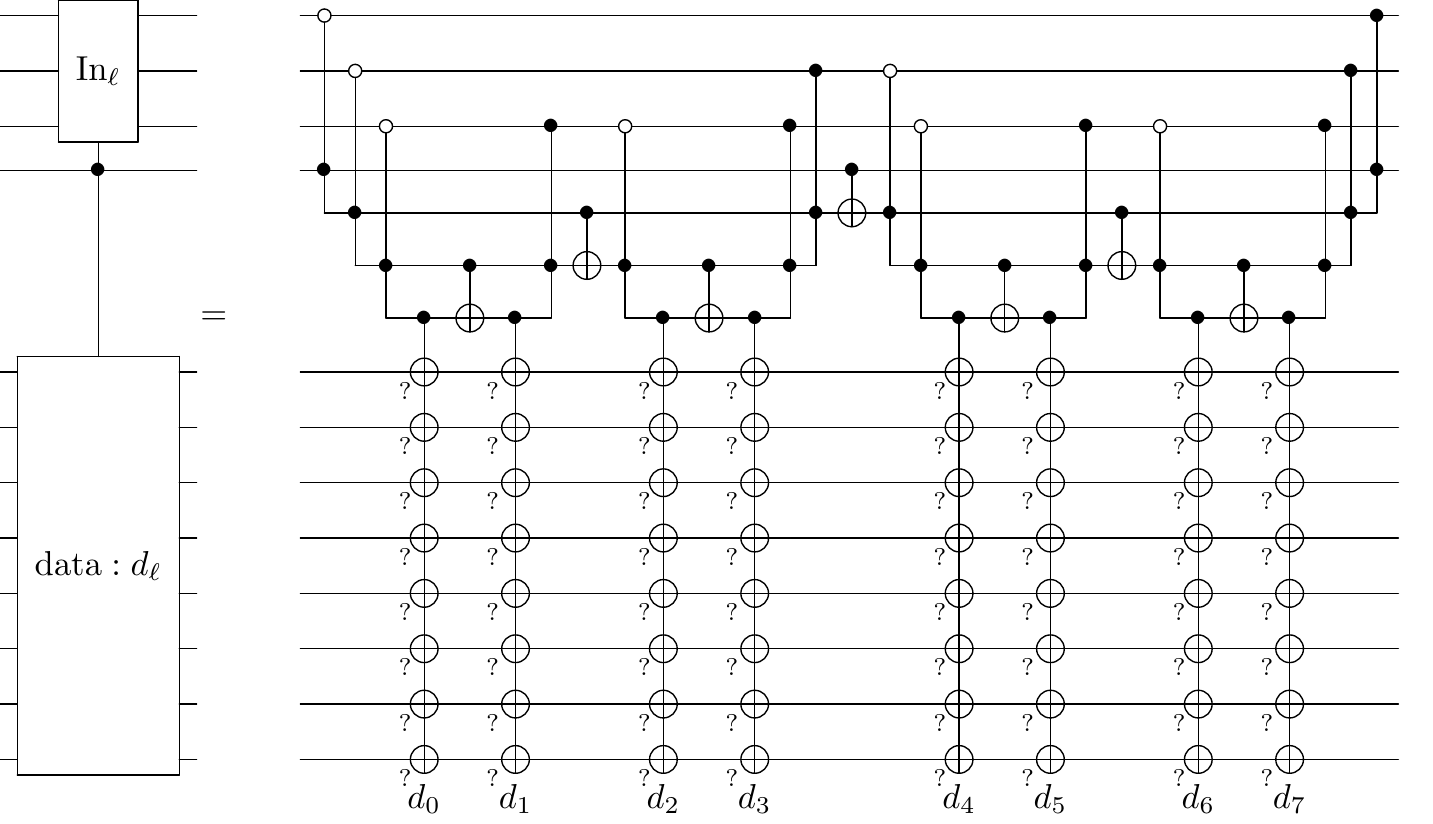}
  }
  \caption{
    \label{fig:qrom}
    Finite-sized example of the ``QROM'' database loading scheme used in our implementation of \textsc{subprepare}. If the index register contains $\ell$, the output register ends up storing $d_\ell$ where $\vec d$ is some precomputed data vector used when constructing the circuit. The top part of the circuit is performing unary iteration as described in \sec{streaming}. The bottom part of the circuit is loading classical data associated with each possible index. The classical data is encoded into the presence or absence of \textsc{CNOT}s on the data lines. The ``?'' marks in the diagram indicate that one should decide whether or not to have a \textsc{cnot} gate on each line, depending on the value of the data to load. This circuit has T-count of $4 L - 4$, which is due entirely to the unary iteration. The T gate cost of this circuit is independent of the number of bits used to store each element of the database.
}
\end{figure}

In this section we explain how one can use the techniques of \sec{streaming} in order to implement a particular efficient form of what we call Quantum Read-Only Memory (QROM) \cite{GidneyQROM}, which is useful in the context of the \textsc{subprepare} routine, a subroutine of the \prep circuit described in \sec{prepare_chem} (in \fig{prepare}). Many quantum algorithms assume the existence of a hypothetical peripheral called ``Quantum Random-Access Memory'' (QRAM) \cite{giovannetti2008qram} which allows classical or quantum data to be accessed via an index under superposition. The purpose of QROM is to read classical data indexed by a quantum register, i.e.\ to perform the following transformation:
\begin{equation}
\textsc{QROM}_d \cdot \sum_{\ell=0}^{L-1} \alpha_\ell |\ell\rangle|0\rangle = \sum_{\ell=0}^{L-1} \alpha_\ell |\ell\rangle |d_\ell\rangle
\end{equation}
where $\ell$ is the index to read, $\alpha_\ell$ is the amplitude of $\ket{\ell}$, and $d_\ell$ is the word associated with index $\ell$ in a classical list $d$ containing $L$ words. Our implementation of QROM is shown in \fig{qrom}. Note that our notion of QROM is unrelated to the discussion of ROM on a quantum computer in Ref.~\cite{Travaglione2001}.

The read-only aspect of QROM makes it distinctly different from QRAM in that one can read from QROM but cannot write to it during the course of a computation. A few algorithms, such as the procedure introduced in \cite{Kerenidis2016} actually do require that one write to QRAM; thus, for such use cases QROM would not be appropriate. A notable difference between this paper and most previous work on QRAM \cite{giovannetti2008qram, giovannetti2008qram2,hong2012robust,arunachalam2015robustness} is that we describe the cost of QROM in terms of a fault-tolerant cost model: the number of T gates performed and the number of ancilla qubits required. Under such cost models, the ``bucket brigade'' QRAM design of Giovannetti \textit{et al.}\ \cite{giovannetti2008qram, giovannetti2008qram2} has T complexity (and thus, also time complexity under reasonable error-correction models) of ${\cal O}(L)$ regardless of the fact that it has depth ${\cal O}(\log L)$ because implementing it as an error-corrected circuit consumes ${\cal O}(L)$ T gates and ${\cal O}(L)$ ancillae qubits. Our implementation of QROM consumes only $4L$ T gates and $\log L$ ancillae, which is a constant-factor improvement in T-count and an exponential improvement in space usage over the construction of Giovannetti \textit{et al.}

\FloatBarrier

\subsection{Subsampling the Coefficient Oracle}
\label{sec:subsample}

In this section we introduce a technique for initializing a state with $L$ unique coefficients (provided by a classical database) with a number of T gates scaling as $4 L + {\cal O}(\log (1/\epsilon))$ where $\epsilon$ is the largest absolute error that one can tolerate in the prepared amplitudes. This result constitutes a general procedure for implementing \prep such that the cost of circuit synthesis is additive, rather than multiplicative (as in most prior schemes). In particular, it improves on the database scheme from \cite{BabbushSparse1} (based on the procedure of \cite{Shende2006}) which requires a number of T gates scaling as ${\cal O}(L \log (L/\epsilon))$. Importantly, our scheme does not increase the value of $L$ or $\lambda$, which would usually be the case for most ``on-the-fly'' strategies for implementing \prep \cite{Berry2015,BabbushSparse1,BabbushSparse2}. 

Generalizing the requirements of \eq{prepare}, we begin with the observation that it would be acceptable to have a $\prep$ circuit which initializes the state
\begin{equation}
\ket{\cal L} \equiv \sum_{\ell = 0}^{L-1} \sqrt{\frac{w_\ell}{\lambda}} \ket{\ell} \ket{\textrm{temp}_\ell}
\end{equation}
where $\ket{\textrm{temp}_\ell}$ is an unspecified junk register entangled with $\ket{\ell}$.
Equivalently, any pure state $\ket{\cal L}$ would suffice if
\begin{equation}
\bra{\cal L} \left(\proj{\ell} \otimes \openone\right) \ket{\cal L} = \frac{w_\ell}{\lambda}
\qquad \qquad
\forall \ell \in [0, L).
\end{equation}
Because \sel only uses $\ket{\ell}$ to {\em control} operations, phase error (including entanglement with the junk register) in the state produced by \prep will commute across \sel and be corrected by $\prep^\dagger$. However, \sel itself necessarily introduces entanglement between its target register $|\psi\rangle$ and the index register $\ket{\ell}$ (plus associated junk register). So $\prep^\dagger$ will not exactly restore $\ket{\ell}$ or the junk register. Although we only specify the action of \prep on the $|0\rangle$ state, \prep will be applied to other states due to this imperfect uncomputation effect. This is accounted for by (i) requiring qubits coming out of $\prep^\dagger$ to be kept and fed back into the next \prep operation and (ii) the reflection step between $\prep^\dagger$ and \prep only affecting the $\ket{0}$ state.

Given the observation that the existence of an entangled junk register is acceptable, we will seek to implement a circuit which effects the transformation,
\begin{align}
\label{eq:swapped}
\ket{0}^{\otimes(1 + 2\mu + 2\log L)}
&\mapsto
\sum_{\ell = 0}^{L-1} \sqrt{\widetilde{\rho_\ell}} \ket{\ell} \ket{\textrm{temp}_\ell}
\end{align}
where $\widetilde{\rho}_\ell \equiv \widetilde{w}_\ell / \lambda$ are probabilities characterizing the approximate Hamiltonian we are encoding. Whereas the exact Hamiltonian would be associated with probabilities $\rho_\ell \equiv w_\ell / \lambda$, the value $\widetilde{\rho_\ell}$ is a $\mu$-bit binary approximation to $\rho_\ell$ such that
\begin{align}
\left | \rho_\ell - \widetilde{\rho}_\ell \right | &=  \frac{\left | w_\ell - \widetilde{w}_\ell \right |}{\lambda} \leq \frac{1}{2^\mu L} \leq  \frac{\delta}{\lambda} =\frac{\sqrt{2}\Delta E}{4 L\lambda\left( 1+\frac {\Delta E^2}{8\lambda^2}\right)}\left( 1-\left\|H\right\|^2/\lambda^2\right), \\ \label{eq:qrom_ancilla}
\mu &= \left\lceil \log\left(\frac{2\sqrt{2} \lambda}{\Delta E}\right)+\log\left( 1+\frac {\Delta E^2}{8\lambda^2}\right)-\log\left( 1-\frac{\left\|H\right\|^2}{\lambda^2}\right) \right\rceil,
\end{align}
where the expression for $\delta$ comes from \eq{delta_bound} of \app{errAnal}, and bounds the largest acceptable deviation in the coefficients of the terms in a Hamiltonian approximating the one we mean to implement.
The second log in \eq{qrom_ancilla} is ${\cal O}(1)$, because we do not take $\Delta E$ larger than $\lambda$.
The Hamiltonians we will be considering will be frustrated, so $\left\|H\right\|/\lambda$ is no larger than a constant (less than 1), and the third log in \eq{qrom_ancilla} is ${\cal O}(1)$ as well.

\begin{figure}[tbh]
\centering
  \resizebox{.6\linewidth}{!}{
    \includegraphics{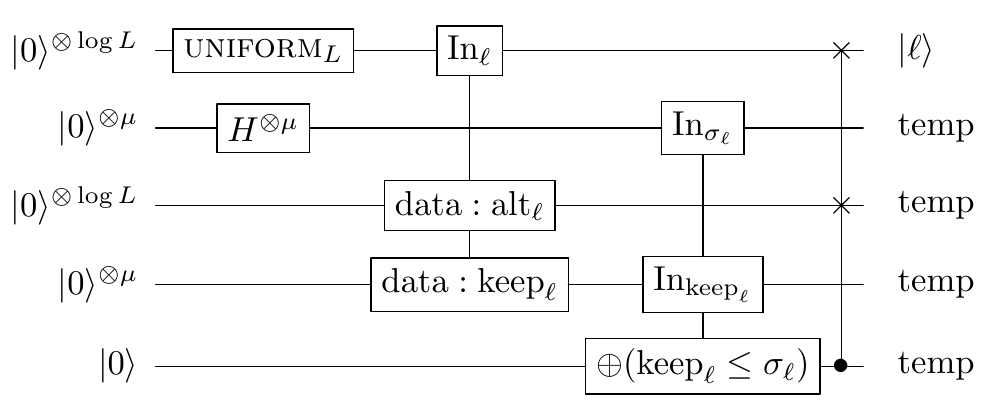}
  }
  \caption{
    \label{fig:reversible-sampling-circuit}
    Generic \textsc{subprepare} circuit for initializing an arbitrary state with $L$ unique amplitudes. Spans $2\mu + 2 \log L + {\cal O}(1)$ qubits and has a T-count of $4 (L + \mu) + {\cal O}(\log L)$ where $\mu$ is calculated as in \eq{qrom_ancilla}. Produces a state $\sum_{\ell=0}^{L-1} \sqrt{w_\ell / \lambda} |\ell\rangle |\text{temp}_\ell\rangle$ where ``temp" is temporary garbage data that will be uncomputed when uncomputing the preparation of this state after application of \textsc{select}. The data loading parts of this circuit use the QROM implementation described in \fig{qrom} in \sec{qrom}. The $\textsc{uniform}_L$ circuit is used to initialize the initial superposition over $\ell$.
}
\end{figure}

\begin{figure}[h]
  \resizebox{\linewidth}{!}{
    \includegraphics{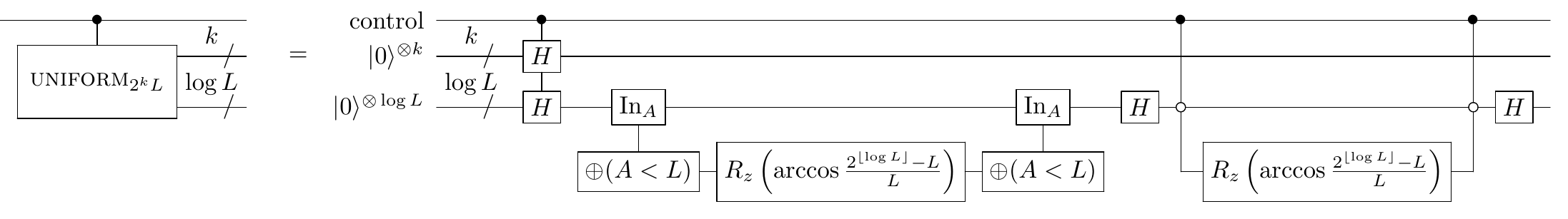}
  }
  \caption{
    A circuit that uses amplitude amplification \cite{Grover2000,Hyer2000} to conditionally and reversibly prepare the uniform superposition $(1/\sqrt{2^k L}) \sum_{\ell=0}^{2^k L - 1} |\ell\rangle$, where $L$ is odd, starting from the $|0\rangle$ state.
    The circuit spans $k + 2 \log L + {\cal O}(1)$ qubits and has a T-count of $2k + 10 \log L + {\cal O}(\log (1/\epsilon))$.
    If the control is omitted, the T-count drops to $8 \log L + {\cal O}(\log (1/\epsilon))$.
    If $L=1$, the $R_z$ rotations are not needed and the T count drops to $2k$.
    If $L=1$ and the control is omitted, the T count is zero.
  }
  \label{fig:uniform}
\end{figure}

The idea behind our scheme will be to create the superposition in an indirect fashion which involves starting in a uniform superposition over an initial index $\ell$ then using a precomputed binary representation of a probability (loaded from QROM), $\textrm{keep}_{\ell}$, to decide whether we should keep $\ell$ or swap it with a classically precomputed alternate index $\textrm{alt}_{\ell}$ which is also loaded from QROM (see \fig{reversible-sampling-circuit}).
Specifically, our procedure creates a uniform superposition in $\ket{\ell}$ over $L$ values and then uses QROM (see \fig{qrom} and \sec{qrom}) to load $\ket{\text{alt}_{\ell}}$ and $\ket{\textrm{keep}_{\ell}}$. Note that if $L$ is not a binary power one can prepare the initial superposition using the amplitude amplification circuit discussed later in \fig{uniform}. The procedure described thus far prepares the state:
\begin{equation}
\sum_{\ell = 0}^{L-1} \sqrt{\frac{1}{L}}\ket{\ell}\ket{\textrm{alt}_{\ell}}\ket{\textrm{keep}_{\ell}}.
\label{eq:preswapped}
\end{equation}
We will then construct a circuit which coherently swaps the registers $\ket{\ell}$ and $\ket{\textrm{alt}_{\ell}}$ with probability $\textrm{keep}_{\ell}$ to create the state in \eq{swapped}.
In order to create the state in \eq{swapped} from \eq{preswapped} we will need to introduce one additional register of size $\mu$, which we refer to as $\ket{\sigma}$.
We will put this entire register into a uniform superposition, and then compare it to the probability represented by $\textrm{keep}_\ell$. If $\textrm{keep}_\ell \le \sigma$, we will swap registers $\ket{\ell}$ and $\ket{\textrm{alt}_{\ell}}$. Thus, after the procedure is finished, i.e.\ in \eq{swapped}, the garbage register will be in the state
\begin{align}
\label{eq:define_subsampling_temp}
\ket{\textrm{temp}_\ell}
=
\frac{1}{\sqrt{2^\mu L \widetilde{\rho_\ell}}} \left(
  \ket{\textrm{alt}_\ell} \ket{\textrm{keep}_\ell} \sum_{\sigma=0}^{\textrm{keep}_\ell-1} \ket{\sigma} \ket{0}
  +
  \sum_{k | \textrm{alt}_k = \ell} \ket{k} \ket{\textrm{keep}_k} \sum_{\sigma=\textrm{keep}_k}^{2^\mu-1} \ket{\sigma} \ket{1}
\right)
\end{align}
where the rightmost qubit is the result of a comparison between $\textrm{keep}_\ell$ and $\sigma$.
For Eq.~\eqref{eq:swapped} to give the correct state, we need $\ket{\textrm{temp}_\ell}$ to be normalized, which means that we require
\begin{equation}\label{requirekl}
\frac{\textrm{keep}_\ell +  \sum_{k | \textrm{alt}_k  = \ell} \left(2^\mu - \textrm{keep}_k \right)}{2^\mu L} = \widetilde{\rho}_\ell = \frac{\widetilde{w}_\ell}{\lambda}
\qquad \qquad \forall \ell \in [0, L).
\end{equation}

We can find the values of $\textrm{keep}_\ell$ and $\textrm{alt}_\ell$ from Eq.~\eqref{requirekl} in a sequential way.
At any step, let ${\cal L}$ denote the set of $\ell$ for which we have already found these values.
Then we can rewrite Eq.~\eqref{requirekl} as
\begin{equation}\label{requirekl2}
\frac{\textrm{keep}_\ell +  \sum_{k\notin{\cal L} | \textrm{alt}_k  = \ell} \left(2^\mu - \textrm{keep}_k \right)}{2^\mu L} = \widetilde{\rho}_\ell  - \frac{\sum_{k\in{\cal L} | \textrm{alt}_k  = \ell} \left(2^\mu - \textrm{keep}_k \right)}{2^\mu L} = \tilde \rho'_\ell.
\end{equation}
This expression involves only known quantities on the right-hand side, which we call $\tilde \rho'_\ell$ for short.
We will show by induction that the average of $\tilde \rho'_\ell$ for $\ell\notin{\cal L}$ is $1/L$, and the $\tilde \rho'_\ell$ are non-negative.
These are clearly true initially, because then $\tilde \rho'_\ell=\tilde \rho_\ell$. Now assume that these conditions are true at some step.
If the values $\tilde \rho'_\ell$ are all equal for $\ell\notin{\cal L}$, then we can just take $\textrm{alt}_\ell=\ell$ and any value of $\textrm{keep}_\ell$, and we will satisfy Eq.~\eqref{requirekl2} for all remaining $\ell$.
Otherwise, there will be one value, $\ell_0$, where $\tilde \rho'_{\ell_0}$ is below the average $1/L$ and another, $\ell_1$, where $\tilde \rho'_{\ell_1}$ is above $1/L$.
For $\ell_0$, we choose $\textrm{keep}_{\ell_0}= 2^\mu L \tilde \rho'_{\ell_0}$ and $\textrm{alt}_{\ell_0}=\ell_1$.
Then $\ell_0$ is added to the set ${\cal L}$ and the values of $\tilde \rho'_\ell$ are updated.
According to Eq.~\eqref{requirekl2},
the only value of $\tilde \rho'_{\ell}$ that is updated is that for $\ell=\ell_1$, where we replace it with $\tilde \rho'_{\ell_1}+\tilde \rho'_{\ell_0}-1/L$.
That ensures the average value of $\tilde \rho'_{\ell}$ for $\ell\notin{\cal L}$ is still $1/L$, and since we had $\tilde \rho'_{\ell_1}>1/L$, the new value is non-negative.

A more intuitive way to understand our approach to preparation is that it is equivalent to classical alias sampling \cite{Walker1974}, which samples $\ell$ with probability $\tilde\rho_\ell$ by the following procedure:
\begin{enumerate}
\item{Select $\ell$ uniformly at random from $[0, L)$.}
\item{Look up $\textrm{alt}_{\ell}$ and $\textrm{keep}_{\ell}$.}
\item{Return $\ell$ with probability $\textrm{keep}_{\ell}/2^\mu$, otherwise return $\textrm{alt}_{\ell}$.}
\end{enumerate}
The procedure for determining the $\textrm{alt}_{\ell}$ and $\textrm{keep}_{\ell}$ is then to work backwards starting from the distribution $\tilde\rho_\ell$, and update this distribution by shifting probabilities from $\ell_1$ to $\ell_0$ until we obtain a uniform distribution \cite{Vose1991}.

\begin{figure}[h]
\includegraphics[width=\linewidth]{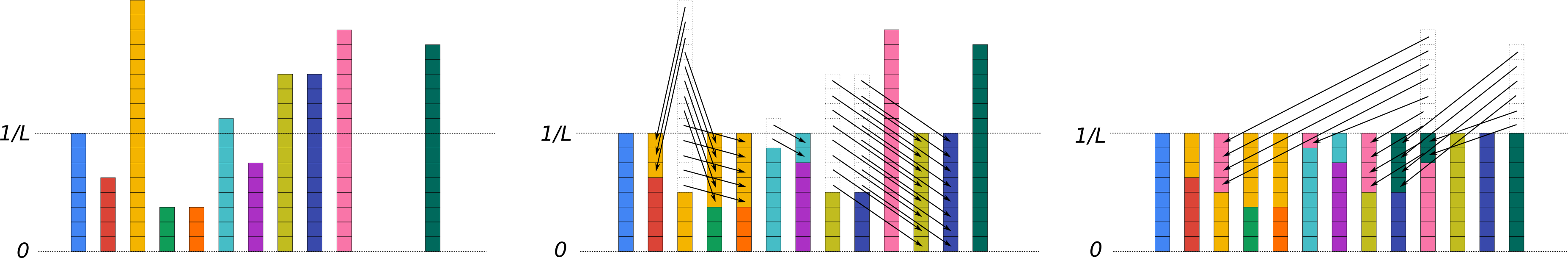}
\caption{\label{fig:subsampling}
Depiction of choosing $\text{alt}_\ell$ and $\text{keep}_\ell$ values for a discretized probability distribution.
The left histogram is the input probability distribution, after discretization into steps of $1/(2^3 L)$.
The squares must be redistributed so that every column has at most two colors and height of $1/L$ (i.e.\ 8 squares as indicated by the dashed line) without moving the bottom square of each column.
The histogram on the right satisfies these constraints, and the histogram in the middle is an intermediate distribution.
Arrows indicate which boxes were moved where.
The color of the top square of each column in the histogram on the right determines the value of $\text{alt}_\ell$, whereas $\text{keep}_\ell$ is determined by where the color transition is within each column.
}
\end{figure}

This procedure is illustrated in \fig{subsampling}.
One starts with a histogram of the desired distribution and looks for a bar that is too small, fixes this by transferring probability from a bar that is too high, and so on until all bars have the correct height. Each probability transfer permanently solves the bar that was too low, and the remaining bars form a smaller instance of the same problem. Thus, it is not possible to get stuck in a loop or a dead-end.
See also the module \path{utils/_lcu_util.py} in version 0.6 of \textsc{OpenFermion} (\url{www.openfermion.org/}) \cite{openfermion} for open source python code that performs this iterative matching process (and also handles discretizing the distribution) in ${\cal O}(L)$ time.

\FloatBarrier

\section{Constructions for the Electronic Structure Hamiltonian}
\label{sec:chemistry}

Using the appropriate discretization into a basis of $N$ spin-orbitals, the electronic structure Hamiltonian can be written as
\begin{equation}
\label{eq:hamiltonian}
H = \sum_{p,q, \sigma} T(p-q)\, a^\dagger_{p,\sigma} a_{q,\sigma}
+ \sum_{p,\sigma} U(p)\, n_{p, \sigma}
+ \sum_{(p, \alpha) \neq (q, \beta)} V(p-q)\, n_{p, \alpha} n_{q, \beta},
\end{equation}
where $a^\dagger_{p, \sigma}$ and $a_{p, \sigma}$ are fermionic creation and annihilation operators on spatial orbital $p \in \{0, \cdots, N/2-1\}$ with spin $\sigma \in \{\uparrow, \downarrow\}$, and $n_{p, \sigma} = a^\dagger_{p, \sigma} a_{p, \sigma}$ is the number operator. These operators satisfy the canonical fermionic anticommutation relations $\{a^\dagger_{p, \alpha}, a^\dagger_{q, \beta}\} = \{a_{p, \alpha}, a_{q, \beta}\} = 0$ and $\{a^\dagger_{p, \alpha}, a_{q, \beta}\} = \delta_{p,q} \delta_{\alpha,\beta}$.

Mapping to qubits under the Jordan-Wigner transformation \cite{jordan1928,Somma2002}, \eq{hamiltonian} becomes
\begin{align}
\label{eq:jw_ham}
H  = & \sum_{p \neq q, \sigma} \frac{T(p-q)}{2} \left( X_{p, \sigma} \overrightarrow{Z} \, X_{q,\sigma} + Y_{p,\sigma} \overrightarrow{Z} \, Y_{q,\sigma} \right)
+ \sum_{(p,\alpha) \neq (q,\beta)} \frac{V(p - q)}{4} Z_{p, \alpha} Z_{q, \beta} \\
& - \sum_{p, \sigma} \left(\frac{T(0) + U(p) + \sum_{q} V(p-q)}{2} \right) Z_{p,\sigma} 
+ \sum_{p} \left(T(0) + U(p) + \sum_{q} \frac{V(p-q)}{2} \right) \openone \nonumber
\end{align}
where we have introduced the notation $\overrightarrow{Z}$ which will be used throughout the paper, which we now explain. The tensor factors on which Pauli operators act can always be interpreted as some integer. For instance, $(p,\sigma)$ is mappable to an integer under a particular choice of canonical ordering in the Jordan-Wigner transformation. When placed between two Pauli operators the notation $A_j \overrightarrow{Z} A_k$ denotes the operator $A_j Z_{j+1} \cdots Z_{k-1} A_k$. The exact mapping between a spin-orbital indexed by $(p,\sigma)$ and a qubit indexed by an integer is discussed later on.

The forms of \eq{hamiltonian} and \eq{jw_ham} encompass a wide range of fermionic Hamiltonians including the molecular electronic structure (aka ``quantum chemistry'') Hamiltonian in any basis that diagonalizes the Coulomb potential \cite{BabbushLow}. The particular coefficients will depend on the discretization scheme and basis functions chosen to represent the system. One such representation, derived for use in quantum simulations in \cite{BabbushLow} prescribes the coefficients
\begin{align}
\label{eq:dualbasishamiltonian}
T(p) = \sum_{\nu} \frac{k_\nu^2 \cos\left(k_\nu \cdot r_p\right)}{2 \, N}, 
\qquad
U(p) = -\! \sum_{j, \nu \neq 0} \frac{4 \pi \, \zeta_j \cos \left(k_\nu \cdot R_j - k_\nu \cdot r_p\right)}{\Omega \, k_\nu^2},
\qquad 
V(p) = \sum_{\nu \neq 0} \frac{2 \pi \cos\left(k_\nu \cdot r_p \right)}{\Omega \, k_\nu^2},
\end{align}
where each spatial orbital $p$ is associated with an orbital centroid $r_p = p \, (2 \Omega / N)^{1/3}$, and $\Omega$ is the computational cell volume. The momentum modes are defined as $k_\nu = 2\pi \nu / \Omega^{1/3}$ with $\nu \in [-(N/2)^{1/3}, (N/2)^{1/3}]^{\otimes 3}$. When dealing with molecular potentials, $R_j$ and $\zeta_j$ are the position and charge of the $j^\text{th}$ nucleus, respectively.

As discussed in \cite{BabbushLow}, the Hamiltonian of \eq{dualbasishamiltonian} corresponds to discretization in a basis composed of rotated plane waves known as the ``plane wave dual'' basis. The basis set discretization error associated with the dual basis is asymptotically equivalent to a Galerkin discretization using any other single-particle basis functions, including Gaussian orbitals \cite{BabbushLow}. Thus, \eq{dualbasishamiltonian} is a general expression of the electronic structure problem that is asymptotically equivalent to any other representation. While well suited for simulating periodic materials, despite asymptotic equivalence, this basis set is not particularly compact for the simulation of molecules. Another basis set compatible with \eq{hamiltonian} and \eq{jw_ham} while being much more appropriate for molecules is the so-called ``Gausslet'' basis \cite{White2017}. Gausslets are derived from a ternary wavelet transformation \cite{Evenbly2016} of Gaussian orbitals and have similar intrinsic basis set discretization error to standard Gaussian orbitals \cite{White2017}.

The simulation procedures here will make use of the structure in \eq{jw_ham}. Specifically, our algorithm will make use of the fact that the Hamiltonian consists of only four types of terms: $Z_p$, $Z_p Z_q$, $X_p \overrightarrow{Z} \, X_q$ and $Y_p \overrightarrow{Z} \, Y_q$ and that there are only $3 N / 2$ unique values of the coefficients. Our algorithms do \emph{not} utilize any particular structure in the dual basis Hamiltonian in \eq{dualbasishamiltonian} beyond the fact that it satisfies the form of \eq{hamiltonian}. This is important since it implies that the techniques of this paper are compatible with other representations of the electronic structure Hamiltonian, such as the finite difference discretization \cite{BabbushLow}, finite element methods, and Gausslet basis sets \cite{White2017} which produce Hamiltonians consistent with \eq{hamiltonian} but not \eq{dualbasishamiltonian}.

\subsection{Electronic Structure Hamiltonian Selection Oracle}
\label{sec:select_chem}

In order to implement the \sel and \prep oracles for the electronic structure Hamiltonian of \eq{hamiltonian}, one must first define a scheme for indexing all of the terms. 
For the case of the general electronic structure Hamiltonian in \eq{hamiltonian} we will index terms with the registers $\ket{\theta}$, $\ket{U}$, $\ket{V}$, $\ket{p}$, $\ket{\alpha}$, $\ket{q}$ and $\ket{\beta}$. The $\ket{p}$ and $\ket{q}$ registers are little-endian binary encodings of integers going from 0 to $N/2-1$, thus using $\log N - 1$ qubits each; the other registers are each a single bit which we use to specify the unitary that $\sel$ should apply to the system register $\ket{\psi}$.

The $\ket{\alpha}$ and $\ket{\beta}$ bits are used to specify the spins $\{\uparrow, \downarrow\}$, which together with the spatial orbital specifications $p$ and $q$, index a spin-orbital. Thus, a register set as $\ket{p}\ket{\alpha}\ket{q}\ket{\beta}$ will be indexing a Hamiltonian term that involves action on the spin-orbitals indexed by $(p,\alpha)$ and $(q,\beta)$. Next, whenever $\ket{U}=\ket{1}$, it will be the case (by construction of our circuits) that $(p,\alpha)=(q,\beta)$ and we will apply the $Z_{p,\alpha}$ terms. If $\ket{V}=\ket{1}$, we will apply the $Z_{p,\alpha} Z_{q,\beta}$ terms. If $\ket{U}\ket{V} = \ket{0}\ket{0}$ and $p < q$, it will also be the case that $\alpha = \beta$ and we will apply the $X_{p,\alpha} \overrightarrow{Z}  X_{q,\alpha}$ terms; if $\ket{U}\ket{V} = \ket{0}\ket{0}$ and $p > q$, it will again be the case that $\alpha = \beta$ and we apply the $Y_{q,\alpha} \overrightarrow{Z}  Y_{p,\alpha}$ terms. Finally, the $\ket{\theta}$ register encodes whether the unitary should have a negative phase (if $\ket{\theta} = \ket{1}$). Thus, our $\sel$ circuit will meet the following specification (where \textsc{undefined} means this case should not occur):
\begin{align}
\label{eq:select_chem}
  \textsc{select}_\textsc{chem}\ket{\theta,U,V,p,\alpha,q,\beta}\ket{\psi} = (-1)^{\theta} \ket{\theta,U,V,p,\alpha,q,\beta} \otimes
  \begin{cases}
    Z_{p,\alpha} \ket{\psi}                                  & \;\;\, U \land \lnot V \land ((p,\alpha) = (q,\beta)) \\
    Z_{p,\alpha} Z_{q,\beta} \ket{\psi}                               & \lnot U \land \;\;\, V \land ((p,\alpha) \neq (q,\beta)) \\
    X_{p,\alpha}\overrightarrow{Z} X_{q,\alpha} \ket{\psi}    & \lnot U \land \lnot V \land (p < q) \land (\alpha = \beta) \\
    Y_{q,\alpha} \overrightarrow{Z} Y_{p,\alpha} \ket{\psi}    & \lnot U \land \lnot V \land (p > q)  \land (\alpha = \beta) \\
    \textsc{undefined} & \text{otherwise}. \\
  \end{cases}
\end{align}

\begin{figure}[h]
  \resizebox{.6\linewidth}{!}{
    \includegraphics{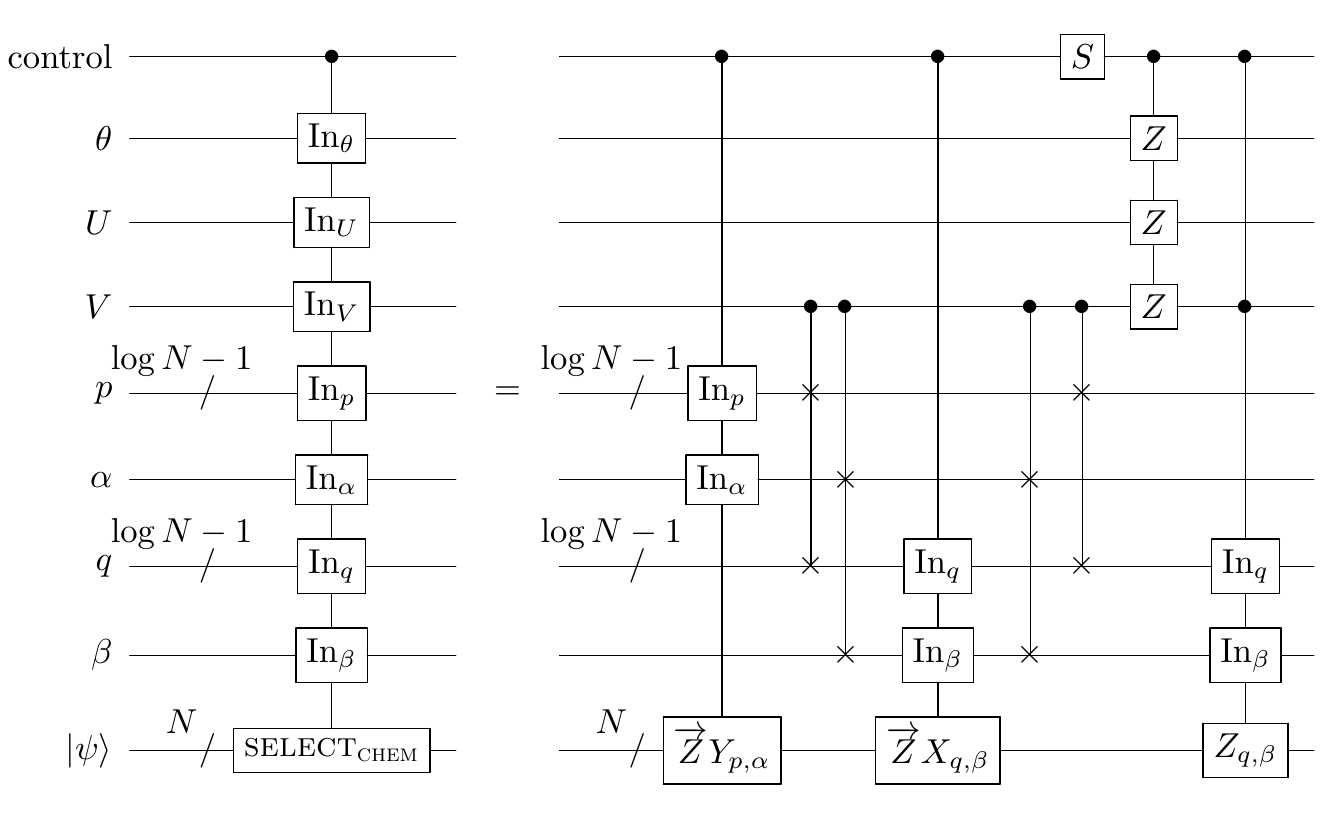}
  }
  \caption{
    \label{fig:selecth}
    $\textsc{select}_\textsc{chem}$ circuit, with a T-count of $12N + 8 \log N + {\cal O}(1)$, which implements the functionality specified by \eq{select_chem}, conditioned on a ``control'' qubit.
    The unitaries performing Majorana and indexed operations (each requiring $4N$ T gates) are explicitly constructed in \sec{streaming}. As described in \fig{majorana}, the unitaries labeled as $\protect\overrightarrow{Z} A_j$ apply the operation $Z_0 \cdots Z_{j-1} A_j$ to the target register, depending on the value from the input $\ket{p}$ register. These operations will require an extra $\log N$ ancillae, so the overall circuit spans $N + 3 \log N + {\cal O}(1)$ qubits. The operation which targets the system register with $Z_q$ is a variant of \fig{unary-iteration-circuit} with the $X_\ell$ gate replaced by $Z_\ell$. All indexed operations reuse the same ancillae.}
\end{figure}

We present our implementation of $\textsc{select}_\textsc{chem}$ in \fig{selecth}. Our circuit relies on subroutines we describe in \sec{streaming} which provide a method for selectively applying strings of Pauli operators to a system register of size $N$, with controls on $\log N$ qubits. Important notation for these subroutines is also defined in \sec{streaming}, and thus that section is necessary for understanding the details of \fig{selecth}.

Since $p$ and $q$ are actually three-dimensional vectors with elements taking integer values $p \in [0, (N/2)^{1/3}-1]$ and $\sigma \in \{\uparrow, \downarrow\}$ we should clarify how the spin orbitals $(p,\sigma)$ are mapped to an integer representing qubits. For ease of exposition, we will define the following mapping function for a $D$ dimensional system,
\begin{equation}
\label{eq:mapping}
M \equiv (N/2)^{1/D}
\qquad \qquad
f(p, \sigma) =  \delta_{\sigma, \downarrow} \, M^D + \sum_{j=0}^{D-1} p_j M^{j}
\end{equation}
where for chemistry $D =3$, for the Hubbard model $D=2$. The $\delta$ function behaves as one might expect; $\delta_{\uparrow, \downarrow} = 0$ and $\delta_{\downarrow, \downarrow}  = 1$. Thus it should be understood that $X_{p, \sigma}$ implies the $X$ operator acting on qubit $f(p, \sigma)$.

\subsection{Electronic Structure Coefficient Preparation Oracle}
\label{sec:prepare_chem}

We see from \eq{hamiltonian} that there are only ${\cal O}(N)$ unique coefficients in the Hamiltonian, despite the Hamiltonian having ${\cal O}(N^2)$ different terms. Based on the indexing in \eq{select_chem} and definition in \eq{prepare}, our \prep initializes
\begin{align}
\label{eq:prepare_chem}
\textsc{prepare}_\textsc{chem} & \ket{0}^{\otimes (3 + 2 \log N)}  \mapsto  \sum_{p,\sigma} \widetilde{U}(p) \ket{\theta_p} \ket{1}_U\ket{0}_V\ket{p,\sigma,p,\sigma}\\
+ & \sum_{p\ne q, \sigma} \widetilde{T}(p-q)\ket{\theta_{p-q}^{(0)}}\ket{0}_U \ket{0}_V \ket{p,\sigma,q,\sigma} 
+\sum_{(p,\alpha) \neq (q,\beta)} \widetilde{V}(p-q)\ket{\theta_{p-q}^{(1)}}\ket{0}_U \ket{1}_V \ket{p,\alpha,q,\beta}\nonumber
\end{align}
where the values of the coefficients and the state of $\ket{\theta}$, related to the coefficients in \eq{jw_ham}, are defined as
\begin{align}
\label{eq:prep_coefficients}
& \widetilde{U}(p) = \sqrt{\frac{\left | T(0) + U(p) + \sum_{q} V(p-q) \right |}{2 \, \lambda}},
\qquad
\widetilde{T}(p) = \sqrt{\frac{\left | T(p) \right|}{\lambda}},
\qquad
\widetilde{V}(p) = \sqrt{\frac{\left | V(p) \right|}{4\, \lambda}},\\
& \theta_p = \frac{1-\textrm{sign}\left(-T(0) - U(p) - \sum_{q} V(p-q)\right)}{2},
\qquad
\theta_{p}^{(0)} = \frac{1 - \textrm{sign}\left(T(p)\right)}{2},
\qquad
\theta_{p}^{(1)} = \frac{1 - \textrm{sign}\left(V(p)\right)}{2}.\nonumber
\end{align}
The $T(p)$ coefficient inside the square root in \eq{prep_coefficients} differs from the coefficient in \eq{jw_ham} by a factor of two since it occurs for each type of term only once depending on whether $p < q$ or $p > q$.

\begin{figure}[h]
  \resizebox{\linewidth}{!}{
    \includegraphics{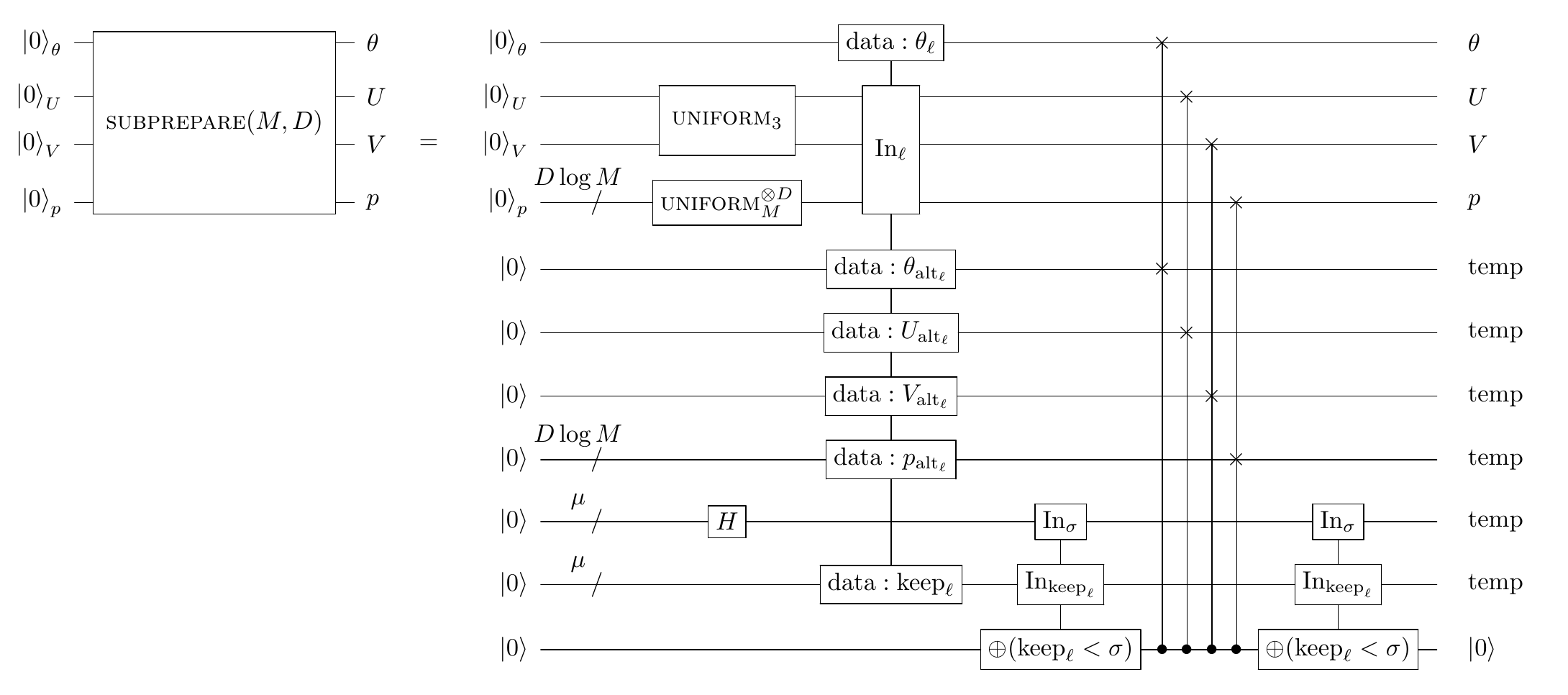}
  }
  \caption{
    \label{fig:subprepare}
    \textsc{subprepare} circuit for the electronic structure Hamiltonian, as called by \fig{prepare}, with a T-count of $6N + {\cal O}(\mu + \log N)$ and a qubit count of $2 \mu + 3 \log N + {\cal O}(1)$ where $\mu$ is defined in \eq{qrom_ancilla}. The data-loading subroutine is implemented as in \fig{qrom}, and has a T-count of $3 \times 4 M^3 - 4 = 6N - 4$. The $\textsc{uniform}$ subroutine is implemented as in \fig{uniform}, and has a T-count of ${\cal O}(\mu)$. The compare-and-swap operations have a negligible ${\cal O}(\log N)$ T-count. As in \eq{mapping}, $D$ denotes system dimension (usually $D=3$) and $M$ refers to the number of values along each dimension such that $N = 2 M^D$. Although we only specify the behavior of the circuit when the $U$, $V$, and $p$ qubits start in the $|0\rangle$ state, the circuit will also be invoked in contexts where this is not the case.
}
\end{figure}

To implement \textsc{prepare} we first synthesize a unitary referred to as \textsc{subprepare} which acts as follows,
\begin{align}
\label{eq:step_one}
\textsc{subprepare} \ket{0}^{ \otimes(2 + \log N)} \mapsto \sum_{d=0}^{N-1} \left(
\widetilde{U}(d)\ket{\theta_{d}}\ket{1}_U\ket{0}_{T} + \widetilde{T}(d)\ket{\theta_{d}^{(0)}}\ket{0}_{U}\ket{0}_{V}
+\widetilde{V}(d)\ket{\theta_{d}^{(1)}}\ket{0}_{U}\ket{1}_{V}
\right)\ket{d}.
\end{align}
Since in this step we initialize a state on ${\cal O}(\log N)$ qubits, the techniques of \cite{Shende2006} would allow one to implement $\textsc{subprepare}$ with a T-count of ${\cal O}(N \log (1/\epsilon))$.
However, in \fig{subprepare} we show an even more efficient method for synthesizing \textsc{subprepare} with T gate complexity ${\cal O}(N + \log (1/\epsilon))$, based on the techniques introduced in \sec{subsample}. Using $\textsc{subprepare}$, we can implement the entire $\prep$ circuit with the same asymptotic T complexity. In our \textsc{subprepare} circuit, $\ell$ is really a vector of integers; thus, we use ``modular vector indices'' such that if $v$ is a 3-dimensional vector within a rectangular space with each dimension having $M$ values then the function application $F(v)$ should be expanded to $F(v) = F(v \mod M) = F(v_x \mod M, v_y \mod M, v_z \mod M)$, consistent with the mapping introduced in \eq{mapping}.

While applying \textsc{subprepare} to create the state in \eq{step_one} we also initialize the $\ket{\alpha}$ qubit in the $\ket{+}$ state with a Hadamard. We then use the $\textsc{uniform}_L$ circuit from \fig{uniform} to initialize the $\ket{q}$ register in an equal superposition in a way that is controlled on the $\ket{U}$ ancilla qubit being in the state $\ket{0}_U$. Subsequent to this step the state becomes 
\begin{align}
\label{eq:step_two}
\textrm{\eq{step_one}} \mapsto & \sum_{d=0}^{N/2-1} \widetilde{U}(d) \ket{\theta_d}\ket{1}_U\ket{0}_{V}\ket{d}\ket{+}_\alpha \ket{0}^{\otimes \log N}\\
& + \sum_{d=0}^{N/2-1}\sum_{q=0}^{N/2-1} \left(
\widetilde{T}(d) \ket{\theta_d^{(0)}} \ket{0}_U \ket{0}_{V}
+\widetilde{V}(d)\ket{\theta_d^{(1)}}\ket{0}_U \ket{1}_{V}\right)\ket{d} \ket{+}_{\alpha} \ket{q} \ket{0} .\nonumber
\end{align}
The register labeled as $\ket{d}$ in \eq{step_one} will ultimately become our $\ket{p}$ register, but immediately after \textsc{subprepare} it is more appropriate to think of it as encoding a value $\ket{p-q}$.
As we can see in \eq{prepare_chem}, when $\ket{V} = \ket{1}$ and $p = q$, it is necessarily the case that $\alpha \neq \beta$. The middle part of our \prep circuit is dedicated to correctly initializing this tricky part of the superposition. To do it, we use an ancilla to apply a Hadamard gate to $\ket{0}_\beta$ only when $\ket{V} = \ket{1}$ and $\ket{d} \neq \ket{0}^{\otimes (\log N - 1)}$. In the event that $\ket{V} = \ket{1}$ and $\ket{d} = \ket{0}^{\otimes (\log N - 1)}$ we apply a CNOT gate with an open control on $\ket{\alpha}$ which targets $\ket{0}_\beta$, thus ensuring that $\ket{\beta} \neq \ket{\alpha}$ when $p - q = 0$. Then, we set $\ket{\beta} = \ket{\alpha}$ for the $U$ and $T$ part of the superposition by applying a Toffoli gate with regular control on $\ket{\alpha}$ and open control on $\ket{V}$, targeting $\ket{\beta}$. After these operations the state can be expressed as
\begin{align}
\label{eq:step_three}
\textrm{\eq{step_two}} \mapsto & \sum_{d=0}^{N/2-1} \sum_{\sigma}\left( \widetilde{U}(d) \ket{\theta_d}\ket{1}_U\ket{0}_{V}\ket{d,\sigma,0,\sigma}
+ \sum_{q=0}^{N/2-1} 
\widetilde{T}(d) \ket{\theta_d^{(0)}} \ket{0}_U \ket{0}_{V}\ket{d,\sigma,q,\sigma}\right)\\
& + \sum_\alpha \left(\widetilde{V}(0)\ket{\theta_0^{(1)}}\ket{0}_U \ket{1}_{V}\ket{0,\alpha,q,\neg \alpha}
+ \sum_\beta \sum_{d=1}^{N/2-1}\sum_{q=0}^{N/2-1} \widetilde{V}(d)\ket{\theta_d^{(1)}}\ket{0}_U \ket{1}_{V}\ket{d,\alpha,q,\beta}\right) .\nonumber
\end{align}

\begin{figure}[h]
  \resizebox{\linewidth}{!}{
    \includegraphics{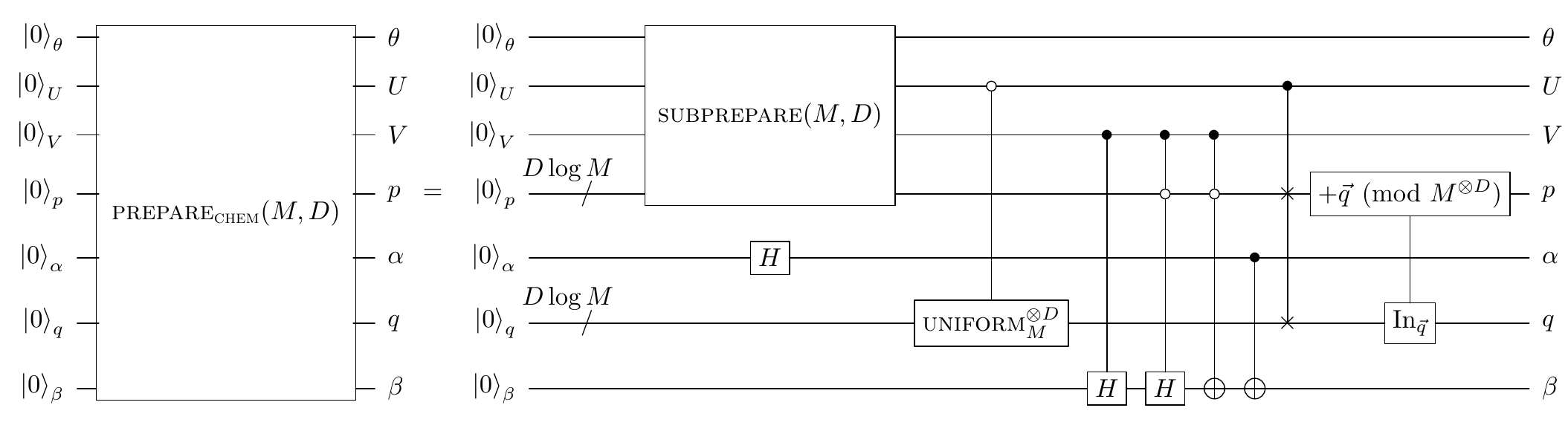}
  }
  \caption{
    $\prep$ circuit for the electronic structure Hamiltonian.
    Implements the unitary in \eq{prepare_chem} with a T count of $6N + {\cal O}(\mu + \log N)$ where $\mu$ is defined in \eq{qrom_ancilla}.
    The $\textsc{subprepare}$ subroutine is the dominant cost, and is implemented as in \fig{subprepare}.
    When $U$ is set by $\textsc{subprepare}$, the Fredkin gates and the addition copy $p$'s value over $q$.
    When $V$ is set, uniform superpositions over $q$ and $\beta$ are prepared except, if $p-q=0$, then $\beta$ is instead set to be opposite to $\alpha$ in order to guarantee $(p,\alpha) \neq (q, \beta)$.
    Beware that this conditional preparation of $\beta$ doubles the weight of $p=q$ cases (relative to $p \neq q$ cases) and must be accounted for in the LCU coefficients given to \textsc{subprepare}.
    When neither $U$ nor $V$ is set, $\alpha$ is copied into $\beta$ and a uniform superposition over $q$ is prepared.
    The remaining operations contribute a negligible ${\cal O}(\mu + \log N)$ T gates.
    The $\textsc{uniform}_M^{\otimes D}$ operation prepares $D$ registers in a uniform superposition of basis states going up to $M$, as in \fig{uniform}.
    Controls on multi-qubit lines are conditioned on every qubit within the line, e.g.\ what appears to be a Toffoli gate is actually a \textsc{not} gate with $1 + D \log M$ controls.
    The action of our $\textsc{prepare}_\textsc{chem}$ circuit is only specified in the case where the inputs are all $|0\rangle$.
    During actual execution this is not the case, because the effects of the \sel operation will prevent $\textsc{prepare}_{\textsc{chem}}^\dagger$ from exactly uncomputing the $U$, $V$, $p$, $q$, $\alpha$, and $\beta$ qubits.
    This is expected behavior and it is accounted for by requiring that the potentially-not-uncomputed qubits be kept and used as inputs for the next $\textsc{prepare}_\textsc{chem}$ circuit.
    }
  \label{fig:prepare}
\end{figure}

The final step consists of converting the $\ket{d}$ register to values representing $\ket{p}$. To do this, we must add the $\ket{q}$ register into the $\ket{d}$ register when $\ket{U} = \ket{0}$ so that $\ket{d + q} = \ket{p-q+q} = \ket{p}$. However, we also want to copy the $\ket{d}$ register into the $\ket{q}$ register when $\ket{U}=\ket{1}$; thus, prior to this operation we also implement a Fredkin gate which swaps $\ket{d}$ and $\ket{q}$, conditioned on $\ket{U}=\ket{1}$. After the Fredkin gate and the addition of $\ket{d}$ into $\ket{q}$,
\begin{align}
\label{eq:step_four}
\textrm{\eq{step_three}} \mapsto & \sum_{d=0}^{N/2-1} \sum_\sigma \left( \widetilde{U}(d) \ket{\theta_d}\ket{1}_U\ket{0}_{V}\ket{d,\sigma,d,\sigma} 
+ \sum_{q=0}^{N/2-1}
\widetilde{T}(d) \ket{\theta_d^{(0)}} \ket{0}_U \ket{0}_{V} \ket{d + q,\sigma,q,\sigma} \right)\\
& + \sum_\alpha \left(\widetilde{V}(0)\ket{\theta_0^{(1)}}\ket{0}_U \ket{1}_{V}\ket{q,\alpha,q,\neg \alpha}
+ \sum_\beta \sum_{d=1}^{N/2-1}\sum_{q=0}^{N/2-1} \widetilde{V}(d)\ket{\theta_d^{(1)}}\ket{0}_U \ket{1}_{V}\ket{d+q,\alpha,q,\beta}\right) .\nonumber
\end{align}
Then, simply by relabeling $d = p-q$ whenever $\ket{U} = \ket{0}$ and $d = p$ whenever $\ket{U} = 1$ we see that our state is identical to the desired one (from \eq{prepare_chem}). We show how to use $\textsc{subprepare}$ to implement $\textsc{prepare}_\textsc{chem}$ in \fig{prepare}. The gate complexity of $\textsc{subprepare}$ is ${\cal O}(N + \log (1/\epsilon))$ and the gate complexity of all other components of this circuit are ${\cal O}(\log N)$. Thus, the overall gate complexity of \prep is ${\cal O}(N +\log 1/\epsilon)$.

\FloatBarrier

\subsection{Resources Required for Electronic Structure Simulation}
\label{sec:chem_t_complexity}

The parameter $\lambda$ from \eq{prepare} has significant implications for the complexity of our algorithm; as seen in \eq{complexity}, our circuit size will scale linearly in $\lambda$. For the case of general electronic structure, we can see from \eq{hamiltonian} that $\lambda$ is
\begin{equation}
\lambda = \sum_{p q} \left | T(p-q) \right |
+ \sum_p \left | U(p) \right |
+ \sum_{p \neq q} \left | V(p-q) \right |.
\end{equation}
This expression and the extremely naive assumption that all coefficients are ${\cal O}(1)$ would imply that $\lambda \in {\cal O}(N^2)$. For the case of quantum chemistry in the dual basis, i.e.\ \eq{dualbasishamiltonian}, the work of \cite{BabbushLow} obtains the same bound:
\begin{equation}
\lambda \in {\cal O}\left(\frac{N^{7/3}}{\Omega^{1/3}} + \frac{N^{5/3}}{\Omega^{2/3}}\right) \in {\cal O}\left(N^{2}\right),
\end{equation}
where the last relation holds when studying electronic structure systems that grow with fixed density $N \propto \Omega$, which is the usual situation. For encoding the electronic structure Hamiltonian we also determined that $P = 6N + {\cal O}(\log (N/\epsilon))$ and $S = 12 N + {\cal O}(\log N)$ in terms of T complexity. Thus, from \eq{complexity} we can conclude that the overall T complexity of our procedure is roughly
\begin{equation}
\label{eq:chem_cost}
\frac{\sqrt{2} \pi \lambda \left(S + 2 P \right) }{\Delta E}
\approx \frac{24 \sqrt{2} \pi \, \lambda}{\Delta E} N
\end{equation}
which for the electronic structure Hamiltonian is rigorously bounded by ${\cal O}(N^3 / \Delta E)$.

Ancilla required for our electronic structure simulation come from three sources: qubits required for our entanglement-based phase estimation (given by \eq{pea_bits}), qubits required to store coefficient values in QROM (given by \eq{qrom_ancilla}), and ancilla actually required for our implementation of \prep and \textsc{select}, which for the electronic structure Hamiltonian simulation is $5 \log N + {\cal O}(1)$. Putting these sources together, the total ancillae required are
\begin{equation}
\label{eq:chem_ancilla}
\log \left(\frac{\sqrt{2} \pi \lambda}{2\Delta E}\right)
+ 2 \log\left(\frac{2 \sqrt{2} \lambda}{\Delta E}\right)
+ 5 \log N  + {\cal O}(1)
=  \log \left(\frac{4\sqrt{2}\pi\lambda^3 N^5}{\Delta E^3}\right)   + {\cal O}(1)
\end{equation}
where the additive constant is small and can be usually neglected for problem sizes of interest.
This expression gives the ancilla count in \thm{chemistry}.

We now estimate resources required for specific problem instances. In practice, we find that $\lambda$ scales better than ${\cal O}(N^2)$, but exactly how much better is system dependent. For a particular material the value of $\lambda$ can be influenced by a number of factors. These factors include the particulars of the bases used, the geometry and atomic composition of the material, and whether one scales toward continuum or thermodynamic limits.

Perhaps the simplest chemistry system which is classically intractable is a molecule without nuclei: the uniform electron gas, also known as jellium. Jellium is a system of $\eta$ electrons with real kinetic energy and Coulomb interactions confined to a box of finite volume $\Omega$ with periodic boundary conditions. Plane waves are a near-ideal basis for the simulation of jellium; the system is naturally expressed using the discretization of \eq{dualbasishamiltonian} with a constant external potential, i.e.\ $\zeta_j = 0$. Jellium is an interesting system to simulate on early quantum computers due to its simplicity, classical intractability \cite{BabbushLow}, historical significance tied to breakthroughs in density functional theory \cite{Hohenberg1964} as well as the fractional quantum Hall effect \cite{stone1992quantum}, and tradition as a benchmark for classical electronic structure calculations.

\begin{figure}[b]
\includegraphics[width=.4\linewidth]{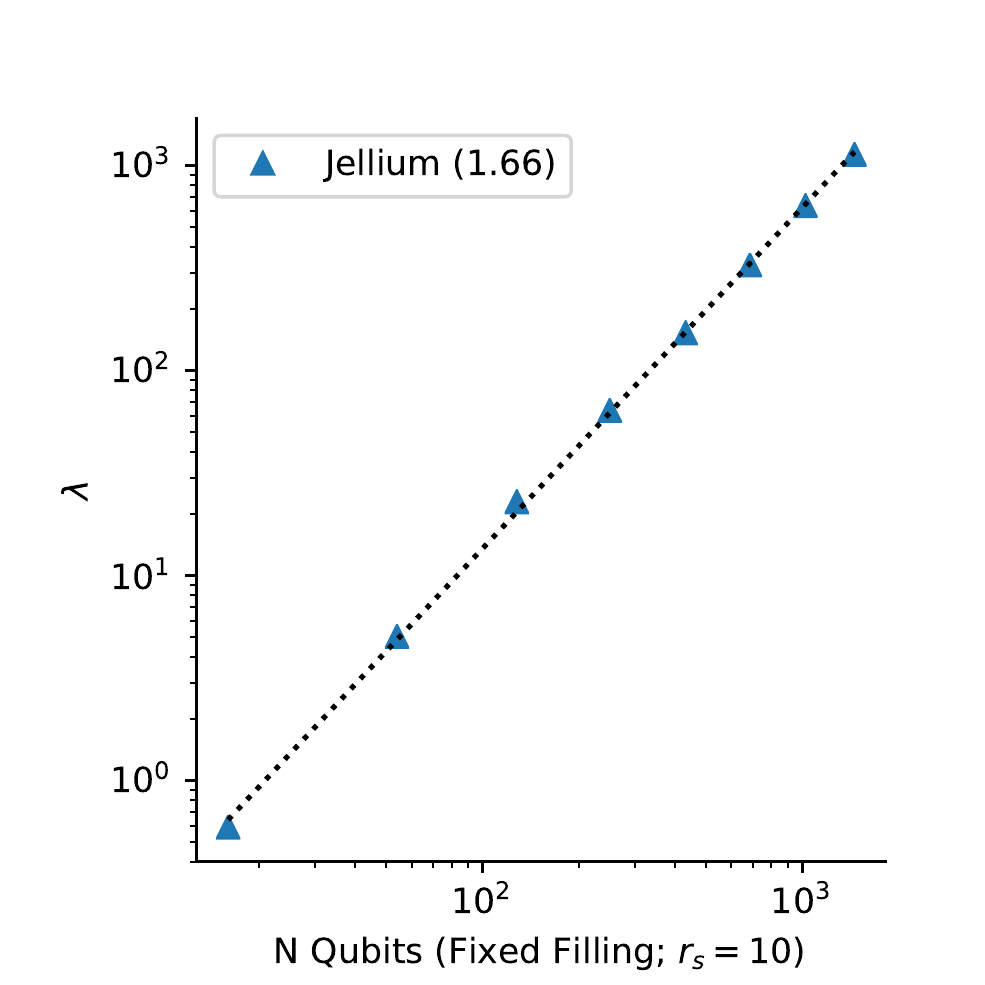}
\caption{\label{fig:lambda_dual} The value $\lambda$ as a function of number of qubits $N$ for the 3D spinful dual basis jellium Hamiltonian at a Wigner-Seitz radius of 10 Bohr radii (assuming the system is initialized at half-filling so $N = \lceil 2 \eta \rceil$), which corresponds to an increasing cell volume as the number of basis functions increases.  This density was chosen for study since it is in the classically challenging regime for jellium \cite{BabbushLow}. The bound is in atomic units of energy (Hartree). The number in parenthesis corresponds to the best fit exponent for the trend line (the dotted line) which suggests that $\lambda = {\cal O}(\sim N^{5/3})$.}
\end{figure}

The phase diagram of jellium is typically parameterized in terms of the Wigner-Seitz radius which characterizes the electron density in three dimensions as $r_s = (3 \Omega / (4\pi \eta))^{1/3}$ where $\eta$ is number of electrons. Although the ground state of jellium at high densities (metallic, $r_s \sim 1$ Bohr radii per particle) and at very low densities (insulating, $r_s \sim 100$ Bohr radii per particle) is well known, the phase diagram in the intermediate density regime is less certain \cite{ceperley1980ground,tanatar1989ground,zong2002spin,attaccalite2002correlation,drummond2009phase,spink2013quantum}. Whereas perturbation theory performs well in the high-density regime \cite{gell1957correlation,freeman1977coupled}, quantum Monte Carlo has been the most competitive simulation tool in the low- to intermediate-density regimes \cite{shepherd2012full,shepherd2012investigation,wilson1995constrained,motta2015imaginary}. For systems with more than fifty electrons, quantum Monte Carlo simulations of jellium typically introduce a bias to control the sign problem, such as the fixed node approximation, full configuration quantum Monte Carlo with initiators, or auxiliary field quantum Monte Carlo with a constrained phase bias. The systematic error from these biases is thought to be as large as half a percent in the energy~\cite{tanatar1989ground,shepherd2012full}, which is on a similar scale to the energy difference between competing phases in the intermediate density regime. Even for modest system sizes such as fifty electrons and twice as many spin-orbitals, quantum simulations can offer bias-free results that cannot be obtained by quantum Monte Carlo.

We include numerics in \fig{lambda_dual} which empirically estimate a tighter bound on $\lambda$ for jellium in the classically challenging regime corresponding to $r_s = 10$ Bohr radii at half-filling $N = \lceil 2 \eta \rceil$. Those numerics, shown in \fig{lambda_dual}, indicate an empirical scaling of $\lambda = {\cal O}(\sim N^{5/3})$. If we target chemical accuracy of $\Delta E = 0.0016$ Hartree then from \eq{chem_cost} we see that roughly twenty-million T gates would be required for jellium with 54 orbitals, two-hundred million T gates would be required for jellium with 128 orbitals, and about a billion T gates would be required for jellium with 250 orbitals. While these numbers are promising, for small sizes these simulations require a number of ancilla comparable to $N$. T counts and ancilla resources are tabulated for several jellium problem instances in \tab{jellium_ts}.

\begin{table}[t]
\begin{tabular}{c|c|c|c|c}
Spin-Orbitals & $\lambda$ Value & Logical Ancilla & Total Logical & T Count \\
\hline
54 & $5$ & 69 & 123 & $1.8 \times 10^{7}$ \\
128 & $23$ & 82 &  210 & $1.9 \times 10^{8}$ \\
250 & $64$ & 91 & 341 & $1.1 \times 10^{9}$ \\
1024 & $640$ & 112 & 1136 & $4.3 \times 10^{10}$ \\
\end{tabular}
\caption{Resources required for quantum simulation of 3D spinful jellium in the dual basis at Wigner-Seitz radius of 10 Bohr radii where the cell volume is calculated assuming the system is at half filling. The units for $\lambda$ are Hartree. The number of logical ancillae is computed as \eq{chem_ancilla} and the number of T gates is computed using \eq{chem_cost}. These estimates assume (rather conservatively in comparison to classical limitations) that we should target additive chemical accuracy of $\Delta E = 0.0016$ Hartree. These problem sizes are large enough that contemporary classical methods cannot reliably provide unbiased estimates with low enough systematic error to resolve competing phases within the fixed-size basis.}
\label{tab:jellium_ts}
\end{table}

The dual basis of \eq{dualbasishamiltonian} is also a natural choice for periodic condensed phase systems (e.g.\ solids) besides jellium. Considering only this basis, there are two parameters that determine the accuracy of the simulation with respect to the true material. First, is the number of plane waves used to discretize the cell, which determines the spacing of the quasi-points in the dual basis \cite{BabbushLow}. That is, more plane waves equate to a finer grid and more accurate discretization. The second parameter is the size of the supercell, which determines the error one incurs by representing an infinite system with a finite, periodic one, also known as the finite-size error. There are different ways of reducing the finite-size error for a physical system.  One common method used in density functional theory utilizes Bloch's theorem to divide the sampling problem into so-called ``k-points'' within the first Brillouin zone \cite{Martin2004}.  The smoothness of the energy with respect to the k-points and additional symmetry provided can offer advantages in certain approaches at the cost of increased complexity often resulting in a complex Hamiltonian representation at non-zero k-points.  The origin k-point, also called the gamma point, maintains a real Hamiltonian for the appropriate basis functions.  An alternative to k-point sampling is increasing the size of the supercell, which increases the relevance of the gamma point.  For simplicity, here we only consider the gamma point, so the natural parameter to change is repetitions of the unit cell that fix the size of the supercell being simulated. A larger supercell tends to incur less finite-size error as the system is scaled to the thermodynamic limit.

It is clear that these two parameters are not entirely independent with respect to accuracy of representation of the true system.  For example, a much larger supercell with the same number of grid points clearly offers a more coarse and less accurate representation of the true system.  Moreover, in classical methods it is common practice to extrapolate along both parameters to increase accuracy for a given computational cost \cite{Martin2004}.  We do not introduce such complexities here, but rather show empirically how the choice of these parameters influences the parameter $\lambda$ which determines the cost of our algorithms, leaving optimizations such as extrapolation schemes to future work.

\begin{figure}[h]
\begin{minipage}[h]{.32\linewidth}
\includegraphics[width=\linewidth]{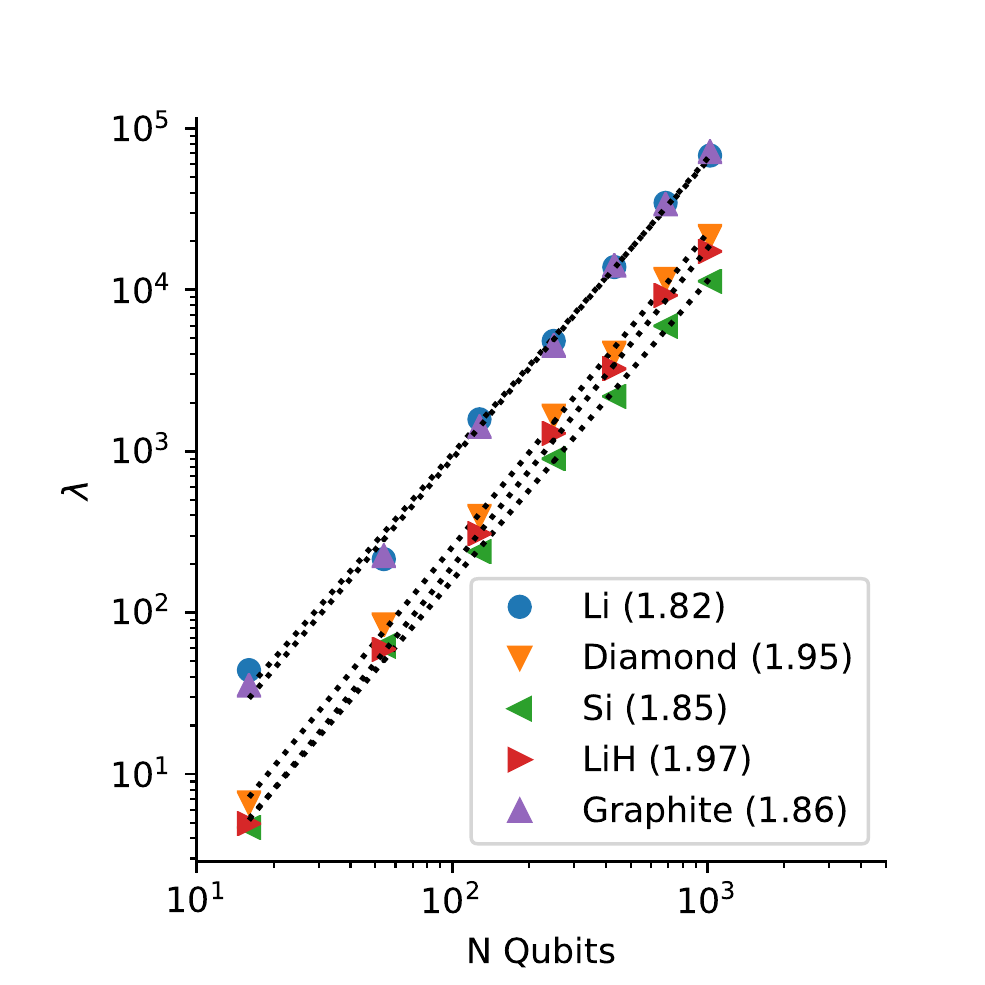}
\end{minipage}
\hfill
\begin{minipage}[h]{.32\linewidth}
\includegraphics[width=\linewidth]{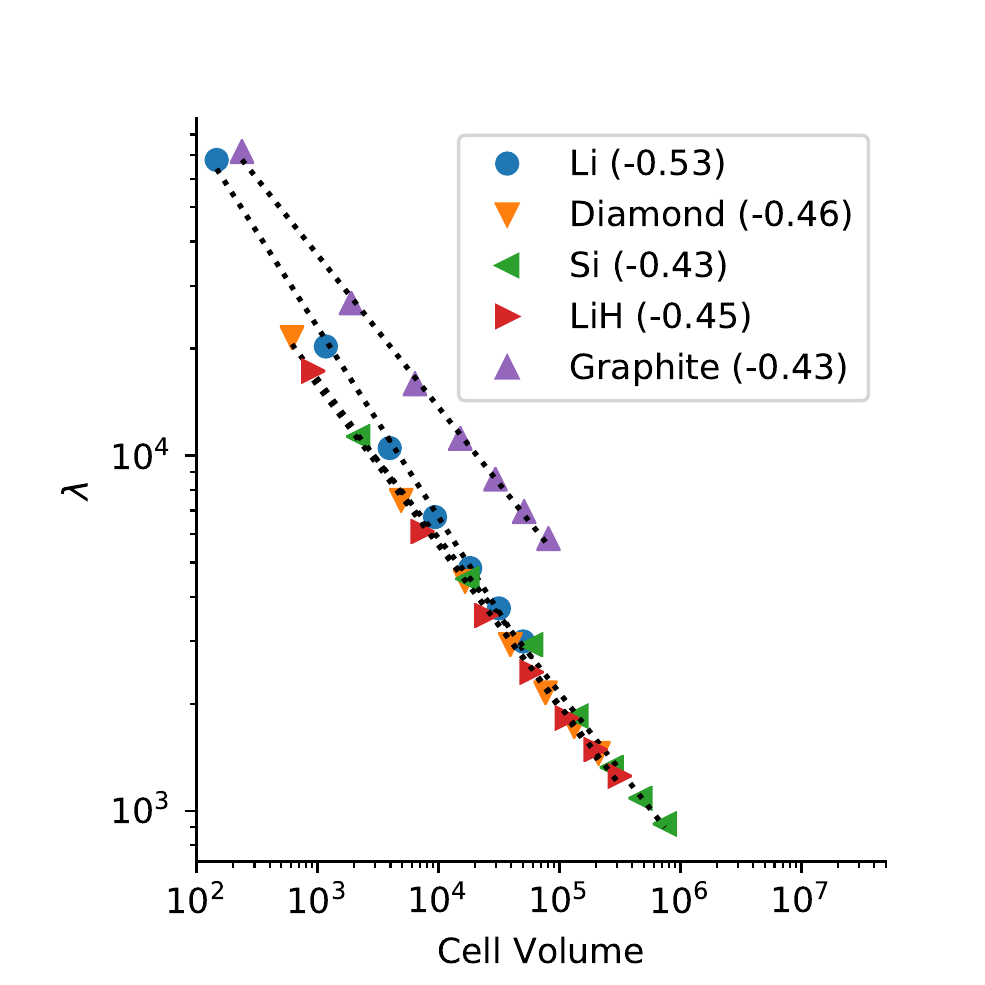}
\end{minipage}
\hfill
\begin{minipage}[h]{.32\linewidth}
    \includegraphics[width=\linewidth]{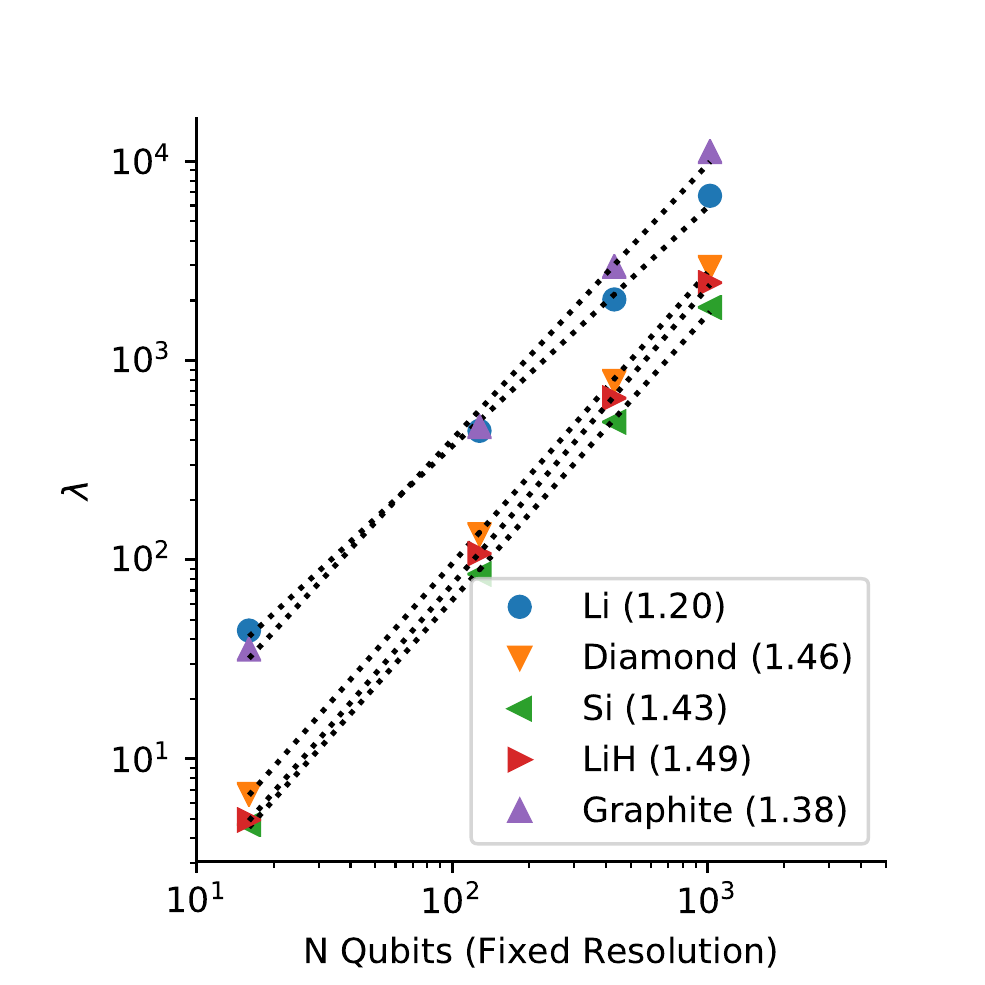}
\end{minipage}
\caption{\label{fig:material-lambda} (Left): A plot of the $\lambda$ value for one unit cell of the listed material near equilibrium bond length as a function of the number of qubits used to discretize the cell. A plane wave dual basis is used with an equal number of points along each of the axes. The value given after the material name corresponds to the best fit scaling for that particular material. The materials display roughly similar scaling in $\lambda$ as a function of the number of qubits. (Center): A plot of the $\lambda$ value as a function of the cell volume for a fixed number of qubits, 1024 in this case. We observe that fixing the number of qubits while increasing the cell size tends to decrease $\lambda$ as expected. (Right): A plot of the $\lambda$ value as a function of the number of qubits where the supercell size is scaled proportionally to the number of points along each axis. In this case, two basis functions are used along each axis in the unit cell, which is scaled proportionally as the cell volume grows.}
\end{figure}

\fig{material-lambda} shows the value of $\lambda$ as a function of the number of qubits being used to discretize the material cell at a fixed supercell size for several real materials.  The number of qubits here is equal to twice the number of plane waves since spin is being treated explicitly.  Equal numbers of plane waves along each of the reciprocal lattice vectors of the supercell are used, as opposed to the more common spherical energy cutoff schemes \cite{Martin2016}, as this enables the use of the plane wave dual basis \cite{BabbushLow}. The exponent (slope on the log-log plot) of a least-squares linear regression to the data is listed alongside the material, and we see empirically that the value of $\lambda$ scales just under $\lambda = {\cal O}(N^2)$ in the number of basis functions while keeping the size of the supercell fixed, which matches the analytical bound rather closely. From the form of the Hamiltonian in \eq{dualbasishamiltonian}, one can see that at a fixed number of plane waves, increasing the volume of the supercell $\Omega$ tends to decrease $\lambda$ such that $\lambda = {\cal O}(\sim \Omega^{-1/2})$, due to representing lower frequency modes with respect to the plane wave representation of the kinetic energy, despite increasing total nuclei present.  We show this effect empirically in the center of  \fig{material-lambda}.

The first two panels of \fig{material-lambda} leave open the question of the impact on $\lambda$ of increasing the supercell size while maintaining a constant density of dual quasi-points or constant density of plane waves, as we expect impact of the last two aspects to compete in some way. Empirically this is shown in the right portion of \fig{material-lambda}, which plots the values of $\lambda$ for a fixed density of points in increasing supercell sizes. Note that this scaling is most comparable to past studies on single molecules since molecular volume tends to grow as one adds electrons.  We observe that the scaling in this case is more favorable as a function of the number of qubits than simply refining the grid alone, and in all cases is better than $\lambda = {\cal O}(\sim N^{3/2})$. From \eq{chem_cost}, this numerical data would suggest that the T complexity of our overall algorithm is empirically bounded by ${\cal O}(\sim N^{5/2} / \Delta E)$ when the goal is simulation of real materials.

To treat molecules properly one should further consider pseudopotentials \cite{tosoni2007comparison}, methods of extrapolation to continuum and thermodynamic limits, and embedding methods \cite{Knizia2012,Bauer2015}. We will leave such a thorough comparison of fault-tolerant resources required for specific instances of real materials other than jellium to future work. However, by comparing \fig{lambda_dual} and the right panel of \fig{material-lambda} it is apparent that jellium is a reasonable proxy for other materials in that $\lambda$ values are comparable. As the rest of the simulation circuit is identical up to the particular angles of certain single-qubit rotations, one can estimate the cost of simulating these materials from our analysis of the fault-tolerant overheads required to simulate jellium in \sec{resources}.

\FloatBarrier

\section{Constructions for the Hubbard Model}
\label{sec:hubbard}

In this section we described specialized implementations of the \sel and \prep oracles for simulation of the planar repulsive-interaction Fermi-Hubbard model and then estimate the overall T complexity of simulating such models. The Hubbard model is a canonical model of a many-electron system often used to model superconductivity in cuprate superconductors. Despite its simplicity, the Hubbard model exhibits a wide range of correlated electron behavior including superconductivity, magnetism, and interaction-driven metal-insulator transitions \cite{SimonsHubbard}.

The Hubbard model is essentially a special case of \eq{hamiltonian} when the model is restricted to a planar grid. The Hamiltonian  can be expressed as
\begin{equation}
\label{eq:hubbard}
H = - t \sum_{\avg{p,q}, \sigma} a^\dagger_{p,\sigma} a_{q,\sigma} + \frac{u}{2} \sum_{p, \alpha \neq \beta} n_{p,\alpha} n_{p,\beta}
\end{equation}
where the notation $\avg{p,q}$ implies that terms exist only between sites that are adjacent on a planar lattice with periodic boundary conditions. This Hamiltonian can be expressed under the Jordan-Wigner transformation as
\begin{align}
H = - \frac{t}{2}\sum_{\avg{p,q},\sigma} \left(X_{p,\sigma} \overrightarrow{Z} X_{p,\sigma} + Y_{p,\sigma} \overrightarrow{Z} Y_{p,\sigma} \right)
+ \frac{u}{8} \sum_{p, \alpha\neq\beta} Z_{p,\alpha} Z_{p,\beta}
- \frac{u}{4} \sum_{p,\sigma} Z_{p,\sigma}
+ \frac{u\, N}{4} \openone .
\end{align}
We will focus on the Hubbard model with periodic boundary conditions (which is a more typical system to study than the Hubbard model with open boundary conditions).

\subsection{Hubbard Model Hamiltonian Selection Oracle}
\label{sec:select_hub}

We see from \eq{hubbard} that there are only three unique coefficients in the Hubbard Hamiltonian: the coefficient of $-XZX$ and $-YZY$ terms is $t/2$, the coefficient of $ZZ$ terms is $u/8$ and the coefficient of local $-Z$ terms is $u/4$. This makes implementation of the \prep circuit exceptionally simple. Ultimately, we will show that the \prep circuit  for the Hubbard model can be implemented at cost ${\cal O}(\log (1/\epsilon))$. This scaling virtually guarantees that for all problem sizes of interest the scaling of the overall algorithms will be dominated by the cost of the \sel circuit.

\begin{figure}[h]
  \resizebox{.7\linewidth}{!}{
    \includegraphics{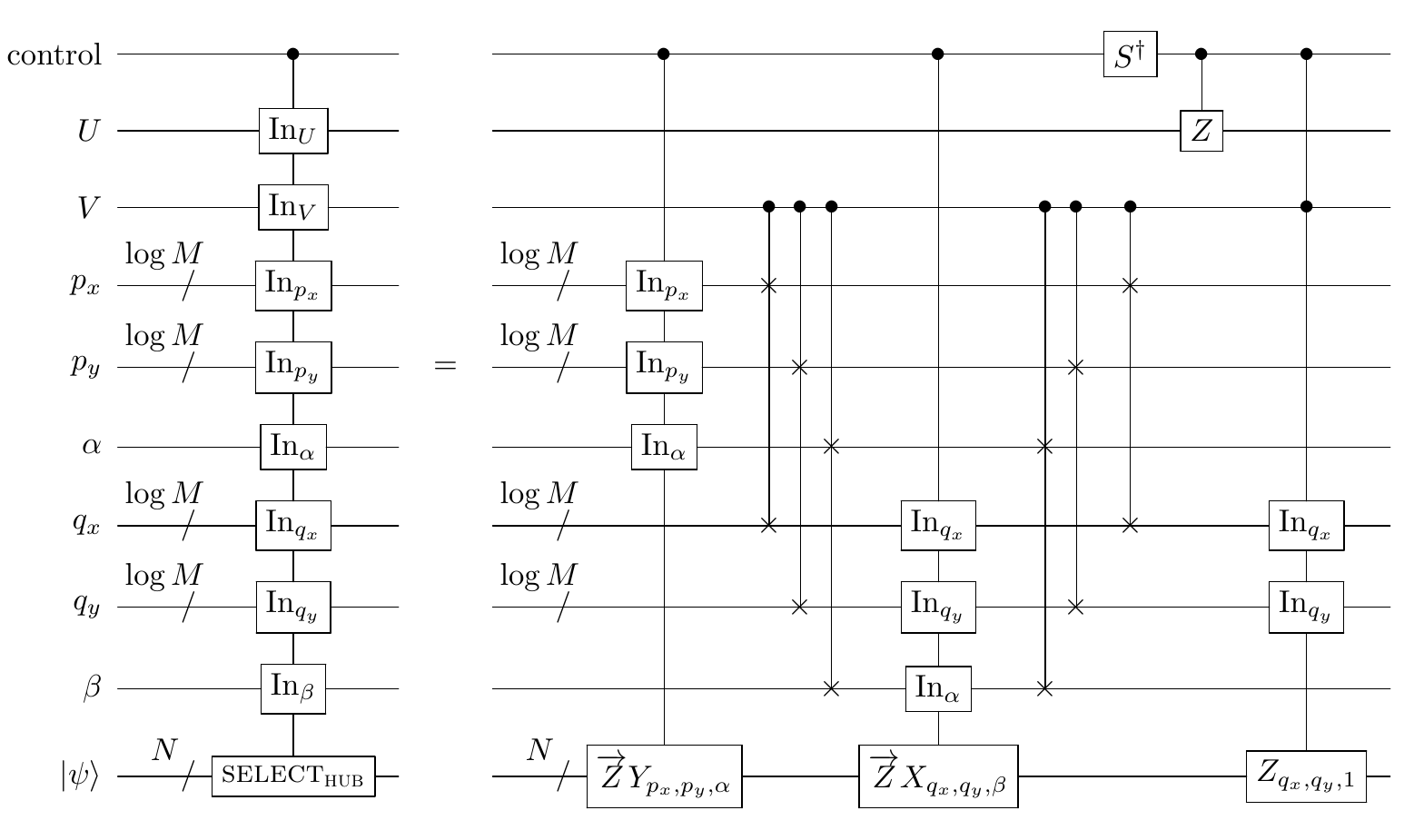}
  }
  \caption{
    \label{fig:selecthub}
    A $\sel$ circuit for the Hubbard model, with function determined by how $p$ relates to $q$. Recall from \eq{mapping} that $M = \sqrt{N/2}$ for the Hubbard model. This circuit has  a T count of $10N + {\cal O}(\log N)$ and spans $N +  3 \log N + {\cal O}(1)$ qubits.
    If $\text{control}$ is \textsc{off}, the circuit has no effect.
    Otherwise, if $\ket{U}\ket{V} = \ket{1}\ket{0}$, it will be the case that $(p,\alpha)=(q,\beta)$, and our circuit applies $-Z_{p,\alpha}$. If $\ket{U}\ket{V} = \ket{0}\ket{1}$, we again have that $p = q$ but this time we also have that $\alpha = 0$ and $\beta=1$ so the circuit applies $Z_{p,0} Z_{q,1}$. If $\ket{U}\ket{V} = \ket{0}\ket{0}$ and $p < q$ the circuit performs $-X_{p,\alpha}\protect\overrightarrow{Z} X_{q,\alpha}$ but if $p > q$ the circuit performs $-Y_{p,\alpha}\protect\overrightarrow{Z} Y_{q,\alpha}$. The larger gates in this circuit are Majorana operators described in \fig{majorana} and an indexed operation explained in \fig{unary-iteration-circuit} (except the $X_\ell$ gate is replaced by a $Z_\ell$ gate).
The Majorana operators each have a T count of $4N$, but the indexed operation has no dependence on $\alpha$ and so has a T-count of $2N$.
All other circuit components have T counts in ${\cal O}(\log N)$.}
\end{figure}

We will index terms in the Hubbard Hamiltonian using the registers $\ket{U}\ket{V}\ket{p_x} \ket{p_y} \ket{\alpha} \ket{q_x} \ket{q_y} \ket{\beta}$. Note that it is important for us to explicitly separate $p_x$ and $p_y$ in our construction of the Hubbard model circuits since this structure is fundamental to the efficiency of our scheme. Our indexing scheme will be nearly identical to the scheme used for the arbitrary chemistry Hamiltonian in \eq{select} but here we will not need the $\theta$ parameter since we know the sign of the parameters in advance. Thus, our $\sel$ circuit for Hubbard will meet the following specification:
\begin{align}
\label{eq:selecthub}
  \textsc{select}_\textsc{hub}\ket{U,V,p,\alpha,q,\beta}\ket{\psi} = \ket{U,V,p,\alpha,q,\beta} \otimes
  \begin{cases}
    -Z_{p,\alpha} \ket{\psi}                                  & \;\;\, U \land \lnot V \land ((p,\alpha) = (q,\beta)) \\
    Z_{p,\alpha} Z_{q,\beta} \ket{\psi}                               & \lnot U \land \;\;\, V \land (p=q) \land (\alpha=0) \land (\beta=1) \\
    -X_{p,\alpha}\overrightarrow{Z} X_{q,\alpha} \ket{\psi}    & \lnot U \land \lnot V \land (p < q) \land (\alpha = \beta) \\
    -Y_{q,\alpha} \overrightarrow{Z} Y_{p,\alpha} \ket{\psi}    & \lnot U \land \lnot V \land (p > q)  \land (\alpha = \beta) \\
    \textsc{undefined} & \text{otherwise}. \\
  \end{cases}
\end{align}
where for ease of exposition $p = p_x + p_y M$ and $q = q_x + q_y M$, consistent with the convention of \eq{mapping} for $D=2$. By exploiting translational invariance in the Hubbard model we are able to implement $\textsc{select}_\textsc{hub}$ in a slightly more efficient fashion, achieving a T count of only $10N + {\cal O}(\log N)$. We show this more efficient implementation in \fig{selecthub}.

\subsection{Hubbard Model Coefficient Preparation Oracle}
\label{sec:prepare_hub}

\begin{figure}[h]
  \resizebox{.9\linewidth}{!}{
    \includegraphics{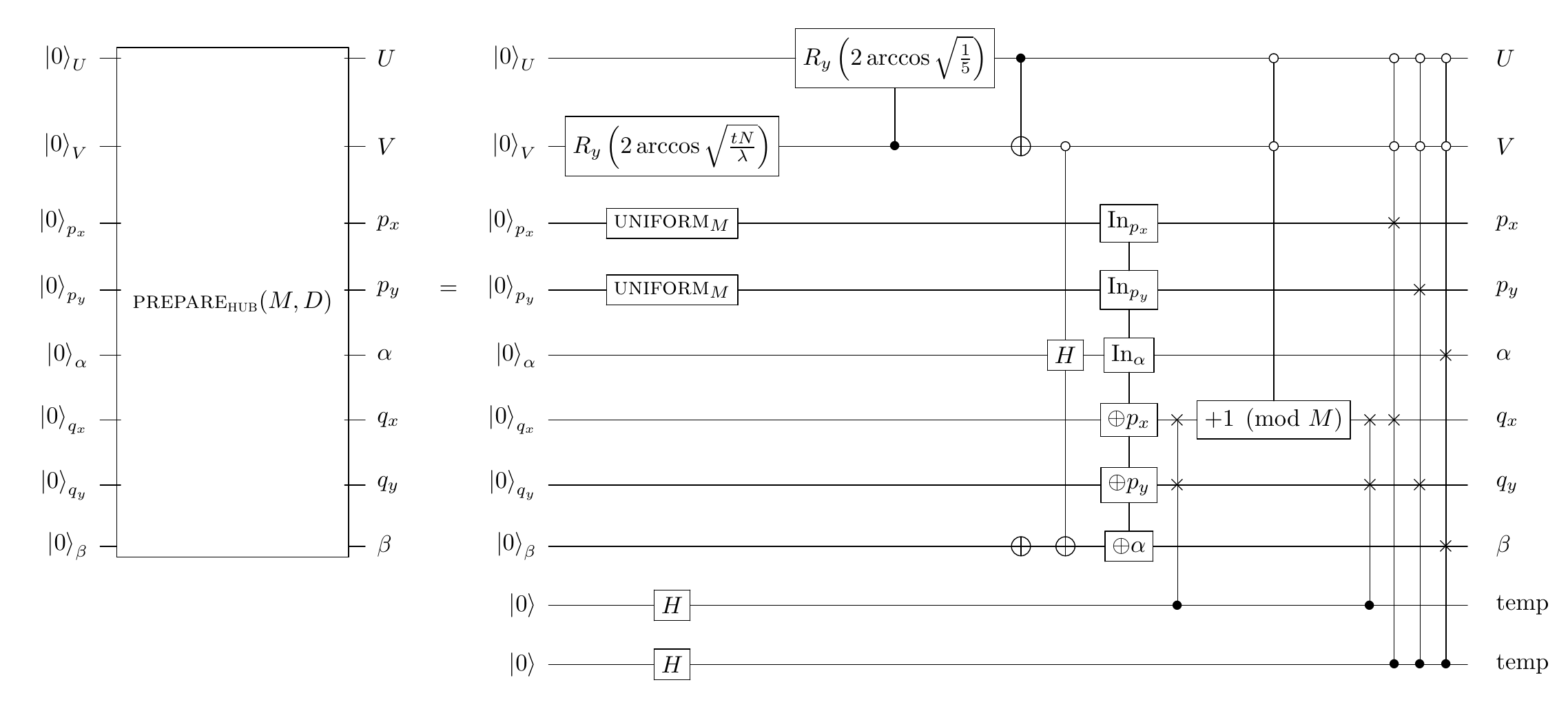}
  }
  \caption{\label{fig:prephub}
    $\textsc{prepare}_\textsc{hub}$ circuit with T-count  ${\cal O}(\log (N/\epsilon))$. The $R_y$ operations are used to prepare the three distinct LCU coefficients which, including multiplicity, are $\sqrt{t N / \lambda}$ (for the $2N$ $T$-type terms), $\sqrt{N u / (4 \lambda)}$ (for the $N=2M^2$ $U$-type terms), and $\sqrt{N u / (8 \lambda)}$ (for the $N/2 = M^2$ $V$-type terms).
    We only specify the action of this circuit in the case where the inputs are all $|0\rangle$.
    During actual execution, the effects of the \sel operation will prevent $\textsc{prepare}^\dagger$ from exactly uncomputing the $U$, $V$, $p_x$, $p_y$, and $\alpha$ qubits as well as the bottom two ancilla qubits. This is expected behavior and it is required that the potentially-not-uncomputed qubits be kept and used as inputs for the next $\textsc{prepare}_\textsc{hub}$ circuit.}
\end{figure}

Our \prep circuit for the Hubbard model will have the following specification
\begin{align}
\label{eq:prephubbard}
\textsc{prepare}_\textsc{hub} & \ket{0}^{\otimes (2 + 2 \log N)}  \mapsto \\
& \sum_{p_x=0}^{M-1} \sum_{p_y=0}^{M-1} \left(\sqrt{\frac{u}{8\lambda}} \ket{0}_U \ket{1}_V \ket{p_x} \ket{p_y} \ket{0}_\alpha\ket{p_x} \ket{p_y} \ket{1}_\beta
 +\sqrt{\frac{u}{4\lambda}} \sum_{\sigma \in \{\downarrow, \uparrow\}} \ket{1}_U \ket{0}_V \ket{p_x} \ket{p_y} \ket{\sigma}\ket{p_x} \ket{p_y} \ket{\sigma}\right. \nonumber\\
& + \sqrt{\frac{t}{2\lambda}}    \sum_{\sigma \in \{\downarrow, \uparrow\}} \left( \ket{0}_U \ket{0}_V  \ket{p_x} \ket{p_y} \ket{\sigma}  \ket{p_x+1} \ket{p_y} \ket{\sigma}
+\ket{0}_U \ket{0}_V  \ket{p_x} \ket{p_y} \ket{\sigma}  \ket{p_x} \ket{p_y+1} \ket{\sigma}\right)\nonumber\\
& \left. + \sqrt{\frac{t}{2\lambda}}    \sum_{\sigma \in \{\downarrow, \uparrow\}} \left( \ket{0}_U \ket{0}_V  \ket{p_x} \ket{p_y} \ket{\sigma}  \ket{p_x-1} \ket{p_y} \ket{\sigma}
+\ket{0}_U \ket{0}_V  \ket{p_x} \ket{p_y} \ket{\sigma}  \ket{p_x} \ket{p_y-1} \ket{\sigma}\right)\right) \nonumber
\end{align}
where the first line above corresponds to $-Z$ and $ZZ$ terms, the second line corresponds to $-X \overrightarrow{Z} X$ terms and the final line corresponds to the $-Y \overrightarrow{Z} Y$ terms. Note that we are looking at a Hubbard model with periodic boundary conditions so wherever something like $\ket{p_x + 1}$ appears we really mean $\ket{(p_x + 1) \mod M}$, which we omitted from the above equation for clarity. Our implementation of $\prep$ begins for the Hubbard model by initializing a two-qubit state containing the three distinct coefficients for the $U$, $V$, and $T$ terms.
This is done with standard circuit synthesis techniques with a T-count of ${\cal O}(\log(1/\epsilon))$ \cite{Shende2006}.
We then spread these coefficients over the various cases.
We depict our implementation in \fig{prephub}.

\FloatBarrier

\subsection{Hubbard Model Resources}
\label{sec:hubbard_t_complexity}

For the case of the planar Hubbard model in \eq{hubbard} it is readily apparent that
\begin{equation}
\lambda = 2 N t + \frac{N u}{2}  \in {\cal O}\left(N\right)
\end{equation}
assuming that we are dealing with the spinful model with periodic boundary conditions. We also determined that $P \in {\cal O}(\log (N/\epsilon))$ and that $S = 10 N + {\cal O}(\log N)$. Thus, the total T cost of the Hubbard algorithm is
\begin{equation}
\label{eq:hubbard_cost}
\frac{\sqrt{2} \pi \lambda \left(S + 2 P \right) }{\Delta E}
= \frac{\sqrt{2} \pi \left(2 t + u / 2\right) N \left(S + 2 P\right)}{\Delta E}
\approx \frac{20 \sqrt{2} \pi t + 5 \sqrt{2} \pi u}{\Delta E} N^2.
\end{equation}
Ancillae required for our Hubbard model simulation come from two sources: qubits required for our entanglement-based phase estimation (see \eq{pea_bits}) and ancillae actually required for our implementation of \prep and \textsc{select}, which for the Hubbard model is $3\log N + {\cal O}(1)$. Putting these sources together, the total ancillae required are
\begin{equation}
\label{eq:hubbard_ancilla}
\log \left(\frac{\sqrt{2} \pi \lambda}{2\Delta E}\right)
+ 3 \log N + {\cal O}(1)
= \log \left(\frac{\sqrt{2} \pi\lambda N^3}{2\Delta E}\right) + {\cal O}(1)
\end{equation}
where the additive constant is small and can usually be neglected for problem sizes of interest.
This expression gives the ancillae count in \thm{hubbard}.

While numerically exact solutions to the Hubbard model are available for one-dimensional \cite{Lieb1968} and infinite-dimensional systems \cite{Metzner1989}, no known polynomial time scaling classical methods can provide reliable solutions to the planar model in all parts of its phase diagram \cite{SimonsHubbard}. For state-of-the-art approximate methods the most challenging low temperature phase of the model appears to be the intermediate interaction regime due to the presence of many competing phases, around $u / t = 4$ \cite{SimonsHubbard}. Accordingly, we will focus our analysis on this regime. If $u = 4t$ then $\lambda = 4 N t$. An interesting and classically challenging-to-obtain accuracy (beyond the agreement of state-of-the-art numerical methods \cite{SimonsHubbard}) for this regime would be in the vicinity of $\Delta E \approx t / 100$ \cite{Jiang2017}; these choices would suggest a T complexity of approximately
\begin{equation}
\label{eq:hubbard_estimate}
\frac{40 \sqrt{2} \pi t}{t / 100} N^2
<
\left(1.8 \times 10^4\right) N^2
\end{equation}
and an ancilla count of approximately
\begin{equation}
\label{eq:hub_ancilla_est}
\log \left(\frac{8 \sqrt{2} \pi t N^3}{t / 100}\right) < 12 + 3 \log N .
\end{equation}
We summarize these resources for various interesting sizes of Hubbard model simulation in \tab{hubbard_ts}.

\begin{table}[h]
\begin{tabular}{c|c|c|c|c}
Dimension & Spin-Orbitals & Logical Ancilla & Total Logical & T Count \\
\hline
$6 \times 6$ & 72 & 33 & 105 & $9.3 \times 10^7$ \\
$8 \times 8$ & 128 & 33 & 161 & $2.9 \times 10^8$ \\
$10 \times 10$ & 200 & 36 & 236 & $7.1 \times 10^8$ \\
$20 \times 20$ & 800 & 42 & 842 & $1.2 \times 10^{10}$\\
\end{tabular}
\caption{Resources required for quantum simulation of planar Hubbard model with periodic boundary conditions and spin, as in \eq{hubbard}. The dimension of the system indicates how many sites (spatial orbitals) are in each side of the square model. The number of system qubits is thus twice the number of spatial orbitals. The number of logical ancillae is computed as \eq{hub_ancilla_est}. Finally, the number of T gates is computed using \eq{hubbard_estimate} which assumes that $u/t = 4$ and $\Delta E = t / 100$. The first three problem sizes in the table are near the classically intractable regime.}
\label{tab:hubbard_ts}
\end{table}

\subsection{Exploiting Locality in Simulations of Lattice Hamiltonians}
\label{sec:lieb}

Looking forward, another way that our circuits can be applied is to accelerate the recent Lieb-Robinson simulation method of~\cite{Haah2018}.  Lieb-Robinson bounds reveal an intriguing fact about local Hamiltonians: interactions spread out in a light-cone similar in form to the causal diamonds used in relativity to indicate the regions of space-time that can have an impact on an event at a point in spacetime~\cite{lieb1972finite}.   More specifically, Lieb-Robinson bounds show that information propagates at finite speeds (up to exponentially small error) in systems with nearest-neighbor interactions.  The idea behind~\cite{Haah2018} is to exploit this structure to break up the evolution into sub-pieces that can be independently simulated, thus reducing the cost of simulation.

We formalize this by envisioning that we have a lattice of $N$ sites, $\Lambda$, and a Hamiltonian that consists of terms that act upon these sites
$
H=\sum_{X\subseteq \Lambda} h_X.
$
Here each $h_X$ is local in that if $h_X$ and $h_Y$ act on different sites in the lattice then $[h_X,h_Y]=0$ and $h_X$ only has support on sites that are a constant Euclidean distance away from each other.  Note that this definition of locality also incorporates the terms within the Hubbard model. The final concept that we need to explain the method is that of distance between sites.  We assume that for all $X,Y\subseteq \Lambda$ that ${\rm dist}(X,Y)$ yields the minimum Euclidian distance between any two points within the lattice vectors contained within sets $X$ and $Y$.  For example, given a lattice $1$D on $10$ sites ${\rm dist}(\{3,4,5\}, \{8,9,10\})=3$.  The following lemma (a restatement of Lemma 6 in~\cite{Haah2018}) explains the impact that the locality imposed by the Lieb-Robinson bound has on simulation.

\begin{lemma}[Patching Lemma]
\label{lem:LR}
Let $\Lambda$ be a lattice on $N$ sites with a Hamiltonian $H=\sum_{X\subseteq\Lambda} h_X$ where each $h_X$ is a local bounded Hamiltonian for every $X\subseteq \Lambda$.  Let $A,B,C$ be subsets of $\Lambda$ and let $H_{P_1\cdots P_q}$ for any sequence $P:\{1,\ldots, q\} \mapsto \{A,B,C\}^q$ be for integer $q\ge 1$ $H_{P_1\cdots P_q} = \sum_{X \in P_1 \bigcup \cdots \bigcup P_q}h_X$ (for example, $H_{AB} = \sum_{X\subseteq A\bigcup B} h_X$.  There are constants $v\ge 0$, called the Lieb-Robinson velocity, and $\mu>0$ such that 
$$
\left\|e^{-iH_{ABC} t} - e^{-iH_{AB} t} e^{iH_Bt}e^{-iH_{BC} t} \right\|\in \mathcal{O}\left(\sum_{X \subseteq (A\bigcup B\bigcup C) \setminus (A\bigcup B)\setminus C}\|h_X\|e^{vt-\mu~{\rm dist}(A,C)} \right).
$$
\end{lemma}

Note that in the above terminology $(A\bigcup B\bigcup C) \setminus (A\bigcup B)\setminus C$ is the boundary of the sets $AB$ and $C$ meaning the set of all terms within the Hamiltonian that act on sites contained in both $A$ or $B$ and $C$. \lem{LR} is the core of the simulation method.  The central idea behind the proof is to use the patching lemma recursively to break up the evolution into a product of evolution operators, each of which contains terms that act on one or two of the constituent subsets of sites in the problem.  This is conceptually similar to a Trotter decomposition; however, as the error in this approximation can be made exponentially small by choosing the patches in~\lem{LR} to be linearly far apart, the error can be controlled in a tighter fashion without requiring short time steps (unlike Trotter decompositions~\cite{Suzuki1993,Wiebe2008}).  For example, consider regions $A,B,C,D$.  Then we can write
\begin{equation}
e^{-i H_{ABCD} t} \approx e^{-iH_{AB} t}e^{iH_B t} e^{-iH_{BCD}t}\approx e^{-iH_{AB} t}e^{iH_B t} e^{-iH_{BC}t}e^{iH_C t}e^{-iH_{CD}t}.
\end{equation}
In order to achieve scaling that is polylogarithmic in $1/\epsilon$, the evolution of each patch needs to be simulated using a method with poly-logarithmic scaling in $1/\epsilon$ such as the truncated Taylor series simulation result~\cite{Berry2015} or qubitization~\cite{Haah2018}.  Our circuits can be used to optimize this result since qubitization remains the best way to simulate the evolution and our \sel and \prep circuits meet the requirements of qubitization oracles. This result is formally given as Theorem 1 of~\cite{Haah2018}, a special case of which is restated below for convenience.

\begin{theorem}
\label{thm:Haah}
Assume the preconditions of~\lem{LR} and that for every unit ball in $\Lambda$ within the Euclidean metric space $\mathbb{R}^D$ at most $\mathcal{O}(1)$ sites are contained within the ball and $h_X=0$ if the diameter of the set $X$ is greater than $1$.  Additionally, let each $h_X$ be efficiently computable and have norm at most $1$.  Then there exists a quantum algorithm that simulates the evolution of $H$ for time $\tau$ with accuracy $\epsilon$ that uses $\mathcal{O}(\tau N \, {\rm polylog}(\tau N/\epsilon))$ $2$--qubit gates and further has gate depth $\mathcal{O}(\tau \, {\rm polylog}(\tau N/\epsilon))$.
\end{theorem}

We claim that our approach can be used to achieve $\widetilde{\cal O}(N/\epsilon)$ scaling for simulating the Hubbard model with nearest-neighbor interactions. The Hubbard model satisfies the preconditions because each term in the Hamiltonian is local on the fermion lattice \cite{Haah2018}. By using our constructions for the \prep and \sel circuits, we can reduce constant factors (and some log factors in T complexity) involved in the qubitized simulation, while saturating the $\widetilde{\cal O}(\tau N)$ scaling of \thm{Haah}.  We then choose $\tau \in {\cal O}(1/\lambda)$ and apply phase estimation on the result.  In order to estimate the eigenvalue to within error $\epsilon$ with high probability we need $\mathcal{O}(\lambda/\epsilon)$ repetitions of the circuit.  Thus by multiplying the two results gives that the overall scaling for simulating such a Hubbard model is $\widetilde{\cal O}(N/\epsilon)$, as claimed.

This approach requires some follow-up work in order to determine exact T counts. Specifically, we need to implement a full qubitized simulation (rather than $e^{-i \arccos(H/\lambda)}$). This transformation is known to be achievable with a poly-logarithmic sized circuit~\cite{Low2016}.  While our work provides a highly optimized method for implementing the oracles needed in this process, more work remains to estimate constant factors associated with this simulation.

\FloatBarrier

\section{Resource Analysis for Fault-Tolerant Implementation}
\label{sec:resources}

Throughout this work, we have focused on the number of T gates as the primary cost model of interest. The reasons for this are our focus on hardware consisting of a 2D nearest-neighbor coupled array of qubits, the intention to use the surface code \cite{Brav98,Denn02,Raus07,Raus07d,Fowl12f}, and the high relative overhead of T gates compared to all others in that context. In this section, we shall discuss the overhead of the complete algorithm in detail.

When using the surface code, each T gate is implemented by first preparing a magic state \begin{equation}
\ket{\rm T} \equiv {\rm T} \ket{+} = \frac{\ket{0}+e^{i\pi/4}\ket{1}}{\sqrt{2}}
\end{equation} that is consumed during the gate. The gate is probabilistic and 50\% of the time T$^\dagger$ is actually applied instead of T. When the gate implemented is not as desired, an S gate must be inserted to correct. Preparing T states requires a substantial amount of time and hardware, which we shall make precise below. In an effort to minimize the number of physical qubits required, we shall therefore only prepare a single T state at a time. We shall assume the availability of a correlated-error minimum weight perfect matching decoder \cite{Fowl13g} capable of keeping pace with $1 \, \mu$s rounds of surface code error detection, and capable of delivering feedforward in 10--20$\,\mu$s. We shall calculate the qubit and time overhead for physical gate error rates $p=10^{-3}$ and $p=10^{-4}$.

The overhead shall be approximated by considering only the overhead of the Majorana operator circuit from \fig{majorana} and the data lookup circuit from \fig{qrom}. It is expected that these circuits will account for over 90\% of the total algorithm overhead. These circuits break down into a number of common pieces, compute \textsc{and}s, uncompute \textsc{and}s, naked CNOTs, and active subcircuits. The number of these pieces, in terms of the number of algorithm target qubits $N$, is shown in \tab{piece-count}.

\begin{table}[h]
\begin{tabular}{c|c|c|c|c}
 & compute \textsc{and}s & uncompute \textsc{and}s & naked CNOTs & subcircuits \\
\hline
\fig{majorana} & $N - 1$ & $N - 1$ & $0.5 \, N$ & $0.5 \, N$ \\
\fig{qrom} & $1.5 \, N - 1$ & $1.5 \, N - 1$ & $0.75 \, N$ & $0.75 \, N$ \\
\end{tabular}
\caption{Breakdown of the various elements that make up the Majorana operator circuit from \fig{majorana} and the data lookup circuit from \fig{qrom}. $N$ is the number of spin orbitals in the system the circuits are being applied to.}
\label{tab:piece-count}
\end{table}

\begin{figure*}[b]
\begin{center}
\includegraphics[width=\linewidth]{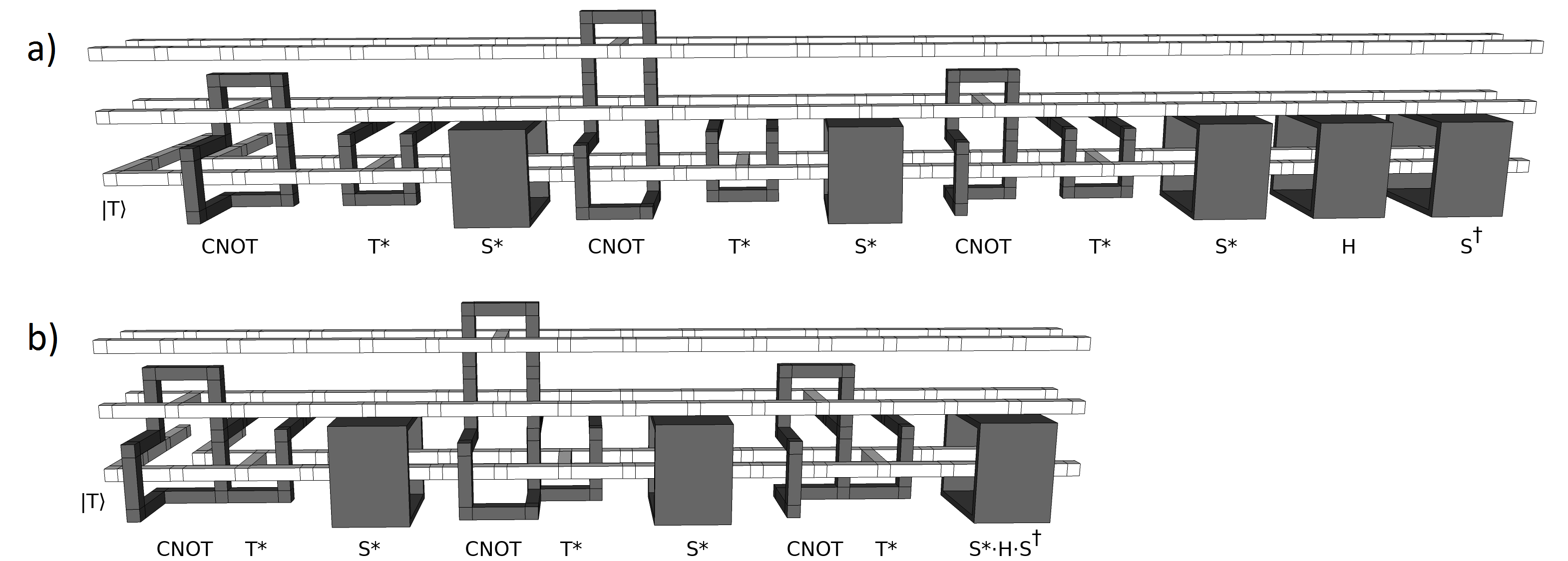}
\end{center}
\caption{a) Canonical surface code \textsc{and} gate computation (\fig{temporary-and-notation}) circuit. The bottom pair of white lines (coming in from the back instead of from the left) represents the injection of a $\ket{\textrm{T}}$ state. Each dark ring is a CNOT. Each dark U-shape labeled T* is a random T or T$^\dagger$ gate (as determined by a measurement in a location not shown in the diagram connecting to the dark U-shape). The boxes labeled S* immediately following each T* are S or S$^\dagger$ gates that are included if the random T or T$^\dagger$ gate results in the incorrect gate. The final two boxes are Hadamard and S$^\dagger$ operations, respectively.  b) Compressed version. The final three boxes can be compressed to a single box as an arbitrary single-qubit Clifford can be performed inside using twists \cite{Brow16} and other techniques \cite{Fowl12a}. Distinct dark structures can be made to touch provided this occurs in a single place \cite{Fowl12h}.}
\label{fig:open-Toffoli}
\end{figure*}

A surface code implementation of compute \textsc{and} \fig{temporary-and-notation} is shown in \fig{open-Toffoli}. The regular geometric structure can be decomposed into plumbing pieces, namely cubic volumes each containing a single small light-colored cube. The compressed \cite{Fowl12h} version has a depth of 15 plumbing pieces. The circumference of each string-like structure (defect) is the surface code distance $d$, and the minimum separation of defects of the same color is also $d$. In the temporal direction (left-right), each unit of $d$ is a round of error detection. In the spatial directions (plane perpendicular to temporal), each unit of $d$ corresponds to two qubits. Note that a single CNOT, after compression, takes depth 1 plumbing piece as drawn. The overhead of any algorithm ultimately needs to be expressed as some number of qubits (space) and seconds (time). A plumbing piece is a convenient device physics and code distance independent measure of space-time volume. As described above, the $(5d/4)^3$ cubic volume of a plumbing piece can easily be converted to qubits and seconds given a code distance $d$ and single-round error detection time.

\begin{figure}
\begin{center}
\includegraphics[width=0.4\linewidth]{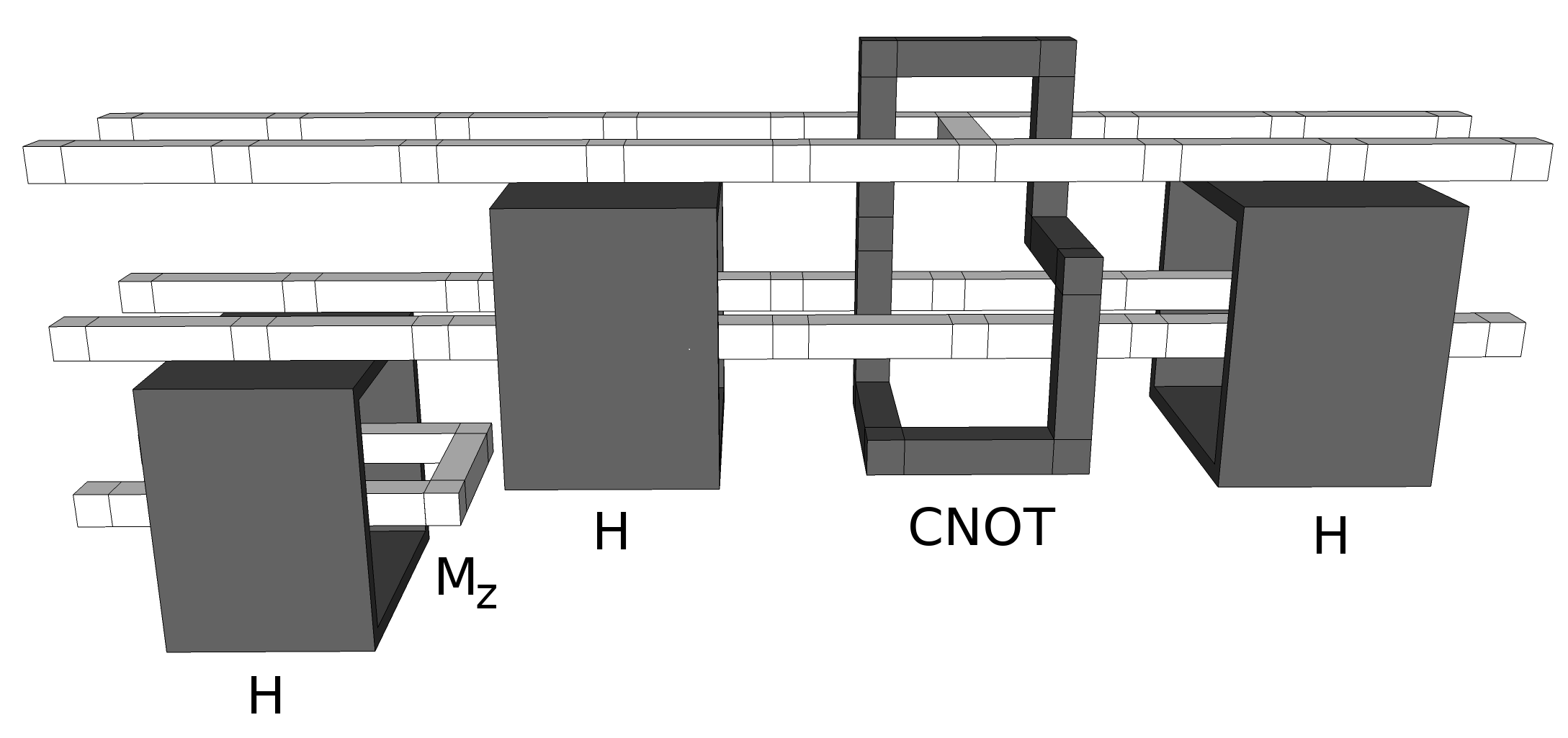}
\end{center}
\caption{Uncompute \textsc{and} (\fig{temporary-and-notation}) implemented directly as a Hadamard operation followed by a measurement on the bottom qubit. The outcome of this measurement determines if the sequence of operations on the top two qubits (a CZ implemented as a CNOT framed by Hadamard operations) are included or omitted.
}
\label{fig:close-Toffoli}
\end{figure}

A plumbing piece depth 8 surface code implementation of uncompute \textsc{and} \fig{temporary-and-notation} is shown in \fig{close-Toffoli}. This could be compressed by performing the measurement differently so no initial Hadamard would be necessary, but the current surface code form is more easily identified with the original abstract form, and further compression is not necessary as the execution time of the algorithm, as we shall see, is limited by our serial preparation of T states.

\begin{figure*}[b]
\begin{center}
\includegraphics[width=0.5\linewidth]{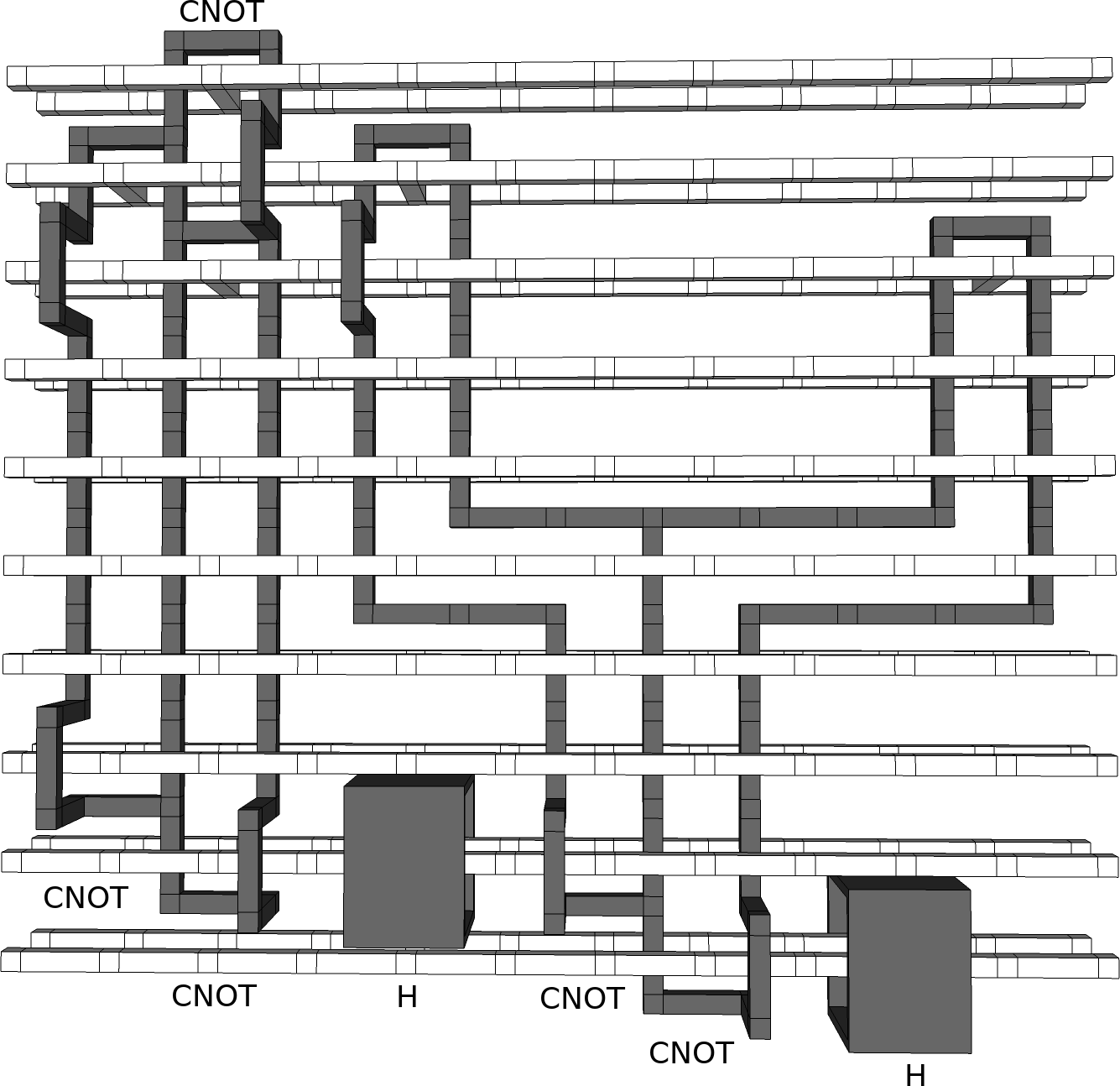}
\end{center}
\caption{Surface code implementation of the inner loop of the Majorana operator circuit.
Contrast with \fig{majorana}, noting that the circuit has been somewhat modified to reduce its surface code spacetime volume.
In particular, the controlled-$Y$ operations from \fig{majorana} have been propagated through the controlled-$Z$ operations, producing \textsc{cnot} operations which are cheaper to perform. This creates phase error which must be corrected by an S gate on the control of the entire Majorana operator. Also not shown are initial Hadamard gates on every target qubit.}
\label{fig:YZ-subcircuit}
\end{figure*}

An effective plumbing piece depth 5 surface code implementation of the Majorana operator active subcircuit is shown in \fig{YZ-subcircuit}. The unusual pair-of-horns structure is to permit an uncompute \textsc{and} circuit to fit in before the final CNOT. The inner loop of the data circuit (\fig{qrom}) is just a pair of single-control multiple-target CNOTs, and a single additional CNOT. This can be implemented in plumbing piece depth 4 and is not shown.

\begin{figure*}
\begin{center}
\includegraphics[width=0.6\linewidth]{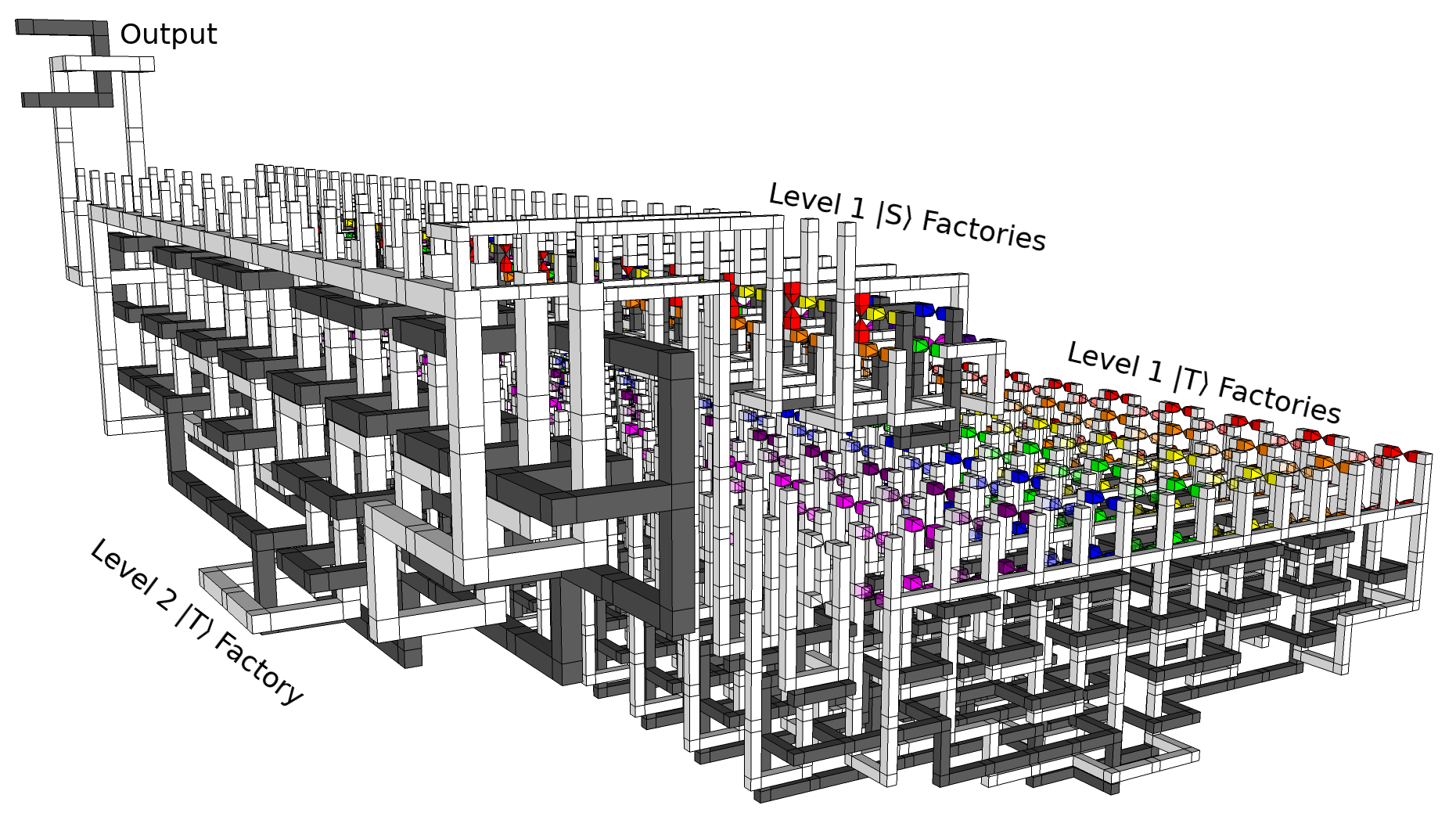}
\end{center}
\caption{Preparing a T state in the surface code. Physical $\ket{\textrm{T}}_0$ states are injected into the 16 lower factories in the lower rear right, which distill them into less noisy $\ket{\textrm{T}}_1$ states. Fifteen of the successful distillations are forwarded to the larger factory towards the front and left, which distills them into a $\ket{\textrm{T}}_2$ state with low enough error. The output is the top left dark U-shape.}
\label{fig:T-state-distillation}
\end{figure*}

Preparing a T state is an involved process \cite{Brav05,Reic05} shown for discussion purposes in \fig{T-state-distillation}. The important features for our purposes are the fact that this can be tiled vertically (meaning in time) every 6 plumbing pieces, and the whole structure occupies an area of 160 plumbing pieces. A significant amount of fast classical feedforward is required as many T gates are followed potentially by S gates, and the paths of the connections from the first (small) level of distillation to the second (large) level must be determined based on which succeed. Our assumption of a 10--20$\, \mu$s latency decoder is sufficient to make this work.
We are interested in the overhead of solving instances of the electronic structure and Hubbard Hamiltonians discussed in prior sections. We must choose a target inaccuracy $\epsilon$ to fix the number of data logical qubits and gates required. To first order, the dependence of the gate count on $\epsilon$ can be ignored. We shall choose $\epsilon=10^{-3}$. \tab{circuit-overview} summarizes the circuit input parameters we will study.

\begin{table}
\begin{tabular}{c|c|c|c|c|c|c}
System  & Spin-Orbitals ($N$) & ${\cal W}$ Queries & \#$\text{Majorana}_N$ & \#$\text{Majorana}_{N/2}$ & \#$\text{QROM}_{3N/2}$ & Max Qubits \\
\hline
Hubbard model & 72           & $1.3 \times 10^5$ & $2.5 \times 10^5$ & $1.3 \times 10^5$ & 0 & 105 \\
Hubbard model & 128         & $2.3 \times 10^5$ & $4.6 \times 10^5$ & $2.3 \times 10^5$ & 0 & 161 \\
Hubbard model & 200         & $3.6 \times 10^5$ & $7.2 \times 10^5$ & $3.6 \times 10^5$ & 0 & 236 \\
Hubbard model & 800         & $1.4 \times 10^6$ & $2.8 \times 10^6$ & $1.4 \times 10^6$ & 0 & 842 \\
\hline
Electronic structure & 54   & $1.4 \times 10^4$ & $4.2 \times 10^4$ & 0 & $2.8 \times 10^4$ & 123 \\
Electronic structure & 128  & $6.3 \times 10^4$ & $1.9 \times 10^5$ & 0 & $1.3 \times 10^5$ & 210 \\
Electronic structure & 250  & $1.7 \times 10^5$ & $5.3 \times 10^5$ & 0 & $3.5 \times 10^5$ & 341 \\
Electronic structure & 1024 & $1.8 \times 10^6$ & $5.3 \times 10^6$ & 0 & $3.5 \times 10^6$ & 1136 \\
\end{tabular}
\caption{
    Cases for which we will calculate the fault-tolerant overhead, along with numbers relevant to estimating the non-negligible components of this overhead.
    Column 3 contains the number of times that the ${\cal W}$ oracle is queried.
    For the Hubbard model we consider the system at intermediate coupling ($u/t = 4$) implying that $\lambda = 4 N t$ and we consider an accuracy of $\Delta E = t / 100$. For jellium, values of $\lambda$ are provided in \tab{jellium_ts} and we target chemical accuracy which is defined as $\Delta E=0.0016$ Hartree.
    Each call to ${\cal W}$ includes a call to $\textsc{select}$, $\textsc{prepare}$, and $\textsc{prepare}^\dagger$ which in turn apply QROM and Majorana operations (which are the dominant costs of the algorithm).
    Columns 4 and 5 estimate the number of times the Majorana operator circuit (see \fig{majorana}) must be applied to the entire system or else to half of the system.
    For electronic structure, there are three Majorana operators of size $N$ per query to ${\cal W}$ (see \fig{selecth}).
    For Hubbard, there are two Majorana operators of size $N$ and one of size $N/2$  per query to ${\cal W}$ (see \fig{selecthub}).
    Column 6 estimates the number of times QROM lookups (see \fig{qrom}) of size $L=3N/2$ are performed.
    This does not occur in our Hubbard model circuits, but happens twice per query to ${\cal W}$ in our electronic structure circuits (once in $\textsc{prepare}$ and once in $\textsc{prepare}^\dagger$; see \fig{subprepare}).
    The final column contains the maximum number of data qubits required at any point in the algorithm, which occurs while applying the Majorana operations in $\textsc{select}$.
    This number does not include space to prepare T states, which will be discussed separately.
}
\label{tab:circuit-overview}
\end{table}

Given \tab{piece-count}, \tab{circuit-overview}, and the plumbing piece depths of the various circuit elements, we can calculate the total number of data plumbing pieces $N^{\rm data}_{PP}$ and hence the code distance required to ensure no more than a 1\% chance of logical error in any data plumbing piece using $p_L(d, p) \simeq 2d(50p)^{(d+1)/2} < 1/(100N^{\rm data}_{PP})$. Similarly, knowing the compute \textsc{and} circuit contains 4 T gates, and no other part of the data or Majorana operator circuits contain T gates, we can calculate the total number of T gates $N_{\rm T}$ and hence the target T state error rate from distillation of $1/(100N_{\rm T})$. We also calculate the total number of T distillation plumbing pieces $N^{\rm T}_{PP}$ and a code distance to ensure the chance of T plumbing piece error is below $1/(100N^{\rm T}_{PP})$. We have elected to keep algorithm error rates low to ensure on average only a few repetitions are required. For both $p=10^{-3}$ and $p=10^{-4}$ and all algorithm instances considered, T state distillation of the form in \fig{T-state-distillation} was sufficient to achieve the target logical error rate. This information is collectively sufficient to calculate the qubit and time overheads, shown for all cases in \tab{overhead}.

\begin{table}
\begin{tabular}{c|c|c|c|c|c}
\multicolumn{2}{c|} {problem} & \multicolumn{2}{c} {physical qubits} & \multicolumn{2}{|c} {execution time (hours)} \\
\hline
System & Spin-Orbitals ($N$) & $p=10^{-3}$ & $p=10^{-4}$ & $p=10^{-3}$ & $p=10^{-4}$ \\
\hline
Hubbard model & 72 & $1.4\times 10^6$ & $4.4\times 10^5$ & 4.6 & 2.6 \\
Hubbard model & 128 & $2.1\times 10^6$ & $6.6\times 10^5$ & 15 & 8.4 \\
Hubbard model & 200 & $3.2\times 10^6$ & $8.9\times 10^5$ & 40 & 21 \\
Hubbard model & 800 & $1.4\times 10^7$ & $3.6\times 10^6$ & $6.7\times 10^2$ & $3.7\times 10^2$ \\
\hline
Electronic structure & 54 & $1.4\times 10^6$ & $3.9\times 10^5$ & 0.82 & 0.43 \\
Electronic structure & 128 & $2.4\times 10^6$ & $8.1\times 10^5$ & 9.9 & 5.6 \\
Electronic structure & 250 & $4.4\times 10^6$ & $1.2\times 10^6$ & 58 & 30 \\
Electronic structure & 1024 & $2.0\times 10^7$ & $4.8\times 10^6$ & $2.7\times 10^3$ & $1.4\times 10^3$ \\
\end{tabular}
\caption{Hand calculation of qubit and time overheads of general chemistry and Hubbard circuits assuming gate error rates of $p=10^{-3}$ and $p=10^{-4}$, a 2D array of nearest-neighbor coupled qubits, and a surface code error detection cycle time of 1$\mu$s. The execution time being estimated is the duration of one complete run of the phase estimation process.
}
\label{tab:overhead}
\end{table}

The previous paragraphs described a manual overhead estimation method. There are a number of approximations that go into such an estimate, in particular assuming that it will always be possible to route gates in 3D spacetime without overhead beyond that of where the data qubits are stored. In order to strengthen the relevance of the presented results, we have also used a software automated approximation method. The software is an improved version of the tool from \cite{paler2017synthesis}. Automated overhead approximation starts from a Clifford+T representation of the circuit to be analyzed (e.g.\ \fig{ranged-operation}), and ends with a full surface code layout, having each gate translated into a corresponding configuration of plumbing pieces. Thus, the automated estimation work flow is similar to the manual one. However, certain circuit particularities were analyzed differently and so similarities and differences between the two methods will be discussed.

The Clifford+T circuit is prepared according to a worst-case scenario, based on the available hardware restrictions, plumbing pieces layout problems and T gate correction mechanisms. The preparation of a single distilled T state at a time is a restriction which influences the resulting surface code layout: the Clifford+T gates have to be scheduled (laid out) in such a way that the T gates will be executed as soon as possible, but not earlier than the availability of distilled T states. The state distillation form in \fig{T-state-distillation} implies that a T gate can be executed on average every 6 plumbing pieces along the time axis. Additionally, T gate implementations are probabilistic and S gate corrections may be necessary. Thus, our scenario considers that all T gates are followed by the corrective S gate, resulting in a synthetic increase of the Clifford+T circuit depth. Circuit preparation is followed by an optimization procedure, where as many Clifford gates as possible are scheduled between two subsequent T gates. The software simulates the availability of the distilled T states, and places T gates whenever their execution is possible. If no T states are available, the T gates are delayed, which will increase later the approximated time overhead.

Finally, the Clifford+T circuit is translated into the surface code layout. From a resource estimation perspective, the complexity of this task is increased because the software currently includes only partially the optimization strategy presented in \fig{open-Toffoli}{\color{red} b}: final boxes can be compressed to a single one, but distinct dark structures are not allowed to touch (for verification/debugging purposes). Due to this fact, the automated approximation used a slightly different Clifford+T realization of the computing \textsc{and} gate (cf.\ \fig{temporary-and-notation}), which had the advantage of being more suitable for automatic placement in stairway-structured circuits (i.e.\ the arrangement of \textsc{and} gates in \fig{majorana}). Automatic placement of those Clifford+T sub-circuits resulted in a shorter depth of the generated surface code layouts. Overhead of the two basic circuits in units of plumbing pieces can be found in \tab{volume_majorana}. This data converted into qubits and time can be found in \tab{auto_overhead}. The automatically generated estimations are comparable to the manual, though generally slightly higher, exceeding the manual estimates by 10--20\%. While the automated method is penalized by missing optimization strategies that are possible when analyzing circuits manually, and the need to provide explicit communication paths for long-range gates, some of this penalty is canceled by using algorithmic methods too complex to perform manually. The fact that both approximation methods lead to such comparable qubit and time overheads strengthens our confidence in these estimates. The time estimates, in particular, are practically identical.

\begin{figure*}
\centering
\includegraphics[width=0.5\linewidth]{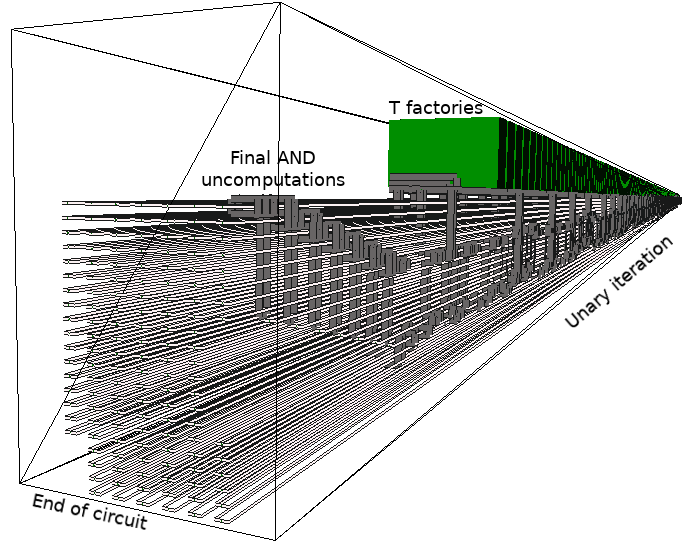}
\caption{Illustration of the circuit layout used for resource estimation. The figure shows a portion of a circuit towards the outputs (the time axis indicating circuit execution runs from the back to the front).  The figure sketches the arrangement of the logical qubits (light gray lines), the CNOTs (dark gray lines), the boxes (green) abstracting the multi-level distillation circuits from \fig{T-state-distillation} and the bounding box (black wire-frame) expressing the resources estimated to lay out the circuit.}
\label{fig:plumbpaler}
\end{figure*}

\begin{table}
\begin{tabular}{c | c | c | c | c | c | c }
System & Circuit$(N)$ & T-count & Area ($\text{PP}^2$) & Time (PP) & Volume ($\text{PP}^3$) & Braided Volume ($\text{PP}^3$) \\
\hline
Hubbard model        & Majorana$_{72}$   &  284 &  17 x 16 &  1,840 &    500,480 &    429,624 \\
Hubbard model        & Majorana$_{128}$  &  508 &  25 x 16 &  3,252 &   1,300,800 &   1,155,817 \\
Hubbard model        & Majorana$_{200}$  &  796 &  36 x 16 &  5,080 &   2,926,080 &   2,637,504 \\
Hubbard model        & Majorana$_{800}$  & 3,196 & 123 x 16 & 20,262 &  39,875,616 &  37,451,032 \\
\hline
Electronic structure & Majorana$_{54}$   &  212 &  20 x 16 &  1,382 &    442,240 &    365,717 \\
Electronic structure & Majorana$_{128}$  &  508 &  32 x 16 &  3,252 &   1,665,024 &   1,473,563 \\
Electronic structure & Majorana$_{250}$  &  996 &  51 x 16 &  6,342 &   5,175,072 &   4685164 \\
Electronic structure & Majorana$_{1024}$      & 4,092 & 165 x 16 & 25,932 &  68,460,480 &  64,114,531 \\
\hline
Electronic structure & QROM$_{\frac{3}{2} 54}$   &  320 &  20 x 16 &  2,068 &    661,760 &    558,098 \\
Electronic structure & QROM$_{\frac{3}{2} 128}$  &  764 &  32 x 16 &  4,872 &   2,494,464 &   2,273,711 \\
Electronic structure & QROM$_{\frac{3}{2} 250}$  & 1,496 &  51 x 16 & 9,508 &   7,758,528 &   7,272,549 \\
Electronic structure & QROM$_{\frac{3}{2} 1024}$ & 6,140 & 165 x 16 & 38,892 & 102,674,880 & 100,399,903 \\
\end{tabular}
\caption{Automatically generated resource estimates of the Majorana and QROM circuits (\fig{majorana} and \fig{qrom}). The area width, height, and time columns give the dimensions of bounding box (e.g. \fig{plumbpaler}) in units of plumbing pieces. The last column is the number of plumbing pieces estimated to be actively used within the bounding box.
The volume numbers do not include the volume of the T factory, but do include idle qubits that are present in the algorithm as a whole but not the individual circuits.
The ``braided volume" refers to the amount of actively used volume, i.e. non-empty space with defects used to encode qubits and operations.
QROM circuits are indexed like QROM$_{\frac{3}{2} N}$ because in context the QROM index size $L$ is always 50\% larger than the number of orbitals $N$.
}
\label{tab:volume_majorana}
\end{table}

\begin{table}
\begin{tabular}{c|c|c|c|c|c}
\multicolumn{2}{c|} {problem} & \multicolumn{2}{c} {physical qubits} & \multicolumn{2}{|c} {execution time (hours)} \\
\hline
System & $N$ Spin-Orbitals & $p=10^{-3}$ & $p=10^{-4}$ & $p=10^{-3}$ & $p=10^{-4}$ \\
\hline
Hubbard model &72& $1.7\times 10^6$ & $5.3\times 10^5$ &4.6& 2.6 \\
Hubbard model &128& $2.4\times 10^6$ & $7.8\times 10^5$ &15& 8.4 \\
Hubbard model &200& $3.8\times 10^6$ & $1.0\times 10^6$ &40& 21 \\
Hubbard model &800& $1.5\times 10^7$ & $4.2\times 10^6$ & $6.7\times 10^2$ & $3.7\times 10^2$ \\
\hline
Electronic structure &54& $1.7\times 10^6$ & $4.7\times 10^5$ &0.85& 0.44 \\
Electronic structure &128& $2.9\times 10^6$ & $9.5\times 10^5$ &10& 5.7 \\
Electronic structure &250& $5.1\times 10^6$ & $1.4\times 10^6$ &58& 30 \\
Electronic structure &1024& $2.3\times 10^7$ & $5.6\times 10^6$ & $2.8\times 10^3$ & $1.4\times 10^3$ \\
\end{tabular}
\caption{Automatically generated qubit and time overheads of general chemistry and Hubbard circuits assuming gate error rates of $p=10^{-3}$ and $p=10^{-4}$, a 2D array of nearest-neighbor coupled qubits, and a surface code error detection cycle time of 1$\,\mu$s. The execution time being estimated is the duration of one complete run of the phase estimation process.
}
\label{tab:auto_overhead}
\end{table}

The data is highly encouraging, with physical qubit counts of order a million and times in hours for all but the largest cases considered. Significant further reduction is expected to be possible. For example, the $N$ qubits in the $|\psi\rangle$ register are only operated on by the Majorana operator circuit and this circuit targets just one of these qubits at a time. This implies the remainder can be stored more compactly in square surface code patches while not being interacted with. This could easily reduce the overhead of these $N$ qubits, which account for 70--80\% of the physical qubits, by a factor of six. This would conservatively lower overall physical qubit requirements by a factor of two.

\FloatBarrier

\section{Conclusion}
\label{sec:conclusion}

In this work we introduced especially efficient fault-tolerant quantum circuits for using phase estimation to estimate the spectra of electronic Hamiltonians. Unlike past work which has focused on realizing phase estimation unitaries encoding $e^{-i H \tau}$, corresponding to time-evolution under $H$ for duration $\tau$, we focused on a recent idea that one might more cheaply realize phase estimation unitaries encoding the quantum walk $e^{i \arccos(H / \lambda)}$ where $\lambda$ is a parameter closely related to the induced 1-norm of the system Hamiltonian \cite{Berry2018,Poulin2017}. We construct explicit quantum circuits for realizing this quantum walk with T complexity linear in basis size for both the planar Hubbard model and electronic structure Hamiltonians in second quantization. We showed that phase estimation projects these systems to an eigenstate and estimates the associated eigenvalue to within additive error $\epsilon$ by querying the quantum walk operator an optimal number of times scaling as ${\cal O}(\lambda / \epsilon)$. To accomplish this we introduced general techniques that we conjecture are near-optimal for streaming bits of a unary register and for implementing a quantum read-only memory. We introduced a new form of Heisenberg limited phase estimation specialized to linear combinations of unitaries based simulations and provided bounds on T complexity and ancilla count which remain tight even at small finite sizes.

In addition to providing explicit Clifford + T circuits we compiled the bottleneck components of these simulations to fault-tolerant surface code gates in order to rigorously determine the resources that would be required for error correcting interesting problem instances. We performed this compilation both by hand and by using automatic tools and found similar overheads in both cases. We found that classically intractable instances of jellium and the Fermi-Hubbard model could be simulated with under one-hundred million T gates and would require about one million physical qubits in the surface code with two-qubit error-rates on the order of $10^{-3}$. At error rates of $10^{-4}$ about an order of magnitude fewer physical qubits would be required. We also priced out simulations of realistic solid-state materials such as diamond, graphite, silicon, metallic lithium and crystalline lithium hydride and found that only slightly more than one billion T gates and a few million physical qubits would be required.

Despite focusing on different systems, our results are most readily comparable to the previous state-of-the-art results from \cite{Reiher2017}. Even though \cite{Reiher2017} sought empirical estimates of the T complexity rather than rigorous upper bounds as we did, they estimated that approximately $10^{15} - 10^{16}$ T gates would be required for a 108 qubit simulation of the FeMoco molecule active space. By comparison, our upper bounds on the T complexity required to solve the classically intractable electronic structure problems studied here were roughly a million times less.
The low T complexity is the result of designing a lean algorithm from the ground up with insights matched to the Hamiltonian and with innovative algorithmic subroutines. The improvements are distributed across several parts of our approach, each of which provides one or two orders of magnitude improvement.
Because our simulations require only a few times more physical qubits than is required by a single T factory, it is reasonable to expect that the simulations we outline here will become practical on the first universal fault-tolerant quantum devices, many years before the simulations discussed in \cite{Reiher2017} would be viable.

Several important directions for future research pertain to the extension of these simulation techniques to representations that would be more effective for single molecules. While the dual basis described in \cite{BabbushLow} is well suited to treating solid-state materials such as the ones explored here (e.g.\ jellium, solid-state silicon, graphite, diamond, lithium and lithium hydride), by combining our techniques with the ``Gausslet'' basis sets of \cite{White2017}, we should also be able to simulate single molecules with similar resolution to Gaussian orbitals -- thus extending our results to systems such as FeMoco with similar overheads to those observed in this work. However, deploying the Gausslet basis functions to systems with large atomic nuclei such as iron (as in FeMoco) will require further research. If basis errors are a concern, then future work should combine results from this paper, \cite{Kivlichan2016} and \cite{Berry2018} in order to determine the cost of encoding first-quantized electronic spectra in quantum circuits; in first quantization, basis errors are suppressed exponentially in the number of qubits used to represent the system.

Another remaining challenge is to compute a tighter upper bound on the number of physical qubits required by the algorithm.
At the first moment that it becomes technologically possible to distill magic states, the number of physical qubits available on one machine will still be extremely limited.
Getting a meaningful computation to fit at all will be difficult. Fortunately, the qubit count estimates of this paper were fairly conservative; in this paper we only explored surface code constructions which we were able to validate and implement in software; these constructions are not necessarily optimal.
For example, the logical qubit representation used in lattice surgery \cite{Horsman2012} requires fewer physical qubits than the double-defect representation used in the estimates of this paper. Furthermore, there are several places in our circuits where we preferred small multiplicative improvements in T-count over small additive improvements in logical qubit count. For example, we delay uncomputing QROM lookups in order to avoid recomputation and, when performing phase estimation, we minimize the number of oracle queries by using a full-size phase register instead of a single phase qubit. Since we have managed to show that with error rates of $10^{-3}$ one can solve interesting problems in chemistry using on the order of a million physical qubits within the surface code, a next natural goal would be to try to further reduce the resources required to be on the order of a hundred-thousand physical qubits.

\subsection*{Acknowledgements}

The authors thank Yuval Sanders, Artur Scherer, M\'{a}ria Kieferov\'{a} and Guang Hao Low for helpful discussions about linear combinations of unitaries based simulation methods. We thank Garnet Kin-Lic Chan and Kostyantyn Kechedzhi for discussions pertaining to the regimes in which the Hubbard model would be interesting to simulate.  We thank Ian Kivlichan, Zhang Jiang and Dave Bacon for helpful comments on an early version of this manuscript. Dominic Berry is funded by an Australian Research Council Discovery Project (Grant No.\ DP160102426).

\bibliographystyle{apsrev4-1_with_title}
\bibliography{Mendeley,References}

\appendix

\section{Propagating Errors from Hamiltonian Coefficients into Phase Estimate}
\label{app:errAnal}

In this appendix we address the question of how accurately coefficients of the Hamiltonian must be prepared in the \prep oracle in order to estimate the Hamiltonian eigenvalues to precision $\epsilon$. As discussed in \sec{walk}, our phase estimation scheme involves estimating the phases induced by the operator $e^{i \arccos(H / \lambda)}$. If one is near the singularity of arccos then a small error in the Hamiltonian can have a significant impact on the phase. Let us define
\begin{align}
\widetilde{H} \equiv \sum_{\ell=0}^{L-1}\widetilde{w}_\ell H_{\ell}
\label{eq:defnsw}
\end{align}
for our approximate encoding of $H$.
Using the state preparation technique in \sec{subsample}, we obtain
\begin{align}
\lambda = \sum_{\ell=0}^{L-1} \widetilde{w}_\ell \, .
\end{align}
We denote by $\delta$ an upper bound on the approximation in any of the $\widetilde{w}_\ell$, so
\begin{equation}
\delta \ge \left |\widetilde{w}_\ell -w_\ell \right | .
\end{equation}
Next note that the error in the eigenphase obeys
\begin{align}
\label{eq:epsbnd}
\epsilon_{\textsc{prep}} & \le \left \|e^{i\arccos(H/\lambda)}- e^{i\arccos({\widetilde{H}}/\lambda)}\right \|\le \ \left\|\arccos(H/\lambda) - \arccos({\widetilde{H}}/\lambda)\right \|\\
&\le \sum_{p=0}^\infty \frac{(2p-1)!!}{\lambda^{2p+1}(2p+1)(2p)!!}\left\|H^{2p+1}-\widetilde{H}^{2p+1} \right\| \nonumber
\end{align}
where $!!$ is the double factorial $z!! = z \cdot (z-2) \cdot (z-4) \cdots 1$ assuming $z$ is a natural number. It is straightforward to show inductively that for any $p>0$ 
\begin{align}
\left\|H^{2p+1}-\widetilde{H}^{2p+1} \right\|
\le \left(2p+1\right)\left(\max\left\{\left\|H\right\|,\left\|\widetilde{H}\right\|\right\}\right)^{2p}\left\|H - \widetilde{H}\right\|.\label{eq:inductive}
\end{align}
We then have from~\eq{defnsw} that
\begin{align}
\label{eq:term1}
\left\|H - \widetilde{H}\right\|
& \le \sum_{\ell=0}^{L-1}\left |w_\ell- \widetilde{w}_{\ell} \right|
\le L\delta.
\end{align}
We further have that
\begin{equation}
\max\left\{\left\|H\right\|,\left\|\widetilde{H}\right\|\right\}
\le \left\|H\right\| + L \delta .
\end{equation}
Substituting these equations into \eq{epsbnd} then gives
\begin{align}
\epsilon_{\textsc{prep}} &\le \sum_{p=0}^\infty \frac{(2p-1)!!}{\lambda^{2p+1}(2p)!!}(\left\|H\right\| + L \delta)^{2p}L\delta \nonumber \\
&= \frac{L\delta}\lambda\sum_{p=0}^\infty \frac{(2p-1)!!}{(2p)!!}\left(\frac{\left\|H\right\| + L \delta}\lambda\right)^{2p} \nonumber \\
&= \frac{L\delta}\lambda \left[1-\left(\frac{\left\|H\right\| + L \delta}\lambda\right)^2\right]^{-1/2}.
\end{align}
This inequality can be solved for $\delta$ to give
\begin{align}
\label{eq:delta_bound0}
\delta &\ge \frac{\epsilon_{\textsc{prep}}}{(1+\epsilon_{\textsc{prep}}^2)L}\left( \sqrt{\lambda^2(1+\epsilon_{\textsc{prep}}^2)-\left\|H\right\|^2}-\epsilon_{\textsc{prep}}\left\|H\right\| \right)\nonumber \\
&\ge \frac{\epsilon_{\textsc{prep}}\lambda}{(1+\epsilon_{\textsc{prep}}^2)L}\left( 1-\left\|H\right\|^2/\lambda^2\right) .
\end{align}
If we require that $\epsilon_{\textsc{prep}}\le \sqrt{2}\Delta E/(4\lambda)$ as in~\eq{errdist}, then this can be obtained by choosing
\begin{equation}
\label{eq:delta_bound}
\delta = \frac{\sqrt{2}\Delta E}{4 L\left( 1+\frac {\Delta E^2}{8\lambda^2}\right)}\left( 1-\left\|H\right\|^2/\lambda^2\right) .
\end{equation}
\end{document}